\documentclass{article}
\PassOptionsToPackage{numbers}{natbib}

 \usepackage[eandd, final]{neurips_2026}

\usepackage[utf8]{inputenc} 
\usepackage[T1]{fontenc}    
\usepackage{hyperref}       
\usepackage{url}            
\usepackage{booktabs}       
\usepackage{amsfonts}       
\usepackage{nicefrac}       
\usepackage{microtype}      
\usepackage[table]{xcolor}  
\usepackage{graphicx}
\usepackage{subcaption}
\usepackage{amsmath}
\usepackage{amssymb}
\usepackage{mathtools}
\usepackage{amsthm}
\usepackage{ragged2e}
\usepackage{tabularx}
\usepackage{array}
\usepackage{colortbl}
\usepackage{makecell}
\usepackage{threeparttable}
\usepackage[most]{tcolorbox}
\usepackage{multirow}
\usepackage{adjustbox}
\usepackage{siunitx}
\usepackage{fvextra}
\usepackage{soul}
\usepackage{float}
\usepackage{pifont}
\usepackage{wrapfig}
\usepackage{fontawesome5}

\newcommand{\modelname}{\texttt{RVCBench}}
\newcommand{\vc}{\texttt{VCL}}
\newcommand{\substd}[2]{%
  \ensuremath{%
    #1\mkern1mu_{\scriptscriptstyle\pm #2}%
  }%
}
\newcommand{\cmark}{\ding{51}}
\newcommand{\pmark}{\(\triangle\)}
\newcommand{\na}{--}

\definecolor{RVCLinkBlue}{HTML}{1A4E8A}
\hypersetup{
    colorlinks=true,
    urlcolor=RVCLinkBlue,
    linkcolor=RVCLinkBlue,
    citecolor=RVCLinkBlue
}

\usepackage{ragged2e}   

\definecolor{insightbg}{HTML}{F5F8F5}
\definecolor{insightborder}{HTML}{364B3E}

\newtcolorbox[auto counter]{insightbox}{%
  enhanced,
  colback=insightbg,
  frame hidden,
  borderline west={1.4pt}{0pt}{insightborder},
  boxrule=0pt,
  arc=1pt,
  outer arc=1pt,
  boxsep=0pt,
  left=7pt,
  right=7pt,
  top=4pt,
  bottom=4pt,
  before skip=5pt,
  after skip=5pt,
  width=\linewidth,
  fontupper=\normalfont
}

\newcommand{\Insight}[1]{%
\begin{insightbox}
\textbf{\color{insightborder}Takeaway~\thetcbcounter.}~#1
\end{insightbox}
}

\newtcolorbox{PromptBox}[1]{
  colback=gray!5!white,
  colframe=gray!75!black,
  fonttitle=\bfseries,
  title=#1,
  arc=4pt,
  outer arc=4pt,
  breakable,          
  left=6pt, right=6pt 
}

\newcommand{\titlelogo}{%
  \raisebox{-0.40em}{\includegraphics[height=1.6em]{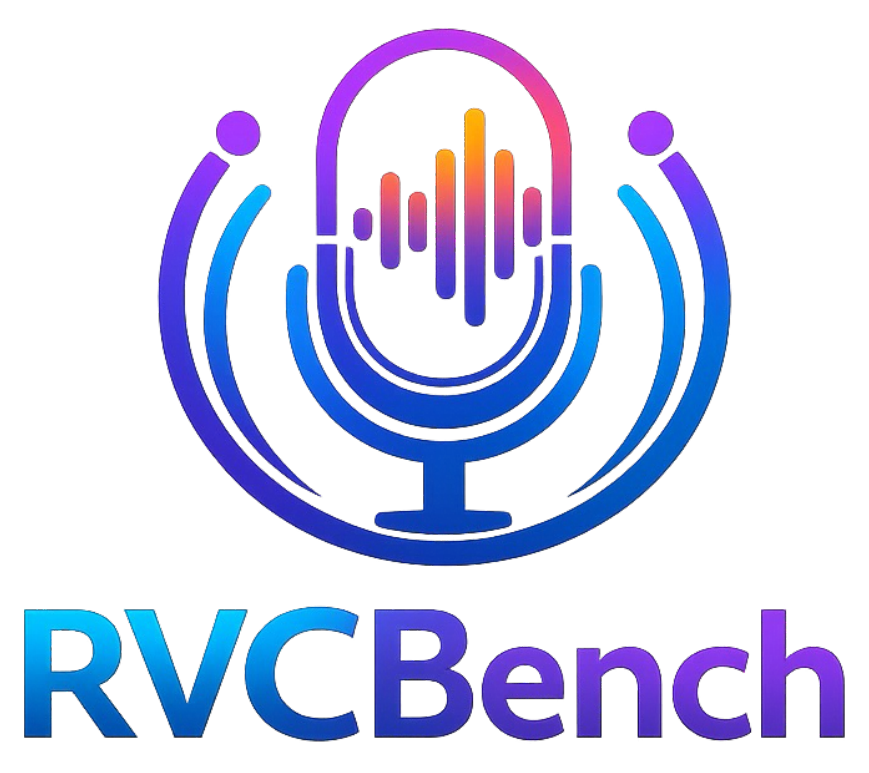}}%
  \hspace{0.6em}%
}
\title{\titlelogo RVCBench: Benchmarking Robustness of Voice Cloning Across Modern Audio Generation Models}

\author{%
  Ruinan Jin\textsuperscript{*,1,2}
  \And
  Xinting Liao\textsuperscript{*,1,2}
  \And
  Hanlin Yu\textsuperscript{1}
  \And
  Deval Pandya\textsuperscript{2}
  \And
  Xiaoxiao Li\textsuperscript{1,2,$\dagger$} \\
  \\
  \textsuperscript{1}The University of British Columbia \\
  \textsuperscript{2}Vector Institute \\
  \texttt{ruinanjin@alumni.ubc.ca}
}

\begin{document}

\maketitle

\begingroup
\renewcommand{\thefootnote}{\fnsymbol{footnote}}
\footnotetext[1]{Equal contribution.}
\footnotetext[2]{Corresponding author.}
\endgroup

\begin{abstract}
  Modern \textbf{V}oice \textbf{Cl}oning (\vc) can synthesize speech that closely matches a target speaker from only seconds of reference audio, enabling applications such as personalized speech interfaces and dubbing. In practical deployments, modern audio generation models inevitably encounter noisy reference audios, imperfect text prompts, multilingual and long-form generation settings, downstream post-processing, and adversarial perturbations, all of which can significantly hurt robustness. Despite rapid progress in \vc{} driven by autoregressive codec-token language models and diffusion-based models, robustness under realistic deployment shifts remains underexplored. This paper introduces \modelname, a comprehensive dataset and benchmark that evaluates \textbf{R}obustness in \textbf{V}oice \textbf{C}lone. \modelname{} contributes a large-scale, task-aligned robustness dataset that instantiates realistic deployment shifts through controlled text-audio pairing, multilingual and long-form scenarios, expressive prompts, post-processing conditions, and passive or proactive audio perturbations. Covering 18 robustness evaluations, 204 unique speakers, and 14,370 utterance-level evaluation items, \modelname{} enables unified evaluation of input sensitivity, generation stability, output resilience, and perturbation robustness. We evaluate 18 representative modern open-source \vc{} models and reveal systematic vulnerabilities in content consistency, speaker similarity, long-form stability, post-processing resilience, adversarial robustness, and detector-facing separability. We open-source the \href{https://github.com/Nanboy-Ronan/RVCBench}{toolkit} and \href{https://huggingface.co/datasets/Nanboy/RVCBench}{dataset} to support reproducible evaluation and future research.
\end{abstract}


\begin{center}
\small
\textbf{Resources:}
\href{https://nanboy-ronan.github.io/RVCBench/}{\faGlobe\ Project Page}
\quad|\quad
\href{https://github.com/Nanboy-Ronan/RVCBench}{\faGithub\ GitHub Repository}
\quad|\quad
\href{https://huggingface.co/datasets/Nanboy/RVCBench}{\raisebox{-0.18em}{\includegraphics[height=1.1em]{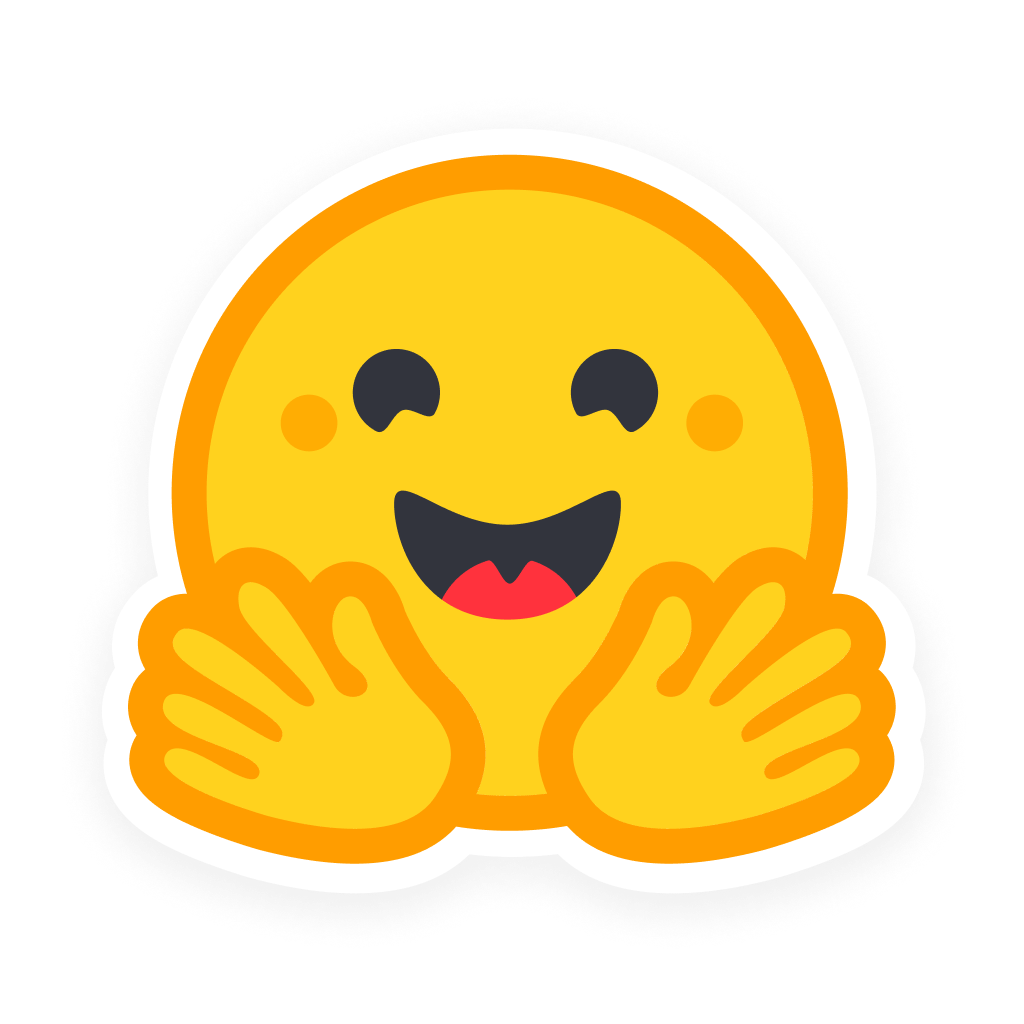}}\ Hugging Face Dataset}
\end{center}

\section{Introduction}
Modern voice cloning (\vc) can synthesize speech that closely resembles a target
speaker from only a few seconds of reference audio. This capability enables applications such as personalized speech interfaces, accessibility tools, and movie dubbing, while also creating risks of impersonation and
fraud~\citep{azzuni2025voice}. Recent progress in audio generation has substantially improved \vc{} quality, driven by autoregressive codec-token language models~\citep{liao2024fish,du2024cosyvoice,higgsaudio2025}, diffusion and flow-based generators~\citep{huynh2025ozspeech,li2023styletts,chen2025f5}, and hybrid systems that combine language-model planning with acoustic
generation~\citep{du2024cosyvoice,cui2025glm}.

Despite this progress, most evaluations still emphasize clean, short-form, and utility-oriented settings. Practically, \vc{} systems face diverse deployment shifts: reference recordings may contain noise or interfering speakers; prompts may be long, irregularly formatted, or multilingual; generated audio may be compressed or re-encoded; and anti-cloning perturbations may be applied to preserve speeches. These shifts make \emph{robustness} a central yet underexplored requirement for modern \vc{}.

\begin{figure}
  \centering
  \includegraphics[width=0.9\linewidth]{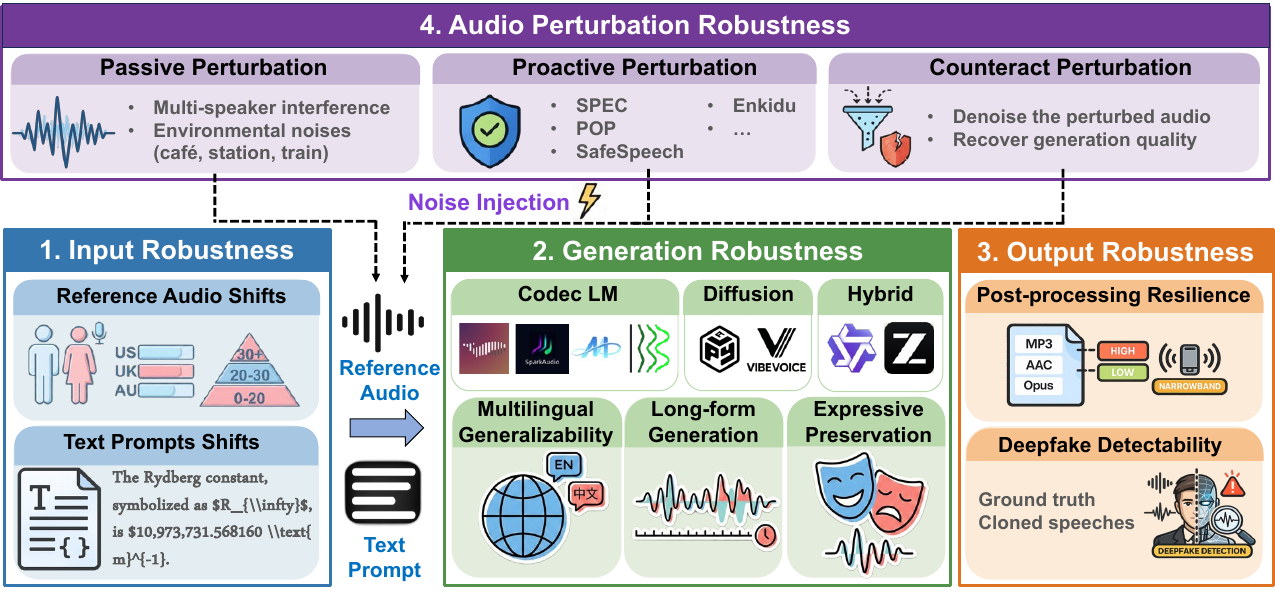}
   \caption{\textbf{Overview of \modelname.} We benchmark robustness on \vc{} across the full generation pipeline along four robustness dimensions, i.e., inputs, generation, outputs, and audio perturbations.}
   \label{fig:main}
   \vspace{-4mm}
\end{figure}

Robustness in \vc{} is distinct from robustness in generic text-to-speech or audio generation. A \vc{} system must preserve textual content, speaker identity, naturalness, and safety-relevant behavior across the full generation
pipeline. Failures may arise from input shifts, generation stressors such as long-form or expressive synthesis, output transformations such as compression, or security-relevant perturbations and countermeasures.

Existing benchmarks do not fully capture this pipeline-level robustness problem. \vc{} benchmarks such as CloneEval~\citep{christop2025cloneval} standardize model comparisons, but primarily under clean or curated conditions and with metrics focused on generation quality. Conversely, general audio robustness and watermarking benchmarks such as AHa-Bench~\citep{cheng2025aha} and AudioMarkBench~\citep{liu2024audiomarkbench} study broader audio-generation or security questions, but are not designed around reference-conditioned speaker preservation. Together, these gaps leave open a fundamental question:

\textbf{\emph{How robust are modern \vc{} systems when realistic deployment
shifts occur across the full generation pipeline?}}

To answer this question, we introduce \textbf{\modelname}, a benchmark for evaluating robustness in voice cloning. As shown in Fig.~\ref{fig:main}, \modelname{} covers four pipeline dimensions: \emph{input robustness},
\emph{generation robustness}, \emph{output robustness}, and \emph{audio-perturbation robustness}. It comprises 18 robustness evaluations across 10 tasks, 204 unique speakers, and 14,370 utterance-level evaluation items. We evaluate 18 modern
open-source \vc{} models spanning autoregressive codec-token models, diffusion/flow-based models, and hybrid language-model-based speech generation
systems.

Our empirical study shows that current \vc{} systems remain fragile under realistic deployment shifts. Reference-audio and text-prompt variations degrade content accuracy and speaker fidelity; multilingual, long-form, and expressive
generation expose failures in identity preservation, prosody, and emotion rendering; compression and narrowband transmission reduce perceptual quality; and passive noise or proactive anti-cloning perturbations substantially reduce cloning performance. These results show that robustness is not merely a by-product of high-quality voice cloning, but a distinct property that must be evaluated explicitly.

Our contributions are fourfold.
\textbf{(1)} We formulate \vc{} robustness as a pipeline-level evaluation problem and introduce \modelname, a unified benchmark covering input, generation, output, and audio-perturbation robustness. \textbf{(2)} We construct a task-aligned robustness dataset by reprocessing public speech corpora into controlled \vc{} stress tests, including re-paired reference audio and target prompts to reduce reliance on canonical corpus examples. \textbf{(3)} We provide a broad empirical study of 18 modern \vc{} models, identifying systematic failure modes that can guide the development of more reliable, robust, and safer voice-cloning systems. \textbf{(4)} We release our \href{https://nanboy-ronan.github.io/RVCBench/}{project page}, \href{https://github.com/Nanboy-Ronan/RVCBench}{code}, and \href{https://huggingface.co/datasets/Nanboy/RVCBench}{dataset}.

\section{Related Works}


\paragraph{Benchmarking \vc{} and audio robustness.}
Existing benchmarks fall into two threads, neither addressing \vc{} robustness \emph{as a pipeline problem}.
\textit{\vc{} and speech-synthesis benchmarks} (e.g., CloneEval~\citep{christop2025cloneval}, EmergentTTS-Eval~\citep{manku2025emergenttts}) standardize mostly curated, quality-centric evaluation; CloneEval covers limited VC models and metrics, while EmergentTTS-Eval targets \emph{TTS} prompt-following rather than reference-conditioned identity preservation. Recently, Song et al.~\citep{song2025degrading} surveyed robustness in voice conversion, underscoring the need to systematically study robustness in modern audio generation–based voice cloning models. \textit{General audio robustness benchmarks} (e.g., AHa-Bench~\citep{cheng2025aha}, AudioWatermarkBench~\citep{liu2024audiomarkbench}) study hallucination or watermark robustness, but omit VC-specific constraints and stressors (reference shifts, long-form identity drift, post-processing, defenses).
Hence, existing work evaluates \vc{} without robustness or robustness without \vc{}. The detailed related work can be found in Appendix~\ref{app:related}.
\section{RVCBench}

\subsection{Benchmark Overview}
\noindent \textbf{Dataset.}
\modelname{} constructs a robustness-oriented benchmark dataset for \vc as shown in Fig.~\ref{fig:main}, addressing the gap between clean-setting \vc{} benchmarks and general audio robustness benchmarks that omit \vc{}-specific stressors such as reference-audio shifts, post-processing, and proactive perturbations. Built from eight public sources, \modelname{} reprocesses them into 18 robustness evaluations across four dimensions, covering 10 tasks, 204 unique speakers, and 14,370 utterance-level evaluation items. Details are introduced in this section later and in Appendix~\ref{app_sec:datasets}. To ensure consistent evaluation and mitigate leakage, we standardize formats, task-specific text-audio pairing, controlled perturbations, and avoid reusing canonical pairs by re-pairing references and prompts while sampling LibriTTS from its original test split. Tasks and dataset construction details are summarized in Table~\ref{tab:rovbench_tasks} and Table~\ref{app_tab:dataset-metadata}.

\noindent \textbf{Voice clone models.}
We operationally include open-source systems that accept target text and a reference waveform and generate speech conditioned on the reference speaker. This scope covers both native zero-shot \vc{} models and adjacent reference-conditioned speech-generation systems, while applying the same text--reference interface and evaluation protocol to every backend. We organize them into three generative families: \ul{\textit{(1) autoregressive codec/speech-token models}}, including FishSpeech~\citep{liao2024fish}, XTTS~\citep{casanova2024xtts}, Spark-TTS~\citep{wang2025spark}, IndexTTS~\citep{deng2025indextts}, MOSS-TTSD v0.7~\citep{zhang2026moss}, and Higgs Audio~\citep{higgsaudio2025}; \ul{\textit{(2) non-autoregressive diffusion/flow and cloning-pipeline models}}, including MaskGCT~\citep{wang2024maskgct}, StyleTTS2~\citep{li2023styletts}, F5-TTS~\citep{chen2025f5}, OZSpeech~\citep{huynh2025ozspeech}, OpenVoice~\citep{qin2023openvoice}, PlayDiffusion, VibeVoice~\citep{peng2025vibevoice}, and ZipVoice; and \ul{\textit{(3) hybrid LM+diffusion/flow models}}, including CosyVoice~2~\citep{du2024cosyvoice}, Qwen3-TTS~\citep{hu2026qwen3}, MGM-Omni~\citep{wang2025mgm}, and GLM-TTS~\citep{cui2025glm}. We evaluate these 18 representative systems across the three families. The appendix additionally reports four extension systems---Fish Audio S2, MOSS-TTS v1.5, dots.tts, and Higgs TTS 3---when complete results are available; further supported systems are listed in Table~\ref{tab:model_comparison} and implemented in our codebase.

\noindent \textbf{Metrics.}
Following common practice in \vc~\citep{higgsaudio2025,cui2025glm,zhang2025safespeech}, we evaluate generation quality from five aspects. \textit{Speaker identity} is measured by Speaker Similarity (SIM), \textit{perceptual naturalness} by Mean Opinion Score (MOS)~\citep{reddy2021dnsmos}, \textit{spectral consistency} by Mel-cepstral Distortion (MCD), \textit{content accuracy} by Word Error Rate (WER), and \textit{inference efficiency} by Real-Time Factor (RTF). Unless otherwise specified, higher SIM and MOS indicate better performance ($\uparrow$), while lower WER, MCD, and RTF are preferred ($\downarrow$). For generation-quality metrics, we report the bootstrap standard error of the mean from 1,000 speaker-level cluster-bootstrap resamples, drawing speakers with replacement and retaining all utterances from each sampled speaker; single-speaker accent groups are descriptive and have no between-speaker uncertainty estimate. Detailed values are reported in Appendix~\ref{app:detailed-results}. We also conducted human studies for MOS and emotion-related evaluation, and discuss how these metrics compensate each other in appendix~\ref{app:metrics} and~\ref{app:additional_results}.

\subsection{Input Robustness}

\noindent \textbf{Reference audio shifts.}
In the real-world deployment, \textit{reference audio shifts} across languages and demographics. 
We design \texttt{RVC-AudioShift}, which varies demographic characteristics of the reference audio to examine which speaker traits are best captured by existing \vc~models. 
Specifically, we assess performance using prompt audios with diverse demographic attributes from VCTK~\citep{yamagishi2019cstr}, spanning 12 accents (ranging from widely represented varieties, e.g., American, to underrepresented ones, e.g., Indian and South African), gender (male and female), and age (from under 20 to over 30).

\noindent \textbf{Text prompt shifts.}
In practical \vc~applications, user queries exhibit variations in both \textit{format} and \textit{content}.
Irregular formatting (e.g., typos or mixed-language fragments) can induce hallucination by disrupting text–audio alignment, while content variations may involve \textit{scam-related} user intent.
Motivated by these observations, we propose \texttt{RVC-TextShift}, which comprises two subsets of prompts: (1) \textit{hallucination} prompts generated by prompting an LLM to introduce realistic format irregularities, and (2) \textit{scam-content} prompts constructed from robocall scripts from a public dataset\footnote{\url{https://github.com/wspr-ncsu/robocall-audio-dataset}}.
Using \texttt{RVC-TextShift}, we evaluate robustness to text prompt shifts by measuring the maintenance of semantic and acoustic consistency.


\subsection{Generation Robustness}

\noindent \textbf{Long-form generation.}
It is crucial to know whether the model can sustain coherent and stable synthesis when either the input text or the reference audio (or both) is lengthy, without accumulating errors or drifting in speaker identity~\citep{zhang2026moss}.
We further evaluate \textit{long-context robustness} (\texttt{RVC-LongContext}) in two settings: \textit{(1) Long-text}, which uses extended text prompts and requires sustained generation over long context, and \textit{(2) Long-audio}, which uses long reference audios that involve richer speaker cues. We measure how performance changes with increasing text length and reference duration, and report both content accuracy and speaker similarity to capture performance change over long spans.

\noindent \textbf{Expressive preservation.}
\texttt{RVC-Expression} evaluates a \vc~model’s ability to preserve the target speaker identity while faithfully rendering paralinguistic attributes, e.g., emotion, emphasis, and speaking style, inferred in the text prompt.
This setting reflects practical voice interfaces (e.g., movie dubbing) and also covers high-risk scenarios such as telephone fraud. 
We construct \texttt{RVC-Expression} by pairing VCTK speakers with robocall-style prompts from an online telephone-fraud dataset and augmenting the prompts with explicit style cues. 
In addition to general metrics of generation quality, we primarily assess \vc~models 
on emotion adherence between output audio and text prompts to determine whether they can successfully infer prosody while maintaining speaker consistency.

\noindent \textbf{Multilingual generalizability.}
Recent progress in audio generation models improves their generalizability across speech-related tasks~\citep{higgsaudio2025,manku2025emergenttts,wang2025mgm}.
%
 %
In \texttt{RVC-Multilingual}, we test whether \vc{} models can reliably synthesize speech under two types of language shifts, i.e., (1) \emph{single-language} \vc~ containing \texttt{English-VC} built from VCTK and LibriTTS, \texttt{Chinese-VC} built from AISHELL-1~\citep{bu2017aishell}, and \texttt{French-VC} built from Common Voice FR, and (2) \emph{cross-lingual} \vc{} instantiated with the EMIME~\citep{wester2011emime} bilingual English-Mandarin set, where the language of the reference audio may differ from that of the input text.
These settings probe whether models preserve speaker identity while maintaining target-language content consistency under language shifts. Due to the presentation limit and not all models support multilingual generalization, we mainly discuss the results of this in Appendix~\ref{app:additional_results}.

\subsection{Output Robustness}
\noindent \textbf{Post-processing resilience.} 
In practical deployment, cloned speech is often subject to ubiquitous post-processing (e.g., compression and re-encoding), which can degrade both semantic intelligibility and speaker-related acoustic information such as timbre and prosody.
To explore post-processing resilience, \texttt{RVC-Compression} transforms cloned speech with a realistic encode--decode pipeline under seven fixed conditions: MP3 at 32/64 kbps, AAC at 32/64 kbps, Opus at 16/24 kbps, and a telephone narrowband simulation~\citep{ren2023speaking}. Codec and bitrate are not fully crossed, so this setting evaluates the specified deployment transforms rather than separately identifying codec and bitrate effects.
%
%
Then, we compute the generation quality of speeches after compression, which verifies whether the cloned speeches maintain acoustic information.

\noindent \textbf{Deepfake detectability.}
As \vc~fidelity improves and synthetic artifacts become less apparent, distinguishing cloned speech from real utterances becomes increasingly challenging.
To quantify detectability, we introduce \texttt{RVC-Detectability} as a safety-oriented detector-based diagnostic, rather than a direct measure of output robustness. It covers 300 ground truth utterances of VCTK, 100 benign cloned speeches with text from VCTK, and 200 scam cloned speeches with text from robocall~\citep{prasad2023robocall}.
We assess the separability between ground truth audios and cloned speeches using four families of zero-shot deepfake detectors, e.g., SpeechLLM-as-Judges models (SQ-LLM)~\citep{wang2025speechllm}.
%
%

\subsection{Audio Perturbation Robustness}

\noindent \textbf{Passive perturbation.}
In real-world scenarios, reference audio inevitably contains ambient noise and multi-speaker interference, which hinders \vc~models from reliably extracting speaker-specific acoustic features.
%
 %
To simulate the scenarios and measure robustness under passive perturbations, \texttt{RVC-PassiveNoise} firstly mixes VCTK utterances with 10 types of environmental noises (e.g., cafeteria, car, and kitchen) from VoiceBank+DEMAND~\citep{valentini2017noisy}.
\texttt{RVC-PassiveNoise} contains 800 multi-speaker mixtures, where clean target utterances are mixed with interfering speech at four SNR levels ($-5$, $0$, $+5$, and $+10$~dB).
We then evaluate \vc~outputs generated from the perturbed references to quantify how well each method preserves speaker identity and content fidelity under passive noise and interference.

\noindent \textbf{Proactive perturbation.}
To address deepfake misuse, we consider
speaker-side proactive defenses (e.g., Enkidu~\citep{feng2025enkidu}, POP~\citep{zhang2023mitigating}, SafeSpeech and its variant SPEC~\citep{zhang2025safespeech}) that add human-imperceptible perturbations to the reference audio to prevent unauthorized cloning.
\texttt{RVC-AdvNoise} evaluates whether the \vc~methods can attack these methods.
First, we apply the proactive defenses to clean LibriTTS reference utterances and compute perceptual metrics between protected and original audio. We additionally evaluate i.i.d. Gaussian noise with standard deviation $\epsilon=0.03137$ as a non-optimized perturbation baseline.
Next, similar to passive perturbation, we assess the generation quality of \vc~methods to reflect their attack effectiveness in \texttt{RVC-AdvNoise}. 

\noindent \textbf{Counteract perturbation.}
Though the adversary perturbation can be used to defend against cloning attacks, several denoising methods are being developed for removing them.
In \texttt{RVC-AntiProtect}, we purify the protected ground truth utterances with adversarial perturbation by DEMUCS\citep{defossez2020real}.
DEMUCS is widely used for removing a variety of background noise, i.e.,  stationary and non-stationary noises, and room reverb.
By using denoised speech as the reference, we re-evaluate generation quality to measure how effectively the target speaker's identity is protected against denoising-based countermeasures.

\section{Results}
This section summarizes key empirical findings from evaluating modern \vc{} models on \modelname{}. Each subsection opens with a concise takeaway and representative observations, while more detailed analyses and per-model, per-task, and metric-level results are deferred to the appendix due to space constraints.

  

\begin{figure}[htbp]
    \centering
    \vspace{-10pt}
\includegraphics[width=\linewidth]{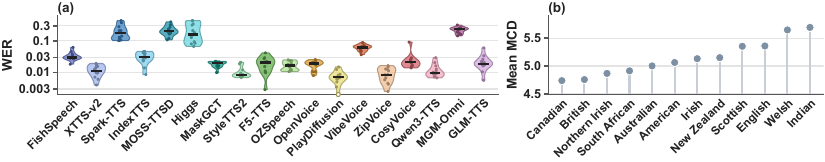}
    \caption{\textbf{Performance over various accents.} (a) WER distribution of each model over accents (violin, log scale; dots: accents, bar: median, hollow: WER$=$0); (b) MCD across accents.}
    \label{fig:acc}
    \vspace{-10pt}
\end{figure}

\subsection{Degradation under Input Shifts}


\Insight{
Reference audios influence generation
quality via varying accents, while text shifts mainly negatively impact the content accuracy.
}




\begin{figure}[htbp]
    \centering
    \includegraphics[width=\linewidth]{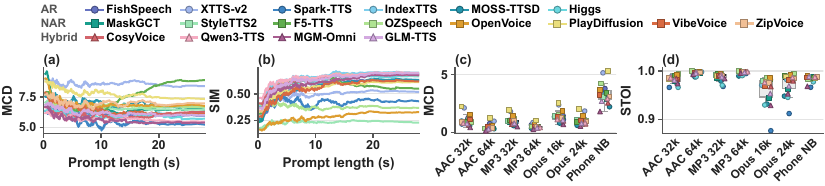}
    \caption{\textbf{Prompt length and compression.} (a) MCD and (b) SIM versus prompt audio length on LibriTTS; (c) MCD and (d) STOI after each codec.}
    \label{fig:prompt-length}
\end{figure}

\paragraph{Accent-conditioned results vary within the evaluated VCTK subset.}
Results on \texttt{RVC-AudioShift} show descriptive variation across accent-labelled groups. As shown in Fig.~\ref{fig:acc} (a), WER varies across models and accents, while Fig.~\ref{fig:acc} (b) shows corresponding variation in spectral distortion. Within this subset, Indian-accented references have the highest average MCD and Canadian-accented references have a lower average MCD. Because the Australian, British, New Zealand, and Welsh groups each contain only one speaker, their means are descriptive and do not support population-level accent rankings. These results identify candidate robustness gaps that require larger, speaker-balanced samples.

\paragraph{Text-prompt shifts compromise content consistency.}
  
\begin{figure}[htbp]
    \centering
    \includegraphics[width=\linewidth]{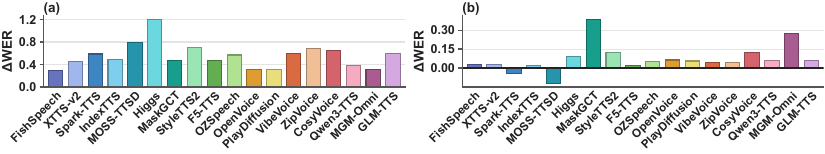}
    \caption{\textbf{Performance under text-prompt shifts.} Relative WER changes compared with standard VCTK prompts under (a) hallucination-style prompts and (b) scam-style prompts.}
    \vspace{-10pt}
    \label{fig:text_shift}
\end{figure}
In \texttt{RVC-TextShift}, we replace standard VCTK prompts with hallucination-style and scam-related prompts to examine whether \vc{} models remain content-consistent under text shifts. Fig.~\ref{fig:text_shift} reports the relative WER change against the original VCTK prompts. Hallucination-style prompts consistently increase WER for most models, with several models showing large degradation, indicating that formatting irregularities and unusual tokens can disrupt text-to-speech alignment. Scam-related prompts lead to smaller but still noticeable changes, suggesting that content-domain shifts also affect robustness, though less severely than hallucination-style perturbations. Overall, current \vc{} models remain sensitive to unconventional prompt formats and out-of-distribution text content, exposing a key weakness in content preservation.

\subsection{Instability in Diversified Generation Tasks}
\Insight{
\vc{} models are generally not robust to language shifts, long-form text prompts, and emotionally expressive prompts.
}

\paragraph{Audio length variations can impact the generation, and long texts break the robustness of content accuracy.}
In the long-form context, we evaluate \texttt{LongAudio} and \texttt{LongText} separately. \emph{First}, in \texttt{LongAudio}, the audio length variations proportionally impact the generation quality. As shown in Fig.~\ref{fig:prompt-length} (a) and (b), both metrics change most rapidly in the short-prompt regime, e.g., MCD (Fig.~\ref{fig:prompt-length} (a)) decreases sharply as prompt length increases from a few seconds to roughly 8-12 seconds, while SIM (Fig.~\ref{fig:prompt-length} (b)) rises accordingly. 
Beyond this range, the margin of performance gains converges, indicating diminishing returns from additional reference audio.
Model sensitivity, however, differs: stronger backends (e.g., CosyVoice, Higgs, and MGM-Omni) continue to benefit with higher SIM and lower MCD, whereas others show limited SIM improvements even with long prompts (e.g., StyleTTS2 and OZSpeech). 

\begin{figure}
    \centering
    \includegraphics[width=\linewidth]{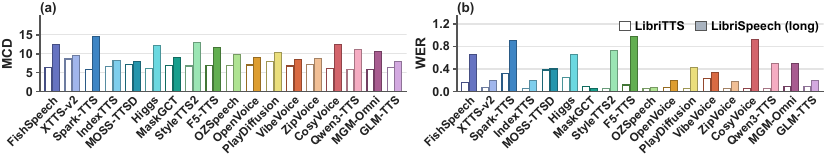}
    \caption{\textbf{Performance on long-text generation.} (a) MCD and (b) WER on LibriTTS and long-form LibriSpeech.}
    \label{fig:long_text}
    \vspace{-10pt}

\end{figure}

\emph{Second}, for the \texttt{LongText}, we observe a big degradation in both content and spectral consistency as sequence length increases. As shown in Fig.~\ref{fig:long_text}, WER rises markedly across all backends (Fig.~\ref{fig:long_text} b), indicating that long-form synthesis makes it harder to preserve textual content without omissions or substitutions. In parallel, MCD also increases (Fig.~\ref{fig:long_text} a), suggesting accumulating acoustic drift and reduced spectral stability over extended generations. The degradation is more pronounced on LibriSpeech than LibriTTS for most models, consistent with a harder long-form setting. While models such as CosyVoice and MGM-Omni remain relatively more stable, no model is immune to the long-text effect, highlighting that maintaining both content fidelity and timbral consistency over long horizons remains a key challenge for \vc{} backends.


\begin{wrapfigure}{r}{0.6\textwidth}
  \centering
  \vspace{-10pt}
  \includegraphics[width=\linewidth]{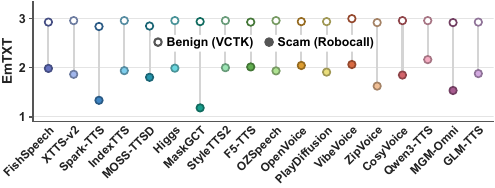}
  \vspace{-8pt}
  \caption{\textbf{Performance on audio-text emotion alignment on \texttt{RVC-Expression}.}}
  \label{fig:expressive}
  \vspace{-10pt}
\end{wrapfigure}

\paragraph{Expressive emotion transfer is unreliable under domain shifts like scams, showing lower alignment and higher variance.}
In Fig.~\ref{fig:expressive}, we examine expressive fidelity by measuring the emotional alignment between the input text and the synthesized speech using Audio LLM-as-Judge, where we prompted Gemini-3-flash to score if the generated audio aligns with the emotion from the text prompt.
Overall, models receive higher EmotionAlignment scores (higher points) on benign prompts (VCTK) than on scam prompts, despite scam texts typically carrying stronger and more explicit affect (e.g., urgency, fear, pressure, or reassurance) designed to elicit an immediate reaction.
From a robustness perspective, this is a clear failure mode. Under a domain shift to emotionally salient content, \vc{} systems do not reliably express the intended emotion into prosody and speaking style. 
The scam condition also shows a larger variance across models, suggesting that expressive rendering is unstable.

\subsection{Sensitivity over Perceptual in Outputs}  
\Insight{
Compression and narrowband transmission lead generated audio to lose acoustic and intelligibility-related information. Complementary deepfake detection results show model-dependent separability between generated and real speech.
}




\paragraph{Compression distorts the spectral and speech consistency.}
We analyze perceptual robustness by comparing cloned speech against its reconstructed counterparts after different compression and transmission processes, and visualize the results of \texttt{RVC-Compression} in Fig.~\ref{fig:prompt-length} (c)-(d).
We focus on perceptual-oriented metrics, i.e., MCD and STOI, to quantify how signal-level distortions introduced by lossy codecs affect the preservation of spectral structure and speech intelligibility.
Across all models, compression consistently increases MCD and reduces STOI, indicating that perceptual degradation is inevitable once cloned speech undergoes aggressive signal processing.
\vc~models have the most noticeable sensitivity in spectral consistency after using narrow band transmission by Phone NB, while having a diverse change of perception after using different low-bitrate compressions.
Overall, different \vc~models exhibit heterogeneous sensitivity to compression, suggesting that perceptual robustness is strongly coupled with model architectures and vocoder designs rather than being a by-product of overall generation quality.
%


\begin{figure}
    \centering
    \includegraphics[width=\linewidth]{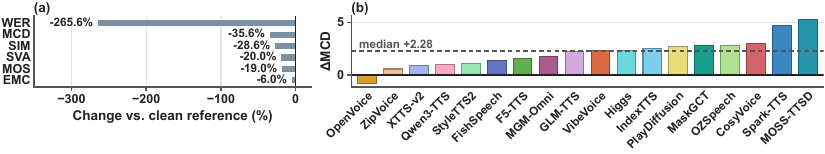}
    \caption{\textbf{Performance degradation from clean to audios with natural noises.} (a) Mean change of each metric vs.\ clean references (\%, negative is worse); (b) per-model change in MCD.}
    \label{fig:bgnoise_sig}
\end{figure}

\paragraph{Generated audio shows model-dependent detector-facing separability.}
\begin{table}[t]
\centering
\scriptsize
\setlength{\tabcolsep}{2.6pt}
\caption{\textbf{Detector-facing separability for the 18 representative main-paper models.} EER (\%, lower is more separable) is summarized as the median [first quartile, third quartile] across nine detectors. The IQR describes heterogeneity among detectors, not uncertainty about a voice-cloning model. Detector-wise results for all evaluated models are reported in the appendix.}
\label{tab:deepfake_m}
\resizebox{\linewidth}{!}{%
\begin{tabular}{lcccccc}
\toprule
Metric & FishSpeech & Spark-TTS & MOSS-TTSD v0.7 & Higgs & OZSpeech & StyleTTS2 \\
\midrule
EER median [IQR] $\downarrow$ & 8.00 [5.00, 13.67] & 36.33 [30.00, 42.33] & 37.00 [26.67, 45.67] & 23.00 [18.67, 33.67] & 0.67 [0.00, 1.00] & 11.67 [6.67, 22.00] \\
\bottomrule
\end{tabular}%
}
\vspace{0.35em}
\resizebox{\linewidth}{!}{%
\begin{tabular}{lcccccc}
\toprule
Metric & PlayDiffusion & VibeVoice & CosyVoice & MGM-Omni & GLM-TTS & F5-TTS \\
\midrule
EER median [IQR] $\downarrow$ & 1.33 [1.00, 3.33] & 22.67 [21.33, 37.67] & 34.00 [30.00, 41.33] & 34.67 [31.67, 52.00] & 31.67 [25.67, 43.00] & 12.33 [7.33, 23.00] \\
\bottomrule
\end{tabular}%
}
\vspace{0.35em}
\resizebox{\linewidth}{!}{%
\begin{tabular}{lcccccc}
\toprule
Metric & OpenVoice & Qwen3-TTS & MaskGCT & XTTS-v2 & IndexTTS & ZipVoice \\
\midrule
EER median [IQR] $\downarrow$ & 3.33 [0.67, 4.33] & 41.33 [28.67, 46.00] & 22.00 [15.33, 34.33] & 5.00 [1.33, 8.33] & 15.33 [12.67, 26.67] & 13.00 [7.67, 18.00] \\
\bottomrule
\end{tabular}%
}
\end{table}

We feed \texttt{RVC-Detectability} to state-of-the-art speech deepfake detectors, including 
(1) self-supervised learning-based detectors (e.g., XLSR-SLS~\citep{zhang2024audio}, Wav2Vec2-ECAPA, HuBERT-ECAPA, WavLM-ECAPA~\citep{kulkarni2024exploring}, and TCM-ADD~\citep{truong2024temporal}),
(2) graph neural network (GNN)-based detectors (e.g., RawGAT-ST~\citep{tak2021end} and AASIST~\citep{jung2022aasist}),
(3) convolutional neural network (CNN)-based detectors (e.g., RawNet2~\citep{tak2021end_raw}), and
(4) an audio-based LLM-as-a-judge, i.e., SQ-LLM~\citep{wang2025speechllm}.
We leverage nine detectors to assess detector-facing separability between ground-truth and generated speech using EER, minDCF, and ACC. 
Table~\ref{tab:deepfake_m} summarizes EER as the median and interquartile range across nine detectors for each of the 18 representative models. Detailed per-detector EER, minDCF, and ACC results are reported in Table~\ref{supp_tab:deepfake1}--\ref{supp_tab:deepfake3}.
Higher ACC and lower EER/minDCF indicate stronger separability under this evaluation protocol.
Table~\ref{tab:deepfake_m} shows that separability is model-dependent: OpenVoice, OZSpeech, and PlayDiffusion are more easily separated from real recordings, whereas CosyVoice, Spark-TTS, MGM-Omni, Qwen3-TTS, and GLM-TTS show lower average separability under conventional detectors.
We interpret detector-facing separability as a complementary safety signal rather than a direct robustness measure.
Combined with the compression results, it broadens output-side evaluation beyond information preservation to include safety-relevant distinguishability.

\subsection{Vulnerability in Perturbation and Counteraction}
\Insight{
Both passive and proactive perturbation prevent \vc~from cloning the acoustic fidelity, which cannot be easily counteracted by the denoisers.
}

\begin{figure}
    \centering
    \includegraphics[width=\linewidth]{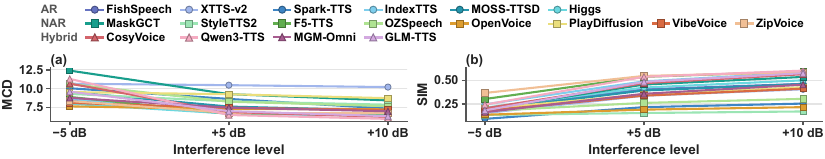}
    \caption{\textbf{Performance on multi-speaker interference.} (a) MCD. (b) SIM.}
    \label{fig:multi_db}
    \vspace{-4mm}
\end{figure}

\paragraph{Passive perturbation degrades the generation performance of some \vc~models.}
\modelname~injects natural background noise and multi-speaker interference at different decibel (dB) levels to evaluate how such \emph{passive perturbations} affect voice-cloning quality.
First, using \texttt{Per-BGNoise} as the reference audio, we compare cloning performance with clean versus noisy references in Fig.~\ref{fig:bgnoise_sig}, and further report the average performance gap across all models in Fig.~\ref{fig:bgnoise_sig}(a).
When background noise is introduced, many models struggle to preserve speaker-specific acoustic cues, resulting in a significant degradation in spectral consistency.
%
Notably, roughly half of the evaluated \vc~models are relatively robust to noisy references, showing insignificant change in WER compared to using clean references.
In contrast, the remaining models exhibit varying degrees of performance drop under noisy reference conditions.
We further investigate the impact of varying interference levels from other speakers on \vc{} performance, with the averaged results of multispeaker interferences in Fig.~\ref{fig:multi_db}. 
In multi-speaker scenarios, all evaluated models exhibit noticeable degradation in speaker identity, naturalness, and spectral consistency as the intensity of interfering increases.

%

\paragraph{Adversarial perturbation hinders the acoustic fidelity.} 
To evaluate the resilience against impersonation risks, \modelname~assesses \vc~performance when subjected to adversarial perturbations from state-of-the-art anti-clone methods, such as SafeSpeech~\citep{zhang2025safespeech} and Enkidu~\citep{feng2025enkidu}.  
The corresponding performance heatmaps for SIM is illustrated in Fig.~\ref{fig:adversial_and_protect} (a), while the MOS is presented in the Appendix~\ref{app:additional_results}.
%
%
%
Generally, Safespeech~\citep{zhang2025safespeech} and its variant SPEC demonstrate superior efficacy in disrupting \vc~outputs across most \vc~models.
%
%
In contrast, models such as MOSS-TTSD v0.7 and Spark-TTS lack resilience, failing significantly under both noisy and adversarial conditions.
These results suggest that simulating high-fidelity acoustic features remains a major challenge for \vc~models when faced with targeted adversarial interference.



\begin{figure}
    \centering
    \includegraphics[width=\linewidth]{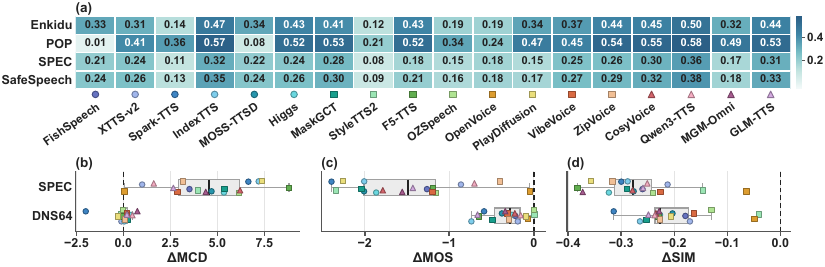}
    \caption{\textbf{Performance degradation by adversarial perturbation and anti-protection robustness.} (a) SIM under each perturbation (darker: better cloning). (b)-(d) MCD, MOS, and SIM change vs.\ clean references after SPEC and after DEMUCS denoising (DNS64); markers follow the keys under (a).}
    \label{fig:adversial_and_protect}
    \vspace{-10pt}

\end{figure}

\paragraph{\vc~models fail to recover acoustic fidelity after protection denoising.}
We use \texttt{RVC-AntiProtect} to quantify changes in generation quality when reference audios are first perturbed by the anti-cloning method SPEC and subsequently processed by a denoising module (DEMUCS~\citep{defossez2020real}), relative to clean reference conditions.
In Fig.~\ref{fig:adversial_and_protect} (b)-(d), we can find that the resulting performance, e.g., MCD, SIM, and MOS, has an evident degradation when referring to audios perturbed by SPEC.
Although denoising by DEMUCS slightly reduces distortion for some \vc~models, the resulting speech remains noticeably inferior to clean-reference generation, with negative shifts in SIM and MOS persisting in most cases.
%
%
%
Overall, these results indicate that anti-cloning defenses impose lasting degradation on acoustic fidelity, and that naive denoising is generally insufficient for protection evasion in current \vc~models.

\section{Conclusion}
In this work, we propose \modelname, a benchmark designed to evaluate the robustness of \vc{} against realistic deployment shifts. By analyzing 10 \vc~tasks with 14,370 utterances across 18 representative models, we uncover critical limitations, including content degradation under input shifts, generation bottlenecks in diverse synthesis settings, weak resilience to post-processing, and vulnerability to passive and proactive perturbations.
These findings suggest that future \vc{} systems should treat robustness as a first-class design objective, with training and evaluation covering noisy and multi-speaker references, OOD text formats, cross-lingual and long-form synthesis, expressive prosody transfer, compression artifacts, and perturbation-aware speaker representations. We hope \modelname{} provides a reproducible testbed for developing more reliable and benign \vc{} technology.

\bibliography{reference}
\bibliographystyle{plain}


\newpage
\appendix
\begingroup
\tableofcontents

\endgroup
\newpage

\section{Related Work}
\label{app:related}
\subsection{Voice clone}

\newcolumntype{L}{>{\raggedrighaudist\arraybackslash}X}
\newcolumntype{C}{>{\centering\arraybackslash}X}

\paragraph{Voice clone.} Voice cloning (\vc) aims to synthesize speech that preserves a target speaker's identity (e.g., timbre and speaker-specific characteristics) while accurately rendering new linguistic content.
Following the standardized terminology proposed by Azzuni and El~Saddik~\citep{azzuni2025voice}, we view \vc{} as an extension of TTS \emph{speaker adaptation} and distinguish:
(1) \emph{speaker adaptation} (fine-tuning a TTS system to a target speaker with limited data),
(2) \emph{few-shot \vc{}} (adaptation with very short reference audio, ranging from seconds to a few minutes),
and (3) \emph{zero-shot \vc{}} (no fine-tuning; the model conditions on a short reference clip via an explicit speaker representation such as a speaker embedding/encoder).
With the maturation of neural TTS, modern \vc{} increasingly targets \emph{zero-shot} cloning from only a few seconds of reference audio while emphasizing high-fidelity identity preservation, natural prosody, and controllability~\citep{azzuni2025voice}.
This paper focuses on benchmarking \emph{zero-shot} \vc{} models under realistic deployment conditions.
Representative system families include (1) sequence-to-sequence and codec/token-based pipelines that map text (and optionally reference speech) to discrete acoustic units followed by neural decoding, and (2) diffusion/flow-based approaches that synthesize mel or latent representations with improved naturalness and stability.
Despite strong progress, most prior \vc{} research and evaluation centers on standard benchmarks and clean reference conditions, leaving gaps in robustness to realistic shifts such as demographics, accents, background noise, multi-speaker interference, and downstream transformations.

\begin{table}[htbp]
\centering
\footnotesize
\setlength{\tabcolsep}{3.0pt}
\renewcommand{\arraystretch}{1.12}

\caption{\textbf{Comparison with prior VC-, TTS-, and audio-robustness evaluation benchmarks.}
Unlike prior resources, \textbf{\modelname{}} explicitly evaluates robustness across the full voice-cloning pipeline,
covering input shifts, challenging generation settings, output robustness, and perturbation-based protection settings.}
\label{tab:benchmark_comparison}

\begin{threeparttable}
\begin{tabularx}{\textwidth}{@{}>{\raggedright\arraybackslash}X *{9}{c}@{}}
\toprule
\multirow{2}{*}{\textbf{Benchmark}} &
\multicolumn{1}{c}{\textbf{Input}} &
\multicolumn{3}{c}{\textbf{Generation}} &
\multicolumn{2}{c}{\textbf{Output}} &
\multicolumn{3}{c}{\textbf{Perturbation}} \\
\cmidrule(lr){2-2}\cmidrule(lr){3-5}\cmidrule(lr){6-7}\cmidrule(l){8-10}
& Shift
& \makecell[c]{Multi-\\lingual}
& \makecell[c]{Long\\form}
& Express.
& Codec
& Detect.
& \makecell[c]{Passive\\noise}
& \makecell[c]{Anti-\\clone}
& Denoise \\
\midrule
\rowcolor{insightbg}
\textbf{\modelname{} (ours)} & \cmark & \cmark & \cmark & \cmark & \cmark & \cmark & \cmark & \cmark & \cmark \\
\rowcolor{gray!4}LibriSpeech-PC {\scriptsize\citep{meister2023librispeech}} & \na & \na & \na & \na & \na & \na & \na & \na & \na \\
ClonEval {\scriptsize\citep{christop2025cloneval}} & \na & \na & \na & \pmark & \na & \na & \na & \na & \na \\
\rowcolor{gray!4}VCC20 {\scriptsize\citep{zhao2020voice}} & \na & \cmark & \na & \na & \na & \pmark & \na & \na & \na \\
Seed-TTS {\scriptsize\citep{anastassiou2024seed}} & \na & \pmark & \na & \cmark & \na & \na & \na & \na & \na \\
\rowcolor{gray!4}CV3-Eval {\scriptsize\citep{du2025cosyvoice}} & \cmark & \cmark & \na & \cmark & \na & \na & \pmark & \na & \na \\
EmergentTTS-Eval {\scriptsize\citep{manku2025emergenttts}} & \pmark & \pmark & \na & \pmark & \na & \na & \na & \na & \na \\
\rowcolor{gray!4}InstructTTS-Eval {\scriptsize\citep{huang2025instructttseval}} & \pmark & \pmark & \na & \pmark & \na & \na & \na & \na & \na \\
TTSDS2 {\scriptsize\citep{minixhofer2025ttsds2}} & \pmark & \cmark & \na & \na & \na & \na & \na & \na & \na \\
\rowcolor{gray!4}ASVspoof/ADD {\scriptsize\citep{liu2023asvspoof,yi2022add}} & \na & \na & \na & \na & \pmark & \cmark & \pmark & \na & \na \\
AudioMarkBench {\scriptsize\citep{liu2024audiomarkbench}} & \na & \na & \na & \na & \pmark & \pmark & \cmark & \na & \na \\
\bottomrule
\end{tabularx}

\begin{tablenotes}[flushleft]
\scriptsize
\item \cmark\ indicates explicit benchmark coverage; \pmark\ indicates partial, analogous, or indirect coverage; \na\ indicates not explicitly covered.
\item Columns cover input-domain shifts, multilinguality, long-form generation, expressiveness, codec resilience, deepfake detectability, passive noise, anti-clone protection, and denoising counteraction. Partial marks outside VC denote related audio/TTS/security coverage rather than full reference-conditioned voice-cloning robustness.
\item For VCC20, the partial mark under \textit{deepfake detectability} reflects related spoofing/objective assessment of submitted voice-conversion systems rather than a dedicated deepfake-detection benchmark.
\end{tablenotes}
\end{threeparttable}

\vspace{-2mm}
\end{table}

\paragraph{Modeling paradigms.}
A useful lens to organize recent \vc{} systems is the \emph{generation paradigm} and where speaker identity is injected into the pipeline (e.g., speaker embedding vs.\ in-context prompting)~\citep{azzuni2025voice}. A notable recent trend is to integrate LLM capabilities directly into speech generation, where an \textit{audio LLM} or LLM-like backbone models speech tokens and aligns text and speech in a unified latent space; compared with classic acoustic-model pipelines, these systems often provide stronger instruction following, longer-context generation, and improved cross-lingual behavior while still supporting zero-shot \vc{} from short references. The model landscape summarized in Table~\ref{tab:model_comparison} spans several representative families. \emph{\ul{(1) Autoregressive codec-, speech-, and audio-token LMs}} generate discrete neural codec or speech tokens conditioned on text and, when supported, reference audio, then decode tokens to waveforms. This family includes early zero-shot neural codec LMs such as VALL-E~\citep{wang2023neural}, expressive and multilingual systems such as Seed-TTS~\citep{anastassiou2024seed}, FishSpeech~\citep{liao2024fish}, XTTS, Spark-TTS~\citep{wang2025spark}, and IndexTTS~\citep{deng2025indextts}, as well as dialogue or foundation-audio models such as MOSS-TTSD v0.7~\citep{zhang2026moss}, FireRedTTS-2~\citep{xie2025fireredtts}, and Higgs Audio~\citep{higgsaudio2025}. \emph{\ul{(2) Non-autoregressive, diffusion, flow-matching, and cloning-pipeline models}} synthesize masked semantic/acoustic tokens or continuous mel/latent features through iterative refinement or learned transport dynamics, improving stability, perceptual quality, and controllability~\citep{azzuni2025voice}; examples include MaskGCT~\citep{wang2024maskgct}, StyleTTS2~\citep{li2023styletts}, F5-TTS~\citep{chen2025f5}, OZSpeech~\citep{huynh2025ozspeech}, OpenVoice~\citep{qin2023openvoice}, PlayDiffusion\footnote{\url{https://github.com/playht/PlayDiffusion}}, VibeVoice for long-form conversational speech~\citep{peng2025vibevoice}, and VoxCPM~\citep{zhou2025voxcpm}. For benchmark inclusion, we require an executable target-text-plus-reference-waveform interface; thus this operational scope includes native zero-shot cloning systems as well as adjacent systems whose public inference stacks expose reference-conditioned generation. \emph{\ul{(3) Hybrid LM + diffusion/flow models}} combine token-level linguistic or coarse-acoustic planning with continuous waveform refinement for scalability, streaming synthesis, and production readiness, such as CosyVoice~2~\citep{du2024cosyvoice}, Qwen3-TTS~\citep{hu2026qwen3}, MGM-Omni~\citep{wang2025mgm}, and GLM-TTS~\citep{cui2025glm}. Table~\ref{tab:tts_model_summary_a} also includes adjacent proprietary instruction-conditioned TTS systems, such as GPT-4o mini TTS and Gemini-TTS, which are useful for contextualizing the broader speech-generation ecosystem but do not expose public reference-conditioned voice cloning in the same way as open \vc{} systems. At the same time, \emph{in-context} cloning via prompting (conditioning on reference audio without fine-tuning) lowers the barrier for fast personalization but can amplify misuse risks, motivating robustness evaluation beyond ``clean MOS/SIM'': modern audio-generation-based \vc{} should tolerate noisy/short references and multilingual or scam-like text, remain stable over long contexts with reduced identity drift, and be accountable under detection and defense mechanisms. Finally, as the voice-cloning model ecosystem continues to grow, it is impractical to exhaustively report all supported systems within the main paper. Under the paper's space constraints, we present results for 18 representative systems in the main text, while integrating the remaining models listed in Table~\ref{tab:tts_model_summary_a} into our toolkit to enable customized benchmark execution.

\paragraph{Evaluation gaps and robustness motivation.}
While prior work reports strong quality and similarity under curated benchmarks (e.g., MOS and speaker-similarity metrics)~\citep{azzuni2025voice},
real deployments face systematic shifts that are rarely stress-tested: noisy or extremely short references, fairness, multi-speaker reference contamination, accent and dialect mismatch, multilingual prompts, long-form generation with identity drift, and post-processing transformations (codec compression, resampling) that affect both perceptual quality and detectability.
The diversity of paradigms above also implies different failure modes (e.g., tokenization artifacts for codec-LMs, oversmoothing or style leakage for diffusion, streaming boundary artifacts for chunk-based hybrids)~\citep{jin2024fairmedfm}.
Our benchmark is designed to capture these practical robustness dimensions across modern zero-shot \vc{} model families rather than optimizing only for clean-condition MOS/SIM.

\paragraph{Related benchmarks for \vc{} and audio robustness.}
Existing evaluation resources related to our study broadly fall into three threads, but none answers the central question of \vc{} robustness as an \emph{end-to-end pipeline problem}. As summarized in Table~\ref{tab:benchmark_comparison}, prior benchmarks typically focus on one isolated aspect of the broader speech-generation ecosystem, such as transcription-format robustness, clean voice conversion or voice cloning, multilingual and expressive TTS, deepfake detection, or audio watermarking. In contrast, \modelname{} explicitly evaluates reference-conditioned \vc{} across the full generation pipeline, covering input shifts, challenging generation settings, output robustness, and perturbation-based protection and counteraction.

\textit{Input and transcription-oriented benchmarks} are useful for evaluating robustness of upstream speech or text processing, but they do not evaluate \vc{} itself. LibriSpeech-PC~\citep{meister2023librispeech}, for example, restores punctuation and capitalization annotations for LibriSpeech and introduces punctuation/capitalization evaluation for end-to-end ASR systems. While such resources are relevant to text-normalization and prompt-format issues, they do not involve reference-conditioned speech generation, speaker-identity preservation, or robustness to reference-audio shifts.

\textit{Voice conversion, voice cloning, and speech-synthesis benchmarks} provide cleaner comparisons of synthesis quality or speaker similarity, but they do not systematically stress-test \vc{} under realistic deployment shifts. ClonEval~\citep{christop2025cloneval} introduces an open benchmark, evaluation library, and leaderboard for voice-cloning models, evaluating systems such as OuteTTS, SpeechT5, VALL-E X, WhisperSpeech, and XTTS-v2 primarily through WavLM speaker-embedding cosine similarity, with additional acoustic-feature and emotion-wise analyses. However, its evaluation remains focused on automatic similarity assessment under curated settings, rather than corrupted references, prompt artifacts, multilingual transfer, long-form identity drift, codec/re-encoding, deepfake detectability, or proactive anti-clone defenses. VCC20~\citep{zhao2020voice} evaluates intra-lingual semi-parallel and cross-lingual voice conversion using a common challenge dataset and listening tests, making it relevant to cross-lingual speaker conversion. Nevertheless, it is not designed as a robustness benchmark for modern zero-shot \vc{} systems conditioned on short reference audio, nor does it cover output post-processing, passive/proactive perturbations, or denoising-based counteraction.

\textit{TTS and speech-generation benchmarks} broaden evaluation toward multilinguality, expressiveness, and instruction following, but they remain only partially aligned with \vc{} robustness. Seed-TTS~\citep{anastassiou2024seed} evaluates large-scale speech generation and in-context TTS capabilities with strong speaker similarity and naturalness, but it is primarily a model evaluation suite rather than a robustness benchmark. CV3-Eval~\citep{du2025cosyvoice} is closer to our setting because it targets zero-shot multilingual speech synthesis in the wild, including multilingual voice cloning, cross-lingual voice cloning, and emotion cloning with noisy real-world references. However, it does not provide a full pipeline robustness view covering long-form stability, post-processing resilience, forensic detectability, anti-clone perturbations, and denoising countermeasures. EmergentTTS-Eval~\citep{manku2025emergenttts} and InstructTTS-Eval~\citep{huang2025instructttseval} evaluate challenging prompt following, prosody, expressiveness, and natural-language style control for TTS systems using model-as-a-judge evaluation, while TTSDS2~\citep{minixhofer2025ttsds2} proposes resources and distributional metrics for comparing human-quality TTS systems across domains and languages. These benchmarks are valuable for assessing controllability and speech-generation quality, but they do not test the defining \vc{} constraint of preserving a target speaker identity from a short reference under realistic reference-audio, prompt, multilingual, long-context, and adversarial shifts.

\textit{Audio forensics, deepfake detection, and watermarking benchmarks} evaluate security and robustness in broader audio settings, but they are not designed to measure \vc{} generation robustness. ASVspoof 2021~\citep{liu2023asvspoof} and ADD 2022~\citep{yi2022add} focus on detecting spoofed, deepfake, or partially fake speech under increasingly realistic conditions, making them relevant to our forensic detectability setting. AudioMarkBench~\citep{liu2024audiomarkbench} systematically evaluates the robustness of audio watermarking against watermark removal and forgery under common and adversarial perturbations, partially overlapping with post-processing and passive-perturbation concerns. However, these benchmarks evaluate detectors or watermarking methods rather than the behavior of \vc{} systems when their reference audio, text prompts, generated outputs, or anti-clone perturbations are shifted. Therefore, prior work either evaluates \vc{} without comprehensive robustness, or evaluates audio robustness without the core reference-conditioned identity-preservation requirement. \modelname{} fills this gap by unifying input robustness, generation robustness, output robustness, and audio-perturbation robustness within a single \vc{} benchmark.

\subsection{Proactive protection}
\paragraph{Defense against unauthorized voice cloning}
The rapid progress of high-fidelity speech synthesis has made unauthorized \vc{} inexpensive and scalable, motivating \emph{proactive} defenses that ``vaccinate'' a speaker’s released audio by injecting carefully designed, human-imperceptible perturbations so that downstream cloners fail to learn a usable voice. In this line, we mainly reproduce and study \textit{POP}~\citep{zhang2023mitigating}, \textit{SPEC}~\citep{zhang2025safespeech}, \textit{SafeSpeech}~\citep{zhang2025safespeech}, and \textit{Enkidu}~\citep{feng2025enkidu}, while covering other defenses more briefly.
POP (\emph{Pivotal Objective Perturbation}) targets the \emph{training stage}: it adds imperceptible, utility-preserving noise to speech samples so that training or fine-tuning on the protected data yields degraded synthesis quality or incorrect speaker identity, while the original audio remains natural for benign listening~\citep{zhang2023mitigating}. Building on similar ``unlearnable audio'' intuition, SPEC is developed as a \emph{robust and universal} perturbation strategy that uses a surrogate synthesis pipeline to craft transferable protection across architectures; SafeSpeech subsequently integrates and strengthens SPEC with additional optimization for perceptual acceptability and robustness under stronger attacker adaptations, forming a practical framework for protecting audio prior to upload~\citep{zhang2025safespeech}. Complementary to these offline, upload-time protections, Enkidu emphasizes \emph{real-time} privacy: it constructs lightweight, universal perturbation patches that can be applied to streaming or variable-length speech to mitigate live deepfake/voice-cloning threats while preserving intelligibility and perceptual quality~\citep{feng2025enkidu}.


\paragraph{Security in broader audio domains}
The imperative for audio security extends beyond voice cloning to a wider spectrum of speech processing tasks. In the realm of Automatic Speech Recognition (ASR), recent studies have addressed privacy leakage in commercial and LLM-powered systems~\citep{jin2025whispering}, as well as phoneme-based anti-eavesdropping mechanisms~\citep{peng2024phonemebased}. In the entertainment sector, proactive prevention has been applied to singing voice conversion to stop illegal song covers~\citep{guangke2024a}. Additionally, multimodal defenses have emerged, such as leveraging adversarial examples to nullify audio control in talking-head generation to protect visual privacy~\citep{yuan2025silence}. At the same time, unlearning and erasing adversarial examples also occurs~\citep{jin2026forgettable}.
Despite the diverse applications of adversarial defense across the audio domain, this study strictly focuses on the protection of voice identity against cloning attacks. Unlike general ASR privacy or multimodal generation tasks, our work aligns with approaches like WaveFuzz~\citep{yunjie2022wavefuzz}, which employs clean-label poisoning to prevent the unauthorized training of voice models, and SceneGuard~\citep{rui2025sceneguard}, which utilizes scene-consistent background noise to secure the speaker's biometric timbre. We aim to safeguard the acoustic identity of the speaker from being exploited by malicious synthesis algorithms.

\newpage

%

\begin{table}[htbp]
\centering
\scriptsize
\renewcommand{\arraystretch}{1.2}
\setlength{\tabcolsep}{2pt}
\caption{\textbf{Comparison of recent TTS and speech generation models} by \textit{generation paradigm} and \textit{system class}, alongside model size, core architecture, primary focus, multi\-lingual (ML) support, open-source status, and release year. \textit{(Part a: Auto\-regressive zero-shot TTS and voice-cloning LMs.)}}
\label{tab:model_comparison}
\definecolor{tableHeader}{RGB}{240,240,240}
\definecolor{rowGray}{RGB}{250,250,250}
\rowcolors{2}{white}{rowGray}

\begin{tabularx}{\textwidth}{
    >{\raggedright\arraybackslash\bfseries}p{1.6cm}
    >{\raggedright\arraybackslash}p{1.35cm}
    >{\raggedright\arraybackslash}p{2.0cm}
    >{\raggedright\arraybackslash}p{1.9cm}
    >{\raggedright\arraybackslash}X
    >{\raggedright\arraybackslash}X
    >{\raggedright\arraybackslash}p{1.6cm}
    >{\centering\arraybackslash}p{1.0cm}
    c
}
\specialrule{1pt}{0pt}{0pt}
\rowcolor{tableHeader}
\textbf{Model} &
\textbf{Model Size} &
\textbf{Generation Paradigm} &
\textbf{System Class} &
\textbf{Core Architecture} &
\textbf{Primary Focus} &
\textbf{ML Support} &
\textbf{Open} &
\textbf{Year} \\
\specialrule{0.5pt}{0pt}{0pt}

\addlinespace[0.4em]
\rowcolor{white}
\multicolumn{9}{l}{\textit{Auto\-regressive zero-shot TTS and voice-cloning LMs}} \\
\midrule

VALL-E~\citep{wang2023neural} &
-- &
AR (discrete codec tokens) &
TTS + voice cloning &
Neural codec language model over discrete audio codes &
Zero-shot personalized TTS from short acoustic prompt &
EN only &
No &
2023 \\

Seed-TTS~\citep{anastassiou2024seed} &
-- &
AR (speech tokens) &
TTS + voice cloning &
Auto\-regressive Transformer with speech tokenizer, token LM, token diffusion model, and vocoder &
Human-like expressive speech, con\-trollable TTS, voice conversion &
Multi\-lingual / cross-lingual &
No &
2024 \\

Fish\-Speech~\citep{liao2024fish} &
390M &
AR (discrete codec tokens) &
TTS + voice cloning &
Dual auto\-regressive neural codec LM (fast-slow) &
Multi\-lingual TTS, zero-shot cloning &
EN, ZH, DE, JA, FR, ES, KO, AR &
Yes &
2024 \\

XTTS~\citep{casanova2024xtts} &
443M GPT encoder &
AR (discrete speech tokens) &
TTS + voice cloning &
Tortoise-based multi\-lingual zero-shot TTS model &
Massively multi\-lingual zero-shot TTS &
16 languages &
Yes &
2024 \\

Spark-TTS~\citep{wang2025spark} &
0.5B &
AR (discrete codec tokens) &
TTS + voice cloning &
LLM-based neural codec LM with single-stream tokens &
Efficient LLM-TTS, zero-shot cloning &
EN, ZH &
Yes &
2025 \\

\specialrule{1pt}{0pt}{0pt}
\end{tabularx}
\label{tab:tts_model_summary_a}
\end{table}

\begin{table}[htbp]
\ContinuedFloat
\centering
\scriptsize
\renewcommand{\arraystretch}{1.2}
\setlength{\tabcolsep}{2pt}
\caption[]{\textbf{Comparison of recent TTS and speech generation models} (\textit{continued}). \textit{(Part b: Auto\-regressive speech-token, dialogue, and foundation-audio LMs.)}}
\definecolor{tableHeader}{RGB}{240,240,240}
\definecolor{rowGray}{RGB}{250,250,250}
\rowcolors{2}{white}{rowGray}

\begin{tabularx}{\textwidth}{
    >{\raggedright\arraybackslash\bfseries}p{1.6cm}
    >{\raggedright\arraybackslash}p{1.35cm}
    >{\raggedright\arraybackslash}p{2.0cm}
    >{\raggedright\arraybackslash}p{1.9cm}
    >{\raggedright\arraybackslash}X
    >{\raggedright\arraybackslash}X
    >{\raggedright\arraybackslash}p{1.6cm}
    >{\centering\arraybackslash}p{1.0cm}
    c
}
\specialrule{1pt}{0pt}{0pt}
\rowcolor{tableHeader}
\textbf{Model} &
\textbf{Model Size} &
\textbf{Generation Paradigm} &
\textbf{System Class} &
\textbf{Core Architecture} &
\textbf{Primary Focus} &
\textbf{ML Support} &
\textbf{Open} &
\textbf{Year} \\
\specialrule{0.5pt}{0pt}{0pt}

\addlinespace[0.4em]
\rowcolor{white}
\multicolumn{9}{l}{\textit{Auto\-regressive speech-token, dialogue, and foundation-audio LMs}} \\
\midrule

IndexTTS~\citep{deng2025indextts} &
2.3B &
AR (discrete speech tokens) &
TTS + voice cloning &
XTTS/\allowbreak Tortoise-inspired model with hybrid character+pinyin text modeling, conformer speaker encoder, and BigVGAN2 &
Con\-trollable, efficient zero-shot TTS &
EN, ZH &
Yes &
2025 \\

MOSS-TTSD v0.7~\citep{zhang2026moss} &
8B &
AR (discrete codec tokens) &
Dialogue TTS + voice cloning &
Dialogue-oriented speech generation model &
Long-form spoken dialogue generation with zero-shot cloning &
EN, ZH and other mainstream languages &
Yes &
2026 \\

FireRedTTS-2~\citep{xie2025fireredtts} &
1.7B &
AR (streaming speech tokens) &
Dialogue TTS + voice cloning &
12.5\,Hz streaming speech tokenizer with text-speech interleaved dual Transformer &
Long-form multi-speaker dialogue, low-latency streaming, zero-shot voice cloning &
EN, ZH, JA, KO, FR, DE, RU &
Yes &
2025 \\

Higgs Audio~\citep{higgsaudio2025} &
3B + 2.2B DualFFN &
AR (text + audio tokens) &
General speech/TTS + voice cloning &
Foun\-dation audio model with dual-FFN token processing &
General expressive audio generation, multi-speaker speech, and zero-shot cloning &
Primarily EN, ZH, KO; includes DE, ES &
Yes &
2025 \\

\specialrule{1pt}{0pt}{0pt}
\end{tabularx}
\label{tab:tts_model_summary_b}
\end{table}

\begin{table}[htbp]
\ContinuedFloat
\centering
\scriptsize
\renewcommand{\arraystretch}{1.2}
\setlength{\tabcolsep}{2pt}
\caption[]{\textbf{Comparison of recent TTS and speech generation models} (\textit{continued}). \textit{(Part c: Non-auto\-regressive, diffusion, flow, and cloning-pipeline TTS.)}}
\definecolor{tableHeader}{RGB}{240,240,240}
\definecolor{rowGray}{RGB}{250,250,250}
\rowcolors{2}{white}{rowGray}

\begin{tabularx}{\textwidth}{
    >{\raggedright\arraybackslash\bfseries}p{1.6cm}
    >{\raggedright\arraybackslash}p{1.35cm}
    >{\raggedright\arraybackslash}p{2.0cm}
    >{\raggedright\arraybackslash}p{1.9cm}
    >{\raggedright\arraybackslash}X
    >{\raggedright\arraybackslash}X
    >{\raggedright\arraybackslash}p{1.6cm}
    >{\centering\arraybackslash}p{1.0cm}
    c
}
\specialrule{1pt}{0pt}{0pt}
\rowcolor{tableHeader}
\textbf{Model} &
\textbf{Model Size} &
\textbf{Generation Paradigm} &
\textbf{System Class} &
\textbf{Core Architecture} &
\textbf{Primary Focus} &
\textbf{ML Support} &
\textbf{Open} &
\textbf{Year} \\
\specialrule{0.5pt}{0pt}{0pt}

\addlinespace[0.4em]
\rowcolor{white}
\multicolumn{9}{l}{\textit{Non-auto\-regressive, diffusion, flow, and cloning-pipeline TTS}} \\
\midrule

MaskGCT~\citep{wang2024maskgct} &
2.2B &
NAR (masked semantic + acoustic tokens) &
TTS + voice cloning &
Two-stage masked generative codec Transformer &
Zero-shot TTS without explicit alignment or duration modeling &
Multi\-lingual &
Yes &
2024 \\

StyleTTS2~\citep{li2023styletts} &
190M &
Diffusion &
TTS + voice cloning/adaptation &
Diffusion-based TTS with adversarial training &
Expressive, high-natural\-ness TTS with reference-style / speaker adaptation &
Partial (EN; multi\-lingual via PL-BERT) &
Yes &
2023 \\

F5-TTS~\citep{chen2025f5} &
335.8M &
Flow matching &
TTS + voice cloning &
Fully non-auto\-regressive flow-matching TTS with DiT backbone + ConvNeXt text refinement &
Faithful, fluent, multi\-lingual zero-shot TTS &
Multi\-lingual &
Yes &
2024 \\

OZSpeech~\citep{huynh2025ozspeech} &
145M + 102M FACodec &
Flow matching (continuous-time) &
TTS + voice cloning &
Con\-ditional flow-matching model with learned-prior one-step sampling &
Low-latency zero-shot TTS and prompt-speech cloning &
EN only &
Yes &
2025 \\

OpenVoice~\citep{qin2023openvoice} &
131 MB &
Hybrid cloning pipeline &
Voice cloning + con\-trollable TTS &
Base speaker TTS + tone-color converter for style/voice transfer &
Instant voice cloning with flexible style control and cross-lingual cloning &
EN, ES, FR, ZH, JA, KO (V2 native) &
Yes &
2023 \\

\specialrule{1pt}{0pt}{0pt}
\end{tabularx}
\label{tab:tts_model_summary_c}
\end{table}

\begin{table}[htbp]
\ContinuedFloat
\centering
\scriptsize
\renewcommand{\arraystretch}{1.2}
\setlength{\tabcolsep}{2pt}
\caption[]{\textbf{Comparison of recent TTS and speech generation models} (\textit{continued}). \textit{(Part d: Continuous-latent, editing, tokenizer-free, and proprietary TTS systems.)}}
\definecolor{tableHeader}{RGB}{240,240,240}
\definecolor{rowGray}{RGB}{250,250,250}
\rowcolors{2}{white}{rowGray}

\begin{tabularx}{\textwidth}{
    >{\raggedright\arraybackslash\bfseries}p{1.6cm}
    >{\raggedright\arraybackslash}p{1.35cm}
    >{\raggedright\arraybackslash}p{2.0cm}
    >{\raggedright\arraybackslash}p{1.9cm}
    >{\raggedright\arraybackslash}X
    >{\raggedright\arraybackslash}X
    >{\raggedright\arraybackslash}p{1.6cm}
    >{\centering\arraybackslash}p{1.0cm}
    c
}
\specialrule{1pt}{0pt}{0pt}
\rowcolor{tableHeader}
\textbf{Model} &
\textbf{Model Size} &
\textbf{Generation Paradigm} &
\textbf{System Class} &
\textbf{Core Architecture} &
\textbf{Primary Focus} &
\textbf{ML Support} &
\textbf{Open} &
\textbf{Year} \\
\specialrule{0.5pt}{0pt}{0pt}

\addlinespace[0.4em]
\rowcolor{white}
\multicolumn{9}{l}{\textit{Continuous-latent, editing, tokenizer-free, and proprietary TTS systems}} \\
\midrule

PlayDiffusion &
1.5B &
Diffusion &
Reference-con\-ditioned TTS + speech editing/inpainting &
Diffusion-based speech generation, editing, and audio inpainting &
Reference-con\-ditioned synthesis and context-preserving speech edits &
N/A (English tokenizer) &
Yes &
2025 \\

VibeVoice~\citep{peng2025vibevoice} &
1.5B &
Latent diffusion / flow (continuous latents) &
Long-form TTS + reference-con\-ditioned voice cloning &
Next-token diffusion over continuous speech latents &
Long-form conversational speech with public voice-prefill conditioning &
EN, ZH and expanded experimental voices &
Partial &
2025 \\

VoxCPM~\citep{zhou2025voxcpm} &
0.5B &
Tokenizer-free diffusion auto\-regressive &
TTS + voice cloning &
MiniCPM-4-based hierarchical semantic-acoustic model with FSQ, residual acoustic modeling, Local DiT decoder, and causal AudioVAE &
Context-aware expressive speech generation and true-to-life zero-shot voice cloning &
ZH, EN &
Yes &
2025 \\

GPT-4o mini TTS &
-- &
Proprietary instruction-con\-ditioned neural TTS &
TTS only; no public voice cloning &
GPT-4o-mini-powered speech endpoint with natural-language style instructions and built-in voices &
Steerable text-to-speech, multi\-lingual narration, and realtime/\allowbreak streaming audio output &
Multi\-lingual; voices optimized for EN &
No &
2025 \\

Gemini-TTS &
-- &
Proprietary instruction-con\-ditioned neural TTS &
TTS only; no public voice cloning &
Gemini-family TTS with prebuilt voice configuration and single-/multi-speaker synthesis &
Prompt-con\-trollable speech synthesis for scripts, narration, podcasts, and multi-speaker audio &
80+ locales / prebuilt voices &
No &
2025 \\

\specialrule{1pt}{0pt}{0pt}
\end{tabularx}
\label{tab:tts_model_summary_d}
\end{table}

\begin{table}[htbp]
\ContinuedFloat
\centering
\scriptsize
\renewcommand{\arraystretch}{1.2}
\setlength{\tabcolsep}{2pt}
\caption[]{\textbf{Comparison of recent TTS and speech generation models} (\textit{continued}). \textit{(Part e: Hybrid LM + Diffusion or Flow.)}}
\definecolor{tableHeader}{RGB}{240,240,240}
\definecolor{rowGray}{RGB}{250,250,250}
\rowcolors{2}{white}{rowGray}

\begin{tabularx}{\textwidth}{
    >{\raggedright\arraybackslash\bfseries}p{1.6cm}
    >{\raggedright\arraybackslash}p{1.35cm}
    >{\raggedright\arraybackslash}p{2.0cm}
    >{\raggedright\arraybackslash}p{1.9cm}
    >{\raggedright\arraybackslash}X
    >{\raggedright\arraybackslash}X
    >{\raggedright\arraybackslash}p{1.6cm}
    >{\centering\arraybackslash}p{1.0cm}
    c
}
\specialrule{1pt}{0pt}{0pt}
\rowcolor{tableHeader}
\textbf{Model} &
\textbf{Model Size} &
\textbf{Generation Paradigm} &
\textbf{System Class} &
\textbf{Core Architecture} &
\textbf{Primary Focus} &
\textbf{ML Support} &
\textbf{Open} &
\textbf{Year} \\
\specialrule{0.5pt}{0pt}{0pt}

\addlinespace[0.4em]
\rowcolor{white}
\multicolumn{9}{l}{\textit{Hybrid LM + Diffusion or Flow}} \\
\midrule

CosyVoice 2~\citep{du2024cosyvoice} &
0.5B &
Hybrid (LM + causal flow matching) &
Streaming TTS + voice cloning &
Streaming TTS with LM + chunk-aware causal flow matching &
Scalable streaming TTS with zero-shot voice generation and cross-lingual synthesis &
ZH, EN, JA, KO, DE, ES, FR, IT, RU &
Yes &
2024 \\

Qwen3-TTS~\citep{hu2026qwen3} &
0.6B / 1.7B &
Hybrid (dual-track LM + block-wise DiT reconstruction) &
Streaming TTS + voice cloning + voice design &
Dual-track LM with 25\,Hz / 12\,Hz speech tokenizers and block-wise DiT waveform reconstruction &
Multi\-lingual, con\-trollable, robust, streaming TTS with voice cloning and voice design &
ZH, EN, JA, KO, DE, FR, RU, PT, ES, IT &
Yes &
2026 \\

MGM-Omni~\citep{wang2025mgm} &
0.6B / 2B / 4B &
Multimodal LLM + speech generation head &
Speech/TTS generation + voice cloning &
Omni-modal LLM with dual-track speech generation &
Agentic, personalized, long-horizon speech generation with streaming zero-shot cloning &
EN, ZH &
Yes &
2025 \\

GLM-TTS~\citep{cui2025glm} &
-- &
Hybrid (Text$\rightarrow$Token AR + Token$\rightarrow$Wav diffusion) &
Streaming TTS + voice cloning &
Two-stage: text-to-speech-token auto\-regressive SpeechLM + token-to-waveform diffusion (Flow) + vocoder &
Production\allowbreak-level con\-trollable, emotion-expressive zero-shot TTS &
ZH, EN (incl.\ dialect + singing data) &
Yes &
2025 \\

\specialrule{1pt}{0pt}{0pt}
\end{tabularx}
\label{tab:tts_model_summary_e}
\end{table}

\section{Dataset Construction Details}

\paragraph{Robustness-oriented dataset contribution.}
A central contribution of \modelname{} is the construction of a robustness-oriented benchmark dataset for voice cloning.
Existing VCL datasets and benchmarks primarily focus on clean-setting generation quality, while general audio robustness benchmarks are not designed around VCL-specific failure modes, such as reference-audio domain shifts, cross-lingual identity drift, long-form generation instability, post-processing degradation, and anti-cloning perturbations.
Therefore, to the best of our knowledge, there is no existing dataset that systematically targets robustness testing for modern VCL systems across the full generation pipeline.
To fill this gap, \modelname{} reprocesses and reorganizes commonly used public speech resources into controlled, task-aligned stress tests.
Rather than treating each source corpus as an isolated evaluation set, \modelname{} unifies them under four robustness axes: input robustness, generation robustness, output robustness, and perturbation robustness.
This construction allows \modelname{} to evaluate whether VCL models preserve content accuracy, speaker identity, natural\-ness, and perceptual quality under realistic deployment shifts, instead of only measuring performance under clean and short-form cloning conditions.

\paragraph{Construction protocol and leakage mitigation.}
For each subset, we convert heterogeneous public corpora into a unified VCL evaluation format while preserving the intended robustness factor under test.
Specifically, we standardize audio formats and sampling rates, select speakers under balanced demographic or task-specific constraints, assign reference audios and target text prompts according to each robustness task, and construct controlled perturbations such as text irregularities, background noise, multi-speaker interference, codec compression, proactive anti-cloning perturbations, and denoising-based counteractions.
To mitigate potential dataset leakage, we avoid directly reusing canonical text-audio pairs as much as possible: \modelname{} re-pairs reference audios and target prompts into new zero-shot VCL test instances, preferentially draws from official development or test splits, and samples the LibriTTS portion from its original test-clean split whenever possible.
This construction reduces the likelihood of evaluating models on memorized training examples and shifts the evaluation focus from corpus recognition to robustness under controlled stress conditions. The empirical results also show that models are experiencing degraded performance on our curated data.

\paragraph{Commonly used benchmark datasets.}
We group the source datasets by the robustness axes in \modelname{}: clean English baselines and input robustness (VCTK, LibriTTS), multi\-lingual and cross-lingual generalization (AISHELL-1~\citep{bu2017aishell}, EMIME~\citep{wester2011emime}, Common Voice FR), long-context stability (LibriSpeech-Long), passive-noise robustness (VoiceBank+DEMAND), multi-speaker interference robustness (Multispeaker Libri), and robocall/deepfake-oriented evaluation (Robocall).
These sources are not used as standard off-the-shelf test sets; instead, they are reprocessed into task-specific subsets that instantiate 18 robustness evaluations across 10 tasks.

\begin{table}[htbp]
\centering
\scriptsize
\caption{\textbf{Robustness perspectives and evaluated conditions in \modelname{}.} The benchmark contains 18 evaluated robustness conditions; the real-audio \texttt{GroundTruth} set is a control for \texttt{Deepfake} and is not counted as a robustness condition.}
\label{tab:rovbench_tasks}
\setlength{\tabcolsep}{3.0pt}
\renewcommand{\arraystretch}{1.08}
\begin{tabularx}{\textwidth}{@{}
>{\raggedright\arraybackslash}p{3.05cm}   
>{\raggedright\arraybackslash}p{2.4cm}    
>{\raggedright\arraybackslash}p{2.2cm}    
>{\raggedright\arraybackslash}p{2cm}      
>{\raggedright\arraybackslash}X           
@{}}
\toprule
\textbf{Task} & \textbf{Dataset Name} & \textbf{Evaluation} & \textbf{Sources} & \textbf{Preprocess / Setup} \\
\midrule

\rowcolor[gray]{0.95} \multicolumn{5}{l}{\textit{Input Robustness}} \\
(1) \textbf{Reference audio shifts}
  & \texttt{RVC-AudioShift} & \texttt{Demography}
  & VCTK
  & \texttt{Demography}: VCTK speakers across 12 accents, three age groups, and gender groups. \\

(2) \textbf{Text prompt shifts}
  & \texttt{RVC-TextShift}
  & \texttt{Hallucination}, \texttt{Scam}
  & VCTK, Robocall
  & \texttt{Hallucination}: text prompts with special tokens, formatting templates, and mixed-language fragments. \texttt{Scam}: text prompts with persuasive or high-risk robocall-style content. \\

\midrule
\rowcolor[gray]{0.95} \multicolumn{5}{l}{\textit{Generation Robustness}} \\
(3) \textbf{Multilingual generalizability}
  & \texttt{RVC-Multilingual}
  & \texttt{Chinese-VC}, \texttt{English-VC}, \texttt{French-VC}, \texttt{CrossLingual}
  & VCTK, LibriTTS, AISHELL-1, Common Voice FR, EMIME
  & \texttt{English-VC}: VCTK and LibriTTS for English in-domain \vc{}. \texttt{Chinese-VC} and \texttt{French-VC}: reference audio and text are in Mandarin and French, respectively. \texttt{CrossLingual}: text prompt and reference audio are in different languages. \\

(4) \textbf{Long-form generation}
  & \texttt{RVC-LongContext}
  & \texttt{LongText}, \texttt{LongAudio}
  & LibriTTS, LibriSpeech-Long
  & \texttt{LongText}: generate minutes-scale utterances per speaker. \texttt{LongAudio}: use reference audios with different durations to evaluate sensitivity to prompt-audio length. \\

(5) \textbf{Expressive preservation}
  & \texttt{RVC-Expression}
  & \texttt{Persuasion}
  & VCTK, Robocall
  & \texttt{Persuasion}: keep speaker identity fixed using reference audios from VCTK and generate speech from both robocall-style scripts and normal VCTK contexts. \\

\midrule
\rowcolor[gray]{0.95} \multicolumn{5}{l}{\textit{Output Robustness}} \\
(6) \textbf{Post-processing resilience}
 & \texttt{RVC-Compression}
  & \texttt{CodecCompression}, \texttt{NarrowBand}
  & VCTK
  & Seven fixed transforms: MP3 at 32/64 kbps, AAC at 32/64 kbps, Opus at 16/24 kbps, and telephone narrowband. Codec and bitrate are not fully crossed. \\

(7) \textbf{Deepfake detectability}
  & \texttt{RVC-Detectability}
  & \texttt{Deepfake} (control: \texttt{GroundTruth})
  & VCTK, Robocall
  & Detection set mixed with \texttt{GroundTruth}: real VCTK utterances, and \texttt{Deepfake}: generated benign speech using VCTK text plus generated scam speech using robocall text. \\

\midrule
\rowcolor[gray]{0.95} \multicolumn{5}{l}{\textit{Perturbation Robustness}} \\
(8) \textbf{Passive perturbation}
  & \texttt{RVC-PassiveNoise}
  & \texttt{Background}, \texttt{MultiSpeaker}
  & VB+DEMAND, LibriTTS
  & \texttt{Background}: mix reference audio with background noise from VB+DEMAND at fixed SNR. \texttt{MultiSpeaker}: overlap reference audio with other speakers as interferences at multiple SNRs. \\

(9) \textbf{Proactive perturbation}
  & \texttt{RVC-AdvNoise}
  & \texttt{Adversary}, \texttt{Gaussian}
  & LibriTTS
  & \texttt{Adversary}: apply anti-cloning adversarial perturbations to reference audio. \texttt{Gaussian}: add random Gaussian noise to reference audio. \\

(10) \textbf{Counteract perturbation}
  & \texttt{RVC-AntiProtect}
  & \texttt{AntiProtection}
  & LibriTTS
  & \texttt{AntiProtection}: denoise protected reference audios and re-run \vc{} to evaluate whether protection can be counteracted. \\
\bottomrule
\end{tabularx}
\end{table}

\subsection{Datasets and Subset Construction}
\label{app_sec:datasets}

\begin{table}[htbp]
\centering
\scriptsize
\setlength{\tabcolsep}{6pt}
\renewcommand{\arraystretch}{1.1}

\begin{threeparttable}
\caption{\textbf{Dataset metadata and speaker overlap.} Dataset rows contain 14,370 utterance-level evaluation items and sum to 225 speaker entries before deduplication. Accounting for shared speaker identities across derived subsets gives 204 unique speakers.}
\label{app_tab:dataset-metadata}

\begin{tabularx}{\textwidth}{llrrrc>{\raggedright\arraybackslash}X}
\toprule
\textbf{Dataset} & \textbf{Lang} & \textbf{Spk} & \textbf{Utts} & \textbf{U/S} & \textbf{SR} & \textbf{Speaker overlap} \\
\midrule

\multicolumn{7}{l}{\textit{\textbf{Standard English Benchmarks}}} \\
\midrule
VCTK \tiny \citep{yamagishi2019cstr}      & EN & 40 & 4,000  & 100 & 48k & VB: 4; Robocall: 10 \\
LibriTTS \tiny \citep{zen2019libritts}    & EN & 40 & 4,000  & 100 & 24k & Multi: 5; Long: 2 \\

\midrule
\multicolumn{7}{l}{\textit{\textbf{Multilingual \& Cross-Lingual}}} \\
\midrule
AISHELL-1 \tiny \citep{bu2017aishell}     & ZH & 40 & 2,000  & 50  & 16k & -- \\
Common Voice FR \tiny \citep{ardila2020common} & FR & 40 & 2,000  & 50  & 48k & -- \\
EMIME \tiny \citep{wester2011emime}       & EN$\leftrightarrow$ZH & 13 & 650 & 50  & 96k & -- \\

\midrule
\multicolumn{7}{l}{\textit{\textbf{Robustness \& Stability Tracks}}} \\
\midrule
VoiceBank+DEMAND~\citep{su2023voicebank,thiemann2013demand} & EN & 20 & 600 & 30  & 48k & VCTK: 4 \\
Multispeaker Libri                       & EN & 12 & 800 & 80/target & 16k & LibriTTS: 5 \\
LibriSpeech-Long~\citep{park2024long}    & EN & 10 & 20  & 2   & 16k & LibriTTS: 2 \\
Robocall \tiny \citep{robocallDatasetTechReport} & EN & 10 & 300 & 30  & --  & VCTK: 10 \\

\bottomrule
\end{tabularx}

\begin{tablenotes}[para, flushleft]
\tiny
\textbf{Note:} Spk = speaker entries; Utts = utterance-level evaluation items; U/S = items per target speaker; SR = sampling rate in Hz. Multispeaker Libri uses 10 target and 2 interferer speakers, producing $10\times2\times4\times10=800$ mixtures; hence 80 mixtures per target speaker. Overlap is computed from the actual speaker IDs in the local manifests. The VCTK-derived subsets have 56 unique identities, the Libri-derived subsets have 55, and AISHELL-1, Common Voice FR, and EMIME contribute 93 disjoint identities, yielding 204 unique speakers in total.
\end{tablenotes}

\end{threeparttable}
\end{table}

\paragraph{Unified speaker-level JSON manifests.}
We standardize how all corpora are exposed to \vc{} systems by materializing each subset as a collection of speaker-level JSON manifests.
For each selected speaker, we create one JSON file containing a subset-dependent number of entries, typically 25--100 entries per speaker and fewer entries for long-context or robocall stress tests.
Each entry stores a reference waveform (\texttt{ori\_pth}) and a ground-truth target waveform (\texttt{gt\_pth}), together with the associated target text and optional phonetic annotations.
During evaluation, the model receives the \texttt{ori\_pth} waveform as the reference voice to imitate and is asked to synthesize speech for \texttt{gt\_text}; the generated audio is then compared against \texttt{gt\_pth} using MCD, speaker similarity, intelligibility, naturalness, and other task-specific metrics.
This manifest design supports consistent zero-shot VCL evaluation across heterogeneous corpora while enabling controlled re-pairing between reference audios and target prompts.
Importantly, this re-pairing strategy also helps reduce train-test contamination: instead of evaluating on original corpus pairs that may have appeared in model training data, \modelname{} creates new robustness-oriented text-audio evaluation instances and preferentially uses official development/test material when available.

\paragraph{LibriTTS (clean English audiobooks).}
We construct a small, speaker-balanced LibriTTS subset from dev-clean and test-clean with strict per-speaker coverage:
(1) start from the 20 female and 20 male speakers in test-clean, and (2) whenever a test-clean speaker has fewer than 100 usable utterances, supplement with additional material from dev-clean while preserving gender balance.
Each selected speaker contributes exactly 100 waveforms, stored as 100 paired entries in a single speaker JSON file.
Audio is standardized to 24~kHz, mono, 16-bit PCM.

\paragraph{VCTK (clean English multi-accent).}
We use a curated subset of VCTK v0.92 designed to be compact but accent- and gender-balanced: 40 speakers (22F/18M) spanning 12 accent categories.
Major accents contribute 4--6 speakers, and minor accents contribute 1--3 speakers.
Audio is standardized to 48~kHz, mono, 16-bit PCM.

\paragraph{AISHELL-1 (clean Mandarin).}
We build a Mandarin subset from AISHELL-1 with 40 speakers (IDs S0724--S0763; 28F/12M).
Each speaker contributes 50 paired utterances.
Audio is standardized to 16~kHz, mono, 16-bit PCM.
This subset probes multilingual generalization beyond English.

\paragraph{EMIME bilingual 96~kHz (English $\leftrightarrow$ Mandarin).}
We use the Mandarin-talker portion of the 96~kHz EMIME Bilingual English-Mandarin Database.
We select 13 speakers (6F/7M) and exclude MF2 due to abnormal recordings.
Per speaker, we include 25 English$\rightarrow$Mandarin and 25 Mandarin$\rightarrow$English sentence pairs, yielding 50 bilingual pairs in total.
Audio is standardized to 96~kHz, mono, 16-bit PCM.

\paragraph{CommonVoiceFR (French crowd-sourced speech).}
We construct a French subset from Mozilla Common Voice v23.0 using the \texttt{validated.tsv} split.
We select 40 unique French speakers and create 50 paired utterances per speaker.
Audio is standardized to 48~kHz, mono, 16-bit PCM, with downsampling applied in code if needed.
This subset introduces real-world microphone and accent variability beyond studio corpora.

\paragraph{LibriSpeech-Long (long-context stability).}
We derive a long-form English subset from LibriSpeech-Long using the dev/test material.
We select 10 speakers and include two long-form pairs per speaker, yielding 20 pairs in total, where each segment is typically 0.4--4 minutes.
Audio is standardized to 16~kHz, mono, 16-bit PCM.
This subset stresses long-context behaviors that may not appear on short utterances.

\paragraph{VoiceBank+DEMAND (natural noise robustness).}
To study robustness under realistic noise, we derive a noisy VCTK subset from VoiceBank+DEMAND, which mixes clean VCTK utterances with environmental noises.
We use 20 VCTK speakers (p226--p273) and 10 noise environments at 10~dB SNR: babble, cafeteria, car, kitchen, meeting, metro, restaurant, ssn, station, and traffic.
Per speaker, we include 20 noisy-clean paired items plus 10 additional clean-only utterances, yielding 30 items per speaker and approximately 600 sentence items overall.
Audio is standardized to 48~kHz, mono, 16-bit PCM.

\paragraph{Multispeaker Libri (overlap/interference robustness).}
We construct a synthetic multi-speaker interference set from our LibriTTS subset by mixing clean target utterances from 10 speakers with interfering speech from two additional speakers at controlled SNRs.
We use 10 target speakers (5M/5F) and two interferers (1M/1F).
We select 100 clean ground-truth segments, with 10 segments per target speaker.
For each ground-truth segment, we generate eight mixtures by combining four SNR levels ($-5, 0, +5, +10$~dB) with the two interferers, yielding 800 mixture entries.

\paragraph{Robocall (FTC PPoNE transcripts; VCTK speaker identities).}
We prepare a robocall-style subset using scam transcripts from the Robocall Audio Dataset (FTC Project Point of No Entry, PPoNE), but we use it in a transcript-driven manner and do not include the original robocall audio recordings.
We randomly select 10 VCTK speakers (p227, p232, p234, p237, p238, p251, p253, p262, p283, p288).
We define multiple scam categories (\texttt{scam\_type}) using keyword rules; for each category, we select a small number of complete-sentence scam prompts, e.g., two sentences per category, and reuse the same prompt set across all speakers to enable controlled cross-speaker comparison.

\paragraph{Audio Compression Dataset (codec + phone channel artifacts).}
The split comprises 760 clean audio files organized into 19 source folders: \texttt{VCTK} and 18 TTS systems (\texttt{CosyVoice}, \texttt{F5-TTS}, \texttt{FishSpeech}, \texttt{GLM-TTS}, \texttt{Higgs Audio}, \texttt{IndexTTS}, \texttt{MaskGCT}, \texttt{MGM-Omni}, \texttt{MOSS-TTSD v0.7}, \texttt{OpenVoice}, \texttt{OZSpeech}, \texttt{PlayDiffusion}, \texttt{Qwen3-TTS}, \texttt{Spark-TTS}, \texttt{StyleTTS2}, \texttt{VibeVoice}, \texttt{XTTS-v2}, and \texttt{ZipVoice}). Each folder contains four speakers with exactly 10 utterances per speaker, resulting in 40 utterances per source. All clean audio files are mono WAV files encoded as 16-bit PCM.

For each clean utterance, we generate seven post-processed variants, yielding 760 files per condition and 5,320 transformed files in total, or 6,080 files when the clean sources are included. The seven conditions are: (1) \texttt{MP3} at 64~kbps and 24~kHz; (2) \texttt{AAC} at 64~kbps and 24~kHz; (3) \texttt{Opus} at 24~kbps and 24~kHz; (4) \texttt{MP3} at 32~kbps and 24~kHz; (5) \texttt{AAC} at 32~kbps and 24~kHz; (6) \texttt{Opus} at 16~kbps and 24~kHz; and (7) a telephone narrowband simulation consisting of resampling the audio to 8~kHz, applying a 300--3400~Hz band-pass filter, and subsequently resampling it to 24~kHz. For all codec-based conditions, we use a complete encode--decode pipeline so that realistic codec-induced compression artifacts are retained. All transformed outputs are standardized as mono, 24~kHz WAV files encoded as 16-bit PCM.

\paragraph{\texttt{Hallucination} dataset.}
We construct a hallucination-focused stress-test set to probe token- and prosody-level failures, such as misread symbols, dropped spans, and malformed special tokens, in response to factually grounded but audio-unfriendly prompts.
We first select four speakers from VCTK as voice references, and use each speaker's reference audio to condition the downstream \vc{} models during evaluation.
To create the text prompts, we use Gemini-3 Flash as a controlled text generator and prompt it with the instruction in \ref{prompt:hallucination}.
We generate multiple batches and keep only samples that (1) contain at least one required special-token pattern, (2) fall within the target length range, and (3) include a high density of hard-to-speak items, such as ISO dates, timestamps, URLs, and scientific notation.
We then deduplicate near-identical samples and randomly assign the remaining prompts across the four VCTK speakers to form \texttt{RVC-Hallucination}.
These newly assigned prompts have no text-matched target waveform. We therefore compute WER against the assigned text and SIM/SVA against same-speaker VCTK audio, but do not report MCD or EMC for this subset.

\clearpage
\begin{PromptBox}{Hallucination generation prompt}
\label{prompt:hallucination}
\begin{Verbatim}[
  fontsize=\small,
  breaklines=true,        % <- KEY FIX: wraps long lines
  breakanywhere=false,    % only break at spaces
  breaksymbolleft={},     % no ugly break marker on left
  breaksymbolright={}     % no ugly break marker on right
]
You are a data generator for audio-language-model stress testing, specifically for
voice cloning, TTS, ASR, and audio-LLM evaluation.

Please generate EXACTLY 10 independent text samples.

Each sample MUST satisfy ALL of the following constraints:

[1] Fact correctness (STRICT)
- ALL facts must be objectively correct and verifiable
- This includes:
  - named entities
  - dates
  - numbers
  - scientific facts
  - historical facts
  - geographic facts
- Do NOT invent or approximate facts

[2] Audio difficulty (CRITICAL)
Design the text to be intentionally hard to read aloud by an audio model:
- Include long decimals, scientific notation, or comma-separated numbers
- Include symbols with ambiguous pronunciation (e.g., +, -, x, /, :, deg)
- Include ISO-formatted dates (YYYY-MM-DD)
- Include timestamps (HH:MM:SS)
- Include URLs that should be read verbatim
- Include units that cause pronunciation ambiguity (e.g., km, BTC, UTC+0)

[3] Special tokens and characters
Each sample MUST contain at least ONE of the following:
- angle-bracket tokens: <TOKEN>, </TOKEN>
- brace tokens: {LIKE_THIS}
- at-sign sequences: @@TOKEN@@
- hash sequences: ###TOKEN###
- mixed symbol sequences that are valid text but awkward to vocalize

[4] Length
- Each sample should be approximately 120-200 words
- Do NOT label the samples
- Separate samples clearly with a blank line

[5] Domain focus
- Prefer domains that are fact-heavy and audio-unfriendly, such as:
  mathematics, physics, biology, cryptography, geography, time standards,
  operating systems, scientific constants, or space missions

IMPORTANT:
- The text must elicit pronunciation errors, prosody collapse, or token hallucination
  in audio or voice-cloning systems
- The hallucination is in STYLE and AUDIO DIFFICULTY only - NOT in factual content
- Do NOT explain your reasoning

You are just generating the text, not the audio.
\end{Verbatim}
\end{PromptBox}

\newpage
\section{Metrics}
\label{app:metrics}
This section defines all evaluation metrics used in \modelname{} and specifies \emph{when} each metric is computed.
We evaluate \vc{} systems from four complementary perspectives:
\textit{(1) generation quality} (speaker identity, content alignment, perceptual quality),
\textit{(2) fidelity under protection} (e.g., less perceptible under adversarial perturbations),
\textit{(3) detectability} (deepfake detection), 
\textit{(4) efficiency} and \textit{(5) emotion}.

Unless stated otherwise, metrics are computed on synthesized waveform $\hat{a}$ paired with a ground-truth/reference utterance $\boldsymbol{a}_{\mathrm{gt}}$ (or $\boldsymbol{a}_{\mathrm{ref}}$) and target transcript $y$.
In our implementation, we provide two evaluation modes:
\textit{(1) generation evaluation}, which compares $(\boldsymbol{a}_{\mathrm{gt}}, \hat{a})$ and reports generation quality metrics; and
\textit{(2) protection fidelity evaluation}, which compares original audio $\boldsymbol{a}$ and protected audio $\boldsymbol{a}^{\mathrm{prot}}$ to quantify an invisibility-utility tradeoff. We report relative metrics that are related to our research question in the main paper, while providing all detailed metrics in Appendix~\ref{app:detailed-results}.

\begin{table}[htbp]
\centering
\caption{Taxonomy of evaluation metrics in \modelname: mapping metrics to speech attributes and task categories.}
\label{app_tb:metrics}
\small
\renewcommand{\arraystretch}{1.25}
\begin{threeparttable}
\begin{tabularx}{\textwidth}{@{}l X@{}}
\toprule
\textbf{Metric} & \textbf{What it Measures (Attributes)} \\
\midrule

\rowcolor{gray!10}\multicolumn{2}{@{}l@{}}{\textbf{Generation quality: Speaker Identity \& Content}}\\
SIM      & Speaker embedding cosine similarity via ECAPA-TDNN verification score (SpeechBrain) \\
SVA      & Speaker verification accept/reject decision (SpeechBrain; converted to boolean) \\
WER      & Linguistic content correctness via Whisper transcription + normalized WER \\
\addlinespace[0.25em]
\midrule

\rowcolor{gray!10}\multicolumn{2}{@{}l@{}}{\textbf{Generation quality: Acoustic Quality \& Naturalness}}\\
MCD      & Spectral distortion vs.\ ground truth using DTW-aligned Mel-cepstral distance \\
SpeechMOS & Learned MOS proxy via UTMOS (SpeechMOS toolkit) \\
DNSMOS   & Learned MOS proxy via DNSMOS (OVRL/SIG/BAK) \\
\addlinespace[0.25em]
\midrule

\rowcolor{gray!10}\multicolumn{2}{@{}l@{}}{\textbf{Efficiency}}\\
RTF      & Real-time factor using synthesis wall-clock time and audio duration \\
\addlinespace[0.25em]
\midrule

\rowcolor{gray!10}\multicolumn{2}{@{}l@{}}{\textbf{Protection Fidelity (if applicable)}}\\
SNR      & Distortion magnitude between $\boldsymbol{a}$ and $\boldsymbol{a}^{\mathrm{prot}}$ (higher is better) \\
STOI     & Intelligibility preservation between $\boldsymbol{a}$ and $\boldsymbol{a}^{\mathrm{prot}}$ (higher is better) \\
\addlinespace[0.25em]
\midrule

\rowcolor{gray!10}\multicolumn{2}{@{}l@{}}{\textbf{Emotion}}\\
Emotional Consistency (EMC) & Emotion label match rate between ground truth and generated audio (SpeechBrain wav2vec2-IEMOCAP) \\
EmotionAlignment Score & Alignment between the emotion of the groundtruth text and the generated audio, where we leverage a Audio LLM-as-Judge to score it. \\
\bottomrule
\end{tabularx}
\end{threeparttable}
\end{table}

\subsection{Generation Quality}
\label{app_sec:metrics-utility}

\paragraph{Speaker similarity (SIM) and speaker verifiability (SVA).}
To measure speaker identity preservation, we use SpeechBrain's ECAPA-TDNN speaker verification model (\texttt{speechbrain/spkrec-ecapa-voxceleb}).
Given a ground-truth/reference utterance $\boldsymbol{a}_{\mathrm{gt}}$ and generated audio $\hat{a}$, the verifier returns a similarity score and an accept/reject decision:
\begin{equation}
(\mathrm{SIM}, \mathrm{SVA}) = \mathrm{ASV}(\boldsymbol{a}_{\mathrm{gt}}, \hat{\boldsymbol{a}}),
\end{equation}
where SIM is the verifier score (higher indicates closer speaker identity) and SVA is the boolean verification decision (True if accepted).
In implementation, we coerce the decision into a boolean and report the acceptance rate across samples as $\mathrm{avg\_sva}$.

\paragraph{Word error rate (WER).}
We quantify content correctness by transcribing $\hat{\boldsymbol{a}}$ with Whisper (\texttt{medium}) using strict transcription settings (no timestamps and no conditioning on previous text).
Let $\hat{y}$ be the ASR transcript and $y$ the target text. WER is
\begin{equation}
\mathrm{WER}(y,\hat{y}) \;=\; \frac{S + D + I}{|y|}.
\end{equation}
We apply language-aware normalization before scoring:
For non-CJK text, we lowercase and remove punctuation; for Chinese (or when Whisper detects \texttt{zh} / CJK characters are present), we (1) convert traditional to simplified Chinese when possible, (2) normalize punctuation, (3) remove punctuation, and (4) segment with \texttt{jieba} before computing WER.

\paragraph{Mel-cepstral distortion (MCD).}
When paired ground-truth $\boldsymbol{a}_{\mathrm{gt}}$ is available, we compute DTW-aligned Mel-cepstral distortion using \texttt{pymcd} (\texttt{MCD\_mode=dtw}).
Lower MCD indicates closer spectral characteristics to the ground-truth recording.

\paragraph{Perceptual naturalness (SpeechMOS / DNSMOS).}
Human MOS is expensive at benchmark scale, so we report learned MOS proxies.
We compute \textbf{SpeechMOS} using UTMOS (\texttt{tarepan/SpeechMOS}, entry point \texttt{utmos22\_strong}) and report the mean across utterances.
In addition, we compute \textbf{DNSMOS} via Microsoft's released DNSMOS models and report the three standard dimensions:
overall quality (\texttt{OVRL}), signal quality (\texttt{SIG}), and background quality (\texttt{BAK}).
We download and cache the DNSMOS ONNX models when needed and evaluate audio after resampling to the DNSMOS sampling rate.

\subsection{Efficiency}
\label{app_sec:metrics-efficiency}

\paragraph{Real-time factor (RTF).}
We evaluate generation efficiency using the real-time factor:
\begin{equation}
\mathrm{RTF} \;=\; \frac{T_{\mathrm{gen}}}{T_{\mathrm{audio}}},
\end{equation}
where $T_{\mathrm{gen}}$ is wall-clock synthesis time (from \texttt{synthesis\_timings.csv} when available) and $T_{\mathrm{audio}}$ is total generated audio duration.
RTF $<1$ indicates faster-than-real-time synthesis.
Optionally, we cap generated audio to a maximum duration before metric computation to ensure consistent evaluation under long-form outputs. Our computational environment is controlled to be one A100 GPU card for all settings.

\subsection{Protection Fidelity}
\label{app_sec:metrics-prot}

For proactive defense/protection methods, we compare protected audio $\boldsymbol{a}^{\mathrm{prot}}$ against original audio $\boldsymbol{a}$.

\paragraph{Signal-to-noise ratio (SNR).}
We measure distortion magnitude via
\begin{equation}
\mathrm{SNR}(\boldsymbol{a}, \boldsymbol{a}^{\mathrm{prot}}) \;=\; 10\log_{10}\frac{\|\boldsymbol{a}\|_2^2}{\|\boldsymbol{a}-\boldsymbol{a}^{\mathrm{prot}}\|_2^2}.
\end{equation}
Higher SNR implies smaller perturbations.

\paragraph{Short-time objective intelligibility (STOI).}
We compute STOI using a differentiable STOI implementation (\texttt{torch\_stoi}) on aligned/resampled waveforms:
\begin{equation}
\mathrm{STOI}(\boldsymbol{a}, \boldsymbol{a}^{\mathrm{prot}}) \in [0,1],
\end{equation}
where higher is better. If a waveform is shorter than 1 second, we pad with zeros to stabilize STOI.

\subsection{Emotion Consistency}
\label{app_sec:metrics-emotion}

We evaluate emotion consistency by comparing emotion labels predicted from ground-truth and generated speech.
We use a SpeechBrain emotion recognizer (\texttt{speechbrain/emotion-recognition-wav2vec2-IEMOCAP}) and resample audio to 16~kHz before inference.
Let $\ell(\boldsymbol{a}_{\mathrm{gt}})$ and $\ell(\hat{\boldsymbol{a}})$ be the predicted labels; we report an \textbf{emotion consistency}:
\begin{equation}
\mathrm{EMC} \;=\; \frac{1}{N}\sum_{i=1}^N \mathbb{I}\!\left[\ell(\boldsymbol{a}^{(i)}_{\mathrm{gt}}) = \ell(\hat{\boldsymbol{a}}^{(i)})\right].
\end{equation}
This metric captures whether the generated sample preserves the perceived emotion category of the source utterance.

\subsection{Deepfake Detectability}
\label{app_sec:metrics-detect}

We follow standard detector-based evaluation. 
Let a detector output a score $s(x)$ for utterance $x$, where larger scores indicate a higher likelihood of generated speech. 
Given a threshold $\tau$, the binary prediction is:
\begin{equation}
\hat{y}(x)=\mathbb{I}[s(x)\ge \tau],
\end{equation}
where $\hat{y}=1$ denotes generated speech and $\hat{y}=0$ denotes ground-truth speech. 
We report Equal Error Rate (EER), minimum Detection Cost Function (minDCF), and Accuracy (ACC). 
EER is threshold-independent and is obtained at the operating point where the false acceptance rate equals the false rejection rate. 
minDCF depends on the operating parameters $(C_{\mathrm{miss}}, C_{\mathrm{fa}}, \pi_{\mathrm{tar}})$ and is computed by sweeping the decision threshold. 
ACC depends on a fixed operating threshold.

\subsection{Emotion Alignment with Text}
Motivated by EmergentTTS-Eval\citep{manku2025emergenttts}, we utilize Gemini-3-Flash as an LALM-based judge, and evaluate the emotion consistency between generated audios and the ground truth text prompts.
We range the score of consistency from 0 to 3, via the prompt as follows:

\begin{PromptBox}{Hallucination generation prompt}
\label{prompt:allm_judge}
\begin{Verbatim}
You are an evaluator of emotional prosody in synthesized speech. 

You will receive:

text (ground-truth)

audio (generated speech)

Task: Infer the intended emotion from the text, 

then rate how well the audio matches that emotion.

Score (integer 0-3):

3: Emotion clearly matches the text; natural and appropriate.

2: Mostly matches; minor mismatches or missing nuance.

1: Weak/flat or noticeably mismatched emotion.

0: Clearly wrong emotion or impossible to judge from the audio.

Ignore audio quality, accent, speaker identity, pronunciation

-unless they prevent judging emotion.

Output only the integer score (0, 1, 2, or 3).

Inputs

text: {{{text}}}

audio: {{{audio}}}

Output:

text: only the integer score (0, 1, 2, or 3)
\end{Verbatim}
\end{PromptBox}

\newpage
\section{Experimental Details}

\subsection{Voice Cloning Implementation}
Across all \vc{} backends, we largely follow each model's recommended inference configuration, and only expose a small set of high-level decoding hyperparameters for reproducibility. Unless otherwise noted, reference prompts are assigned in a round-robin manner across utterances. For CosyVoice~2, we run non-streaming synthesis with a fixed speed factor of \texttt{1.0}, and resample prompt audio to \texttt{16\,kHz} (\texttt{prompt\_sample\_rate=16000}).

For audio-LLM-based models, we use controlled stochastic decoding. FishSpeech uses temperature sampling with nucleus sampling (\texttt{temperature=0.8}, \texttt{top\_p=0.8}), with a maximum generation budget (\texttt{max\_new\_tokens=1024}) and chunked decoding for long inputs (\texttt{chunk\_length=200}). Higgs Audio~v2 adopts low-temperature decoding with both top-$k$ and top-$p$ constraints (\texttt{temperature=0.1}, \texttt{top\_k=50}, \texttt{top\_p=0.95}). MGM-Omni is run with stochastic decoding at \texttt{temperature=0.3} and an instruction-based template, e.g., \texttt{``Please read the following text using the same voice as the provided audio sample: \{text\}.''} For MOSS-TTSD v0.7, we keep default decoding settings but standardize instruction prompting with an explicit system message: \texttt{``You are a speech synthesizer that generates natural, realistic, and human-like audio from text.''} Spark-TTS follows temperature-based sampling with top-$k$/top-$p$ control (\texttt{temperature=0.3}, \texttt{top\_k=50}, \texttt{top\_p=0.95}). OZSpeech uses conservative decoding with a very low temperature (\texttt{temperature=0.01}).

For diffusion-based models, we specify their principal inference controls. PlayDiffusion uses \texttt{init\_temp=1.0}, \texttt{init\_diversity=1.0}, and \texttt{top\_k=25}. StyleTTS2 uses \texttt{5} diffusion steps together with \texttt{alpha=0.3}, \texttt{beta=0.7}, and \texttt{embedding\_scale=1.0}. VibeVoice uses the public \texttt{vibevoice/VibeVoice-1.5B} checkpoint, voice-prefill conditioning, \texttt{10} inference steps, \texttt{bfloat16}, and greedy decoding. GLM-TTS uses its default inference setup with \texttt{bfloat16} and FlashAttention when available.


\subsection{Proactive Perturbation Implementation}

\paragraph{GRNoise (Gaussian Random Noise)}
We apply independent and identically distributed Gaussian white noise to the waveform,  i.e., sampling from $N(0, \epsilon^2)$ with noise radius $\epsilon = 0.03137$ being similar to other adversarial methods. 
Each audio utterance is perturbed independently, and the resulting waveform is clipped to the valid audio range $[-1, 1]$.
This baseline does not rely on any model or optimization and serves as a simple random-noise baseline.

\paragraph{SafeSpeech and SPEC}
We implement the SafeSpeech protection framework using the BERT-VITS2 model as a surrogate TTS model.
In SPEC mode, we optimize perturbations to degrade TTS training via mel reconstruction loss, noise reconstruction loss, and KL divergence (with coefficients $\eta_\text{mel}=45$, $\eta_\text{kl}=1$, and $\beta=10$). 
In SafeSpeech mode, a perceptual loss term (e.g., STOI or STFT-based) is additionally added with weight $\eta_\alpha = 1.0$ to ensure imperceptibility. 
Perturbations are computed via projected gradient ascent (step size $\alpha = \epsilon / 10$, max norm $\epsilon = 0.03137$), and clipped to waveform range $[-1, 1]$. We use a batch size of 8, with 200 perturbation epochs and a sampling rate of 24~kHz.

\paragraph{POP}
We implement the Pivotal Objective Perturbation (POP) strategy by using BERT-VITS2 as a surrogate TTS model. 
We jointly maximize the model’s training losses, i.e., mel reconstruction, generator-discriminator feature loss, duration prediction error, and latent KL divergence with coefficients $\eta_\text{mel}=45$ and $\eta_\text{kl}=1$. 
The perturbation is the same with Safespeech and SPEC.

\paragraph{Enkidu}
Enkidu applies a speaker-specific universal perturbation in the frequency domain (STFT).
We train a complex-valued noise patch using the SpeechBrain SpeakerRecognition model as a surrogate, maximizing speaker embedding divergence while preserving intelligibility. 
During training, we randomly mask a ratio of 0.3 of STFT frames and apply a random offset to vary frame alignment. 
We follow the perturbation approach outlined in the original work and select a learning rate of 0.1 over 10 perturbation epochs, with a decay factor of 0.2.
The perturbation intensity is set as 0.4, and optionally smoothed via Wiener filtering.
Audio is reconstructed via inverse STFT with sampling rate as 16~kHz, window size as 1024, and hop length as 512.

\noindent \textbf{Washer (Purification-based bypass).}
We model an attacker-side bypass that attempts to remove proactive perturbations by treating them as ``noise'' and applying a speech denoiser before cloning.
Specifically, we use the Facebook \texttt{denoiser} toolkit with the pretrained DEMUCS model (\texttt{-DEMUCS}). We run the denoiser at 16~kHz and resample the purified audio back to the target sampling rate required by each \vc~model.

\subsection{Deepfake Detection Implementation and Interpretation}
\label{app_sec:deepfake_detection}

For \texttt{RVC-Detectability}, we construct an evaluation set of 600 audio samples from 10 speakers sampled from VCTK. 
For each speaker, we include 30 ground-truth utterances from VCTK, 10 generated utterances with benign text from VCTK, and 20 generated utterances with scam text from Robocall~\citep{prasad2023robocall}.
We randomly shuffle all utterances and evaluate whether state-of-the-art speech deepfake detectors can separate ground-truth speech from generated speech.

We emphasize that \texttt{RVC-Detectability} is intended as a detector-facing safety diagnostic rather than a direct measure of generation robustness.
It quantifies benchmark-level separability between the ground-truth and generated sets under existing deepfake detection methods.
Thus, higher or lower separability should not be interpreted as directly indicating stronger or weaker \vc~robustness.

We evaluate \texttt{RVC-Detectability} by directly feeding each ground-truth or generated utterance into a set of state-of-the-art speech deepfake detectors~\citep{dowerah2026speech}.
The detectors are grouped into four families: 
(1) self-supervised learning (SSL)-based detectors, including XLSR-SLS~\citep{zhang2024audio}, Wav2Vec2-ECAPA, HuBERT-ECAPA, WavLM-ECAPA~\citep{kulkarni2024exploring}, and TCM-ADD~\citep{truong2024temporal};
(2) graph neural network (GNN)-based detectors, including RawGAT-ST~\citep{tak2021end} and AASIST~\citep{jung2022aasist};
(3) convolutional neural network (CNN)-based detectors, represented by RawNet2~\citep{tak2021end_raw};
and (4) an audio-based LLM-as-a-judge detector, SQ-LLM~\citep{wang2025speechllm}.

For the first three families of detectors, we use their released inference pipelines to obtain a generated-speech score for each utterance. 
All test utterances are processed independently, without speaker-level or utterance-level adaptation. 
Given an utterance $x$, a detector produces a scalar score $s(x)$, where a larger score indicates a higher likelihood of generated speech. 
Since different released detectors may use different score conventions, we standardize all scores so that larger values consistently indicate a higher likelihood of generated speech. 
We then compare the detector scores with the ground-truth/generated labels and compute EER, minDCF, and ACC over the full evaluation split. 
For EER and minDCF, we sweep the decision threshold over the score distribution. 
For ACC, we use the detector's default operating threshold when available; otherwise, we use the threshold specified by the detector's released evaluation protocol.

For SQ-LLM, which performs audio-based LLM-as-a-judge detection rather than conventional classifier-based detection, we follow its original implementation and use the detection prompt shown below. 
For each utterance, SQ-LLM is prompted to determine whether the speech is ground-truth or generated. 
We use the prediction scores produced by its released evaluation script and compute EER, minDCF, and ACC under the same metric protocol as the other detectors.
%
We follow the original implementation of SQ-LLM, which is derived from Qwen2.5-Omni, and perform inference with the detection prompt below:

\begin{PromptBox}{Deepfake detection prompt}
[Audio: \textless AUDIO\_PLACEHOLDER \textgreater] \\
Determine if this speech is real or a deepfake.
\end{PromptBox}

\paragraph{Interpretation and limitations.}
Following conventional speech deepfake detection practice, we report EER, minDCF, and ACC to measure detector-facing separability between ground-truth and generated speech.
These results are intended as an aggregate detector-based safety diagnostic under the constructed \texttt{RVC-Detectability}, rather than as a direct measure of \vc~generation robustness or human perceptual realism.
As with many real-vs-generated detection benchmarks, the scores may reflect not only synthesis artifacts but also dataset-domain factors, such as recording conditions, preprocessing differences, or prompt-domain shifts between VCTK-style and Robocall-style texts.
We therefore interpret the results as benchmark-level separability under existing detector-based protocols, while leaving a more tightly controlled analysis of synthesis-specific artifacts to future work.

\clearpage

\section{Additional Analysis}
\label{app:additional_results}

\subsection{Overall performance of Voice Cloning}
\begin{table}[t]
\centering
\caption{\textbf{\vc{} performance across datasets.}
S/M/W/C/R denote SIM, MOS, WER, MCD, and RTF, respectively.
Cross-Lingual denotes English--Mandarin evaluation on Bilingual-UEDIN.
Higher is better for S and M; lower is better for W, C, and R.
\textbf{Bold} and \underline{underline} indicate best and runner-up per dataset/metric. ``--'' indicates not evaluated.}
\label{tab:all-results}
\scriptsize
\setlength{\tabcolsep}{2.1pt}
\renewcommand{\arraystretch}{0.93}
\begin{adjustbox}{max width=\textwidth}
\begin{tabular}{@{}l *{20}{c}@{}}
\toprule
& \multicolumn{5}{c}{\texttt{LibriTTS}} 
& \multicolumn{5}{c}{\texttt{VCTK}} 
& \multicolumn{5}{c}{\texttt{Chinese-VC}} 
& \multicolumn{5}{c}{\texttt{Cross-Lingual}} \\
\cmidrule(lr){2-6} \cmidrule(lr){7-11} \cmidrule(lr){12-16} \cmidrule(lr){17-21}
\textbf{Model}
& S$\uparrow$ & M$\uparrow$ & W$\downarrow$ & C$\downarrow$ & R$\downarrow$
& S$\uparrow$ & M$\uparrow$ & W$\downarrow$ & C$\downarrow$ & R$\downarrow$
& S$\uparrow$ & M$\uparrow$ & W$\downarrow$ & C$\downarrow$ & R$\downarrow$
& S$\uparrow$ & M$\uparrow$ & W$\downarrow$ & C$\downarrow$ & R$\downarrow$ \\
\midrule
FishSpeech
& 0.47 & \underline{4.37} & 0.17 & 6.47 & 3.61
& 0.43 & 4.27 & 0.03 & 4.11 & 2.51
& 0.61 & 2.90 & 0.47 & 4.30 & 2.15
& 0.57 & 4.22 & 0.24 & 6.05 & 2.71 \\

Fish Audio S2
& 0.54 & \underline{4.37} & \textbf{0.04} & 6.16 & 4.67
& 0.51 & 4.27 & \textbf{0.01} & 4.89 & 4.82
& 0.66 & 2.84 & 0.43 & 4.49 & 7.73
& 0.62 & \underline{4.24} & 0.23 & 5.83 & 11.40 \\

OZSpeech
& 0.39 & 3.21 & 0.06 & 6.87 & 8.75
& 0.25 & 3.22 & \underline{0.02} & 7.25 & 2.81
& 0.00 & 1.69 & 1.02 & 9.61 & 270.04
& 0.17 & 2.40 & 0.54 & 10.61 & 23.82 \\

StyleTTS2
& 0.23 & 4.30 & \underline{0.05} & 6.81 & \underline{0.11}
& 0.24 & \textbf{4.32} & \textbf{0.01} & 4.79 & \underline{0.14}
& 0.11 & \textbf{3.73} & 1.01 & 8.56 & \textbf{0.04}
& 0.21 & \textbf{4.37} & 0.81 & 9.42 & \textbf{0.06} \\

Spark-TTS
& 0.41 & 4.06 & 0.33 & 5.83 & 1.56
& 0.53 & 3.89 & 0.22 & 3.86 & 1.02
& 0.57 & 2.84 & 0.56 & 7.86 & 2.18
& 0.48 & 3.99 & 0.32 & \underline{5.65} & 1.43 \\

MOSS-TTSD v0.7
& 0.49 & 4.10 & 0.38 & 7.09 & 0.62
& 0.44 & 3.95 & 0.20 & 3.88 & 0.81
& 0.44 & 2.06 & 3.92 & 6.89 & 0.82
& 0.44 & 3.24 & 1.20 & 8.50 & 1.03 \\

MOSS-TTS v1.5
& 0.57 & 4.32 & 0.06 & 6.66 & 0.76
& 0.51 & 4.24 & \textbf{0.01} & 6.92 & 1.47
& 0.68 & 2.58 & 0.61 & 4.64 & 0.94
& 0.65 & 4.12 & 0.20 & 6.33 & 1.22 \\

dots.tts
& \underline{0.60} & 4.17 & 0.06 & 6.11 & 0.67
& 0.57 & 4.10 & 0.03 & 4.27 & 0.52
& 0.68 & 2.49 & 0.44 & 4.20 & 0.74
& 0.61 & 4.00 & 0.27 & 5.97 & 1.47 \\

Higgs (0.3)
& 0.56 & 4.30 & 0.25 & 6.06 & 1.42
& 0.52 & 4.16 & 0.19 & 4.06 & 1.44
& 0.58 & 2.72 & 0.78 & \textbf{3.46} & 1.91
& 0.54 & 4.15 & 0.49 & 7.61 & 3.12 \\

Higgs TTS 3
& 0.56 & 4.23 & \underline{0.05} & 6.32 & 0.64
& 0.45 & 4.06 & \textbf{0.01} & 5.75 & 0.76
& 0.63 & 2.37 & 0.38 & 4.45 & 0.58
& 0.30 & 4.17 & 0.25 & 7.74 & 0.74 \\

\midrule
CosyVoice
& 0.58 & \underline{4.37} & \underline{0.05} & 6.02 & 4.81
& \underline{0.58} & 4.21 & \underline{0.02} & \underline{3.68} & 2.38
& \textbf{0.72} & 2.68 & \underline{0.15} & 3.84 & 2.23
& 0.65 & 4.20 & 0.17 & 5.91 & 1.59 \\

GLM-TTS
& 0.57 & 4.08 & 0.09 & 6.41 & 1.74
& 0.57 & 4.01 & \underline{0.02} & 3.94 & 1.21
& 0.69 & 2.33 & 0.35 & 4.13 & 0.96
& \underline{0.66} & 3.96 & 0.30 & 6.07 & 0.76 \\

VibeVoice
& 0.48 & 3.83 & 0.23 & 6.76 & 1.86
& 0.44 & 3.95 & 0.06 & 5.09 & 1.20
& 0.56 & 2.47 & 0.23 & 6.82 & 0.86
& 0.53 & 3.81 & \underline{0.13} & 6.26 & 2.34 \\

MGM-Omni
& 0.54 & 4.28 & 0.09 & \underline{5.82} & 0.84
& 0.45 & 3.76 & 0.24 & 4.58 & 0.95
& \underline{0.71} & 2.69 & 0.18 & \underline{3.76} & 1.03
& 0.63 & 4.07 & 0.17 & 5.85 & 1.00 \\

PlayDiffusion
& 0.51 & 4.15 & \underline{0.05} & 8.06 & 0.73
& 0.43 & 4.19 & \textbf{0.01} & 9.07 & 0.81
& 0.44 & 2.78 & 1.79 & 14.61 & 1.39
& 0.46 & 4.09 & 0.86 & 9.60 & 0.68 \\

\midrule
Qwen3-TTS
& \textbf{0.61} & \textbf{4.39} & \underline{0.05} & \textbf{5.79} & 2.02
& \textbf{0.62} & \underline{4.30} & \textbf{0.01} & \textbf{3.44} & 1.97
& \textbf{0.72} & \underline{2.92} & \textbf{0.13} & 3.84 & 2.88
& \textbf{0.67} & 4.14 & \textbf{0.12} & \textbf{5.54} & 2.01 \\

F5-TTS
& 0.56 & 3.99 & 0.12 & 6.96 & 0.61
& 0.54 & 4.09 & \underline{0.02} & 4.75 & 0.94
& 0.70 & 1.85 & 0.42 & 5.02 & 1.12
& 0.65 & 3.92 & 0.24 & 5.86 & 1.35 \\

MaskGCT
& 0.57 & 3.93 & 0.09 & 6.91 & 1.36
& 0.56 & 3.89 & \underline{0.02} & 4.73 & 1.66
& 0.67 & 2.07 & 0.35 & 4.92 & 2.74
& 0.63 & 3.53 & 0.35 & 6.86 & 1.16 \\

OpenVoice
& 0.24 & 4.30 & 0.07 & 7.06 & \textbf{0.08}
& 0.39 & 4.24 & \underline{0.02} & 5.63 & \textbf{0.06}
& 0.43 & 2.87 & 0.46 & 6.34 & \underline{0.22}
& 0.30 & 4.00 & 0.28 & 6.17 & \underline{0.21} \\

XTTS
& 0.45 & 3.81 & 0.07 & 8.62 & 0.62
& 0.45 & 3.96 & \textbf{0.01} & 9.14 & 0.46
& 0.57 & 2.35 & 0.40 & 11.83 & 0.95
& 0.51 & 3.72 & 0.15 & 8.51 & 0.71 \\

IndexTTS
& \textbf{0.61} & 4.06 & \underline{0.05} & 6.61 & 2.23
& 0.57 & 4.05 & 0.03 & 4.68 & 3.31
& \textbf{0.72} & 2.26 & 0.39 & 3.99 & 4.12
& \textbf{0.67} & 3.90 & 0.21 & 5.79 & 3.95 \\

ZipVoice
& 0.58 & 4.13 & \underline{0.05} & 7.09 & 1.46
& 0.55 & 4.17 & \textbf{0.01} & 4.98 & 2.53
& \underline{0.71} & 2.59 & 0.45 & 4.95 & 2.11
& 0.63 & 3.85 & 0.26 & 7.40 & 10.05 \\
\bottomrule
\end{tabular}
\end{adjustbox}
\end{table}

Table~\ref{tab:all-results} presents the \vc~performance in two languages supported by most of \vc{} models (i.e., English and Chinese).
And we compute all \vc~performance on \texttt{RVC-Multilingual}, including in-language \texttt{English-VC-VCTK}, \texttt{English-VC-LibriTTS}, \texttt{Chinese-VC} and cross-language \texttt{CrossLingual}.
Across different generation architectures, most LLM-based \vc~models competitively maintain spectral consistency and speaker identity.
While diffusion-based \vc~models improve content accuracy, they primarily fail to support multiple languages.
The hybrid \vc~models take advantage of both sides, achieving better performance in all evaluations.
Meanwhile, \vc~models exhibit a clear trade-off between generation quality and inference efficiency.
Specifically, slower inference is typically associated with lower error rates and stronger spectral consistency with the target speaker’s ground-truth utterances.
%
Table~\ref{tab:all-results} reports the corresponding generation-quality results; for example, CosyVoice obtains a SIM of 0.58 on both LibriTTS and VCTK.

\subsection{Language shifts beyond English cause a sharp robustness instability.}
For \texttt{RVC-Multilingual}, we compare the English results in against the \texttt{Chinese-VC} and \texttt{Cross-Lingual} settings in Table~\ref{tab:all-results}, and observe a clear robustness drop when moving beyond English.
On English (\texttt{English-VC-LibriTTS} and \texttt{English-VC-VCTK}), many models achieve high naturalness (MOS around 4) with low WER (often below 0.1), consistent with their heavy exposure to English training data. In contrast, \texttt{Chinese-VC} markedly degrades perceptual quality across all evaluated \vc{} models (MOS shifts to the mid-to-low range), while content accuracy deteriorates dramatically (WER rises for most models). \texttt{Cross-Lingual} cloning further compounds these issues: although some models maintain relatively strong speaker similarity, preserving linguistic content becomes the primary bottleneck, evidenced by higher WER and less consistent spectral quality. Overall, these results suggest that language shifts remain fragile and vary considerably across backends.


\subsection{Demography distributions}
\begin{figure}
    \centering
    \includegraphics[width=\linewidth]{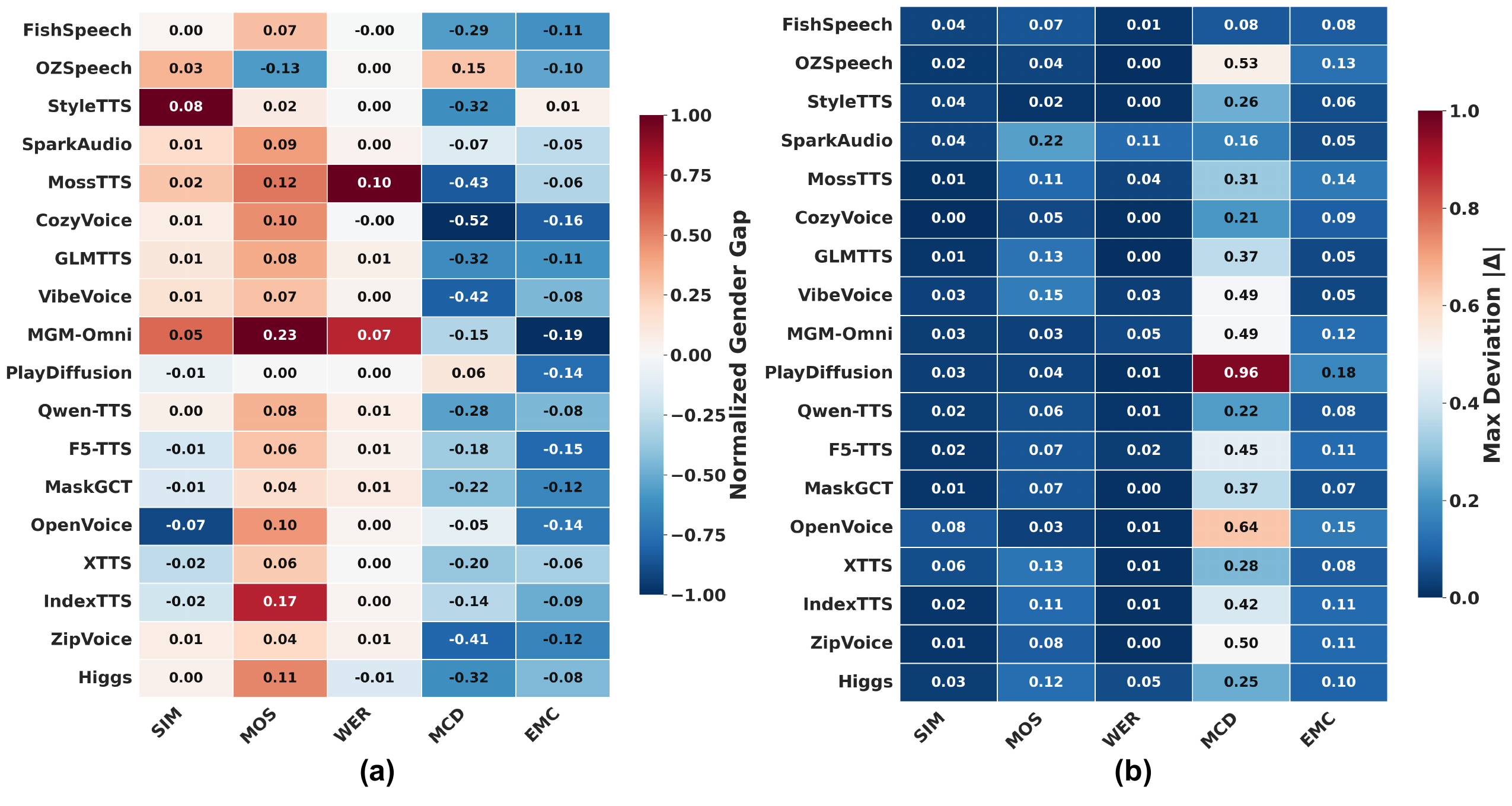}
    \caption{\textbf{Comparison of performance gaps.} (a) gap in gender performance. (b) absolute max deviation over age.}
    \label{fig:gender_gap}
\end{figure}

\textit{In terms of gender}, we compute the gender-gap $\Delta = F-M$ for SIM/MOS/EMC, and $\Delta = M-F$ for WER/MCD, and normalize gaps by metrics in Fig.~\ref{fig:gender_gap}. 
It indicates that there is a consistent male advantage on MCD across almost all models, while speaker similarity and content accuracy remain relatively stable across genders.
\textit{By the change gap} from $\Delta_{\leq20} -\Delta_{20-29}$  to $\Delta_{\geq30} -\Delta_{20-29}$, we can tell that the age of target speaker does not impact \vc.

  

\subsection{User study}
\label{sec:user_study}

Since \modelname{} is designed as a large-scale robustness benchmark, a full human MOS study over all generated samples is prohibitively expensive. Moreover, the goal of this work is not to propose or validate new evaluation metrics. The metrics adopted in our benchmark have been introduced and comprehensively validated in prior work. We therefore use these metrics as scalable evaluation
tools, rather than treating metric validation as the focus of this study. This practice is also aligned with recent TTS and \vc{} evaluation, where objective or model-based metrics are used for large-scale benchmarking and smaller human
studies are included mainly for validation~\citep{zhang2025safespeech,manku2025emergenttts,huang2025instructttseval}. Accordingly, we conduct a pilot human study only as a sanity check for the two perceptual dimensions most relevant to our benchmark: naturalness and text-audio emotion alignment.

\paragraph{Protocol.}
We sample a stratified subset of generated clips covering multiple VCL models and different ranges of automatic scores, where amount of survey data aligns with the existing literature~\cite {zhang2025safespeech} and can be done in a reasonable time for participants. The clips are anonymized and presented in randomized order. Each participant rates each clip along two dimensions: (i) \textbf{naturalness}, using a 1-5 MOS scale, where 1 indicates very unnatural or hard-to-understand speech and 5 indicates highly natural speech; and (ii) \textbf{text-audio emotion alignment}, using a 1--3 scale, where 1 indicates weak or incorrect emotional alignment, 2 indicates partial alignment, and 3 indicates clear alignment with the intended emotion or speaking style. 

\paragraph{Results.}
The pilot results are directionally consistent with our automatic metrics. For emotion alignment, clips assigned higher Gemini-based EmotionAlignment scores also receive higher mean human emotion-alignment ratings on average: 1.83 for the higher-score group versus 1.58 for the lower-score group. This supports the use of the LALM judge as a scalable signal for prompt-level expressive preservation.

For naturalness, the pilot is less informative for fine-grained model ranking because the sampled clips receive high human MOS overall. The average human naturalness rating is 4.67, while the corresponding automatic MOS proxy average is 4.33. This suggests that the automatic naturalness proxy is broadly consistent with human perception at the coarse level, but the human ratings are saturated on this subset and should not be interpreted as validating fine-grained naturalness differences between high-quality models.

Overall, the pilot study is intended as supporting evidence rather than a replacement for a large-scale perceptual evaluation. We therefore use human ratings to contextualize the automatic metrics, while relying on SpeechMOS, DNSMOS, and Gemini-based EmotionAlignment for scalable benchmark-wide evaluation.

\subsection{Architecture and Parameter Scale}
\label{app:architecture_scale}

We further analyze whether robustness is more closely associated with the model architecture or with raw parameter scale. Although parameter count is often used as a coarse indicator of model capacity, our results suggest that it is not a reliable predictor of robustness in \vc{}. Instead, the generation backbone and conditioning design appear to explain the observed robustness trends more clearly.

We group the evaluated models into three families: autoregressive codec-token language models, diffusion/flow-based models, and hybrid LM+diffusion/flow systems. These families show distinct robustness profiles. Autoregressive codec-token models are generally competitive in clean English settings, but several of them degrade substantially under harder language shifts. Diffusion/flow-based systems often achieve strong content accuracy on easier English tasks, but their speaker similarity can remain limited. Hybrid systems provide the most balanced performance across multilingual and cross-lingual settings, suggesting that combining language-level planning with acoustic generation can improve robustness under distribution shift.

For example, on English-VCTK, diffusion-based models such as StyleTTS2 and PlayDiffusion achieve the best WER of 0.01, indicating strong content accuracy in this relatively clean setting. However, their SIM scores remain lower, at 0.24 and 0.43 respectively, than those of strong hybrid systems such as CosyVoice and GLM-TTS, which obtain SIM scores of 0.58 and 0.57 with WER of 0.02. Under harder language shifts, the advantage of hybrid systems becomes more evident. On Chinese-VC, CosyVoice, MGM-Omni, and GLM-TTS achieve SIM scores of 0.72, 0.71, and 0.69, respectively; on Cross-Lingual evaluation, they remain at 0.66, 0.63, and 0.66. In contrast, several autoregressive models show sharper degradation in content accuracy, such as FishSpeech, whose WER increases from 0.03 on English-VCTK to 0.47 on Chinese-VC, and Higgs, whose WER increases from 0.19 to 0.78.

We do not observe a monotonic trend with respect to parameter scale alone. For instance, CosyVoice, with 639M parameters, consistently outperforms Higgs, with 5.97B parameters, across language-shift settings. Similarly, GLM-TTS, with 2.16B parameters, remains more robust than Higgs on both Chinese-VC and Cross-Lingual evaluation. Conversely, some smaller diffusion/flow-based models perform well on easier content-transcription metrics but are less reliable overall: StyleTTS2, with 191M parameters, achieves English-VCTK WER of 0.01, and OZSpeech, with 247M parameters, reaches WER of 0.02, but their SIM scores remain much lower than those of the strongest hybrid systems.

Overall, these results suggest that \vc{} robustness is shaped more by architecture family, conditioning mechanism, multilingual training, and speaker-representation design than by model scale alone. We emphasize that this analysis is descriptive rather than causal, since released models differ not only in parameter count but also in training data, objectives, alignment strategies, and inference pipelines. Nevertheless, the observed trends indicate that future robust \vc{} models should prioritize architectural and conditioning improvements, rather than relying solely on scaling.

\section{Future Directions}
\label{app:future_directions}

\modelname{} is designed primarily as a diagnostic benchmark rather than as a new
robust \vc{} training algorithm. Nevertheless, the systematic failure modes
observed in our evaluation suggest several directions for improving synthesis
robustness. In this section, we discuss how the proposed benchmark can guide
future work on robustness-oriented training, architecture design, and safety
evaluation.

\paragraph{Robust reference conditioning.}
Our results show that reference-audio variations, including accent, background
noise, and multi-speaker interference, can degrade both content accuracy and
speaker fidelity. Future \vc{} systems may benefit from reference-quality
estimation, speaker-disentangled representations, multi-reference aggregation,
and training with corrupted or demographically diverse reference audios. These
techniques could reduce over-dependence on superficial acoustic cues and improve
identity preservation under realistic recording conditions. RVC-AudioShift and
RVC-PassiveNoise provide targeted evaluation settings for measuring progress in
this direction.

\paragraph{Robust text and semantic conditioning.}
Text-prompt shifts, especially formatting irregularities and out-of-distribution
content, can increase content errors in current \vc{} systems. This suggests
that robust \vc{} pipelines should incorporate stronger text normalization,
prompt sanitization, uncertainty-aware generation, and semantic consistency
checking between the input text and synthesized speech. Such methods can be
evaluated directly using RVC-TextShift, which probes whether models preserve
linguistic content under realistic prompt variations.

\paragraph{Long-form and expressive consistency.}
Long-text generation exposes content omissions, substitutions, and acoustic
drift, while expressive prompts reveal unstable emotion and style rendering.
Future models may require explicit long-context planning, hierarchical
generation, speaker-consistency regularization, and prosody-control objectives.
RVC-LongContext and RVC-Expression provide complementary testbeds for evaluating
whether such mechanisms improve long-horizon identity preservation, content
fidelity, and paralinguistic controllability.

\paragraph{Output and channel robustness.}
Generated speech is often compressed, re-encoded, or transmitted through
narrowband channels in practical applications. The degradation observed under
post-processing suggests that future systems could benefit from codec-aware
training, compression augmentation, differentiable audio post-processing, and
vocoder designs optimized for downstream transmission stability. RVC-Compression
can be used to measure whether these techniques preserve intelligibility,
spectral consistency, and speaker-related cues after common signal-processing
operations.

\paragraph{Perturbation-aware robustness and safety.}
Passive noise and proactive anti-cloning perturbations expose a tension between
robustness and safety. In authorized \vc{}, systems should be robust to benign
reference corruption such as background noise or channel artifacts. In contrast,
privacy-preserving defenses intentionally aim to reduce cloneability under
unauthorized use. Future work should therefore distinguish robustness to natural
corruption from robustness against protective perturbations. For authorized
cloning, promising directions include noise-robust speaker encoders,
source-separation-aware conditioning, and denoising-aware training. For
anti-cloning protection, defenses should be evaluated against adaptive
purification and denoising-based countermeasures. RVC-AdvNoise and
RVC-AntiProtect provide initial protocols for studying this trade-off.

\begin{table}[t]
\centering
\small
\setlength{\tabcolsep}{5pt}
\renewcommand{\arraystretch}{1.3}
\caption{\textbf{Potential robustness-improvement directions suggested by \modelname.}}
\label{tab:future_robustness_directions}
\begin{tabular}{p{0.27\linewidth} p{0.44\linewidth} p{0.21\linewidth}}
\toprule
\textbf{Observed vulnerability} &
\textbf{Potential future measure} &
\textbf{\modelname{} component} \\
\midrule
\textit{Accent \& demographic sensitivity} &
Demographically balanced training, speaker-disentangled embeddings, and
group-aware calibration &
\texttt{RVC-AudioShift} \\[3pt]
\textit{Noisy or multi-speaker references} &
Reference-quality scoring, source separation, multi-reference aggregation,
and noise augmentation &
\texttt{RVC-PassiveNoise} \\[3pt]
\textit{Irregular or OOD text prompts} &
Text normalization, prompt sanitization, uncertainty-aware generation,
and semantic consistency checking &
\texttt{RVC-TextShift} \\[3pt]
\textit{Long-form content \& speaker drift} &
Hierarchical generation, explicit memory mechanisms, chunk-level consistency
constraints, and speaker-consistency regularization &
\texttt{RVC-LongContext} \\[3pt]
\textit{Weak expressive or prosodic control} &
Prosody supervision, emotion/style disentanglement, and
style-consistency objectives &
\texttt{RVC-Expression} \\[3pt]
\textit{Compression \& channel sensitivity} &
Codec-aware training, compression augmentation, differentiable
post-processing, and robust vocoder design &
\texttt{RVC-Compression} \\[3pt]
\textit{Vulnerability to adversarial perturbations} &
Perturbation-aware speaker encoders and adaptive robustness evaluation &
\texttt{RVC-AdvNoise} \\[3pt]
\textit{Denoising against anti-cloning protection} &
Adaptive defense evaluation and perturbations robust to purification attacks &
\texttt{RVC-AntiProtect} \\
\bottomrule
\end{tabular}
\end{table}

Overall, these directions indicate that \modelname{} can be used not only to compare existing \vc{} systems, but also to guide the development of models that
are more robust under realistic deployment shifts. We hope that future work will
use \modelname{} to evaluate whether proposed training objectives, architectural
components, and safety mechanisms improve robustness across the full \vc{}
pipeline.

\clearpage
\section{Additional Benchmarking Results}
\label{app:detailed-results}
\subsection{Detailed results}
This subsection provides the detailed numeric results in the paper.

\paragraph{Uncertainty reporting.} Unless a table states otherwise, every $\pm$ value for generation-quality metrics reports the standard error of the displayed mean from 1,000 speaker-level cluster-bootstrap resamples. Each resample draws speakers with replacement and retains all utterances belonging to every sampled speaker, so utterances from the same speaker are not treated as independent observations. For an accent represented by one speaker, between-speaker uncertainty is not estimable and is reported as N/A; these rows are descriptive only. RTF is reported as a point estimate.

\begin{table}[htbp]
\centering
\small
\setlength{\tabcolsep}{2pt}
\caption{\textbf{Performance on LibriTTS and VCTK.} Appendix result tables cover the 18 representative main-paper models plus four extension systems: Fish Audio S2, MOSS-TTS v1.5, dots.tts, and Higgs TTS 3.}
\begin{tabularx}{\textwidth}{l l *{7}{>{\centering\arraybackslash}X}}
\toprule
\textbf{Dataset} & \textbf{Baseline} & \textbf{SIM} $\uparrow$ & \textbf{MOS} $\uparrow$ & \textbf{WER} $\downarrow$ & \textbf{MCD} $\downarrow$ & \textbf{RTF} $\downarrow$ & \textbf{SVA} $\uparrow$ & \textbf{EMC} $\uparrow$ \\
\midrule
\multirow{22}{*}{LibriTTS} & FishSpeech & $\substd{0.47}{0.01}$ & $\substd{4.37}{0.01}$ & $\substd{0.17}{0.01}$ & $\substd{6.47}{0.10}$ & 3.61 & $\substd{0.91}{0.01}$ & $\substd{0.68}{0.02}$ \\
 & Fish Audio S2 & $\substd{0.54}{0.01}$ & $\substd{4.37}{0.01}$ & $\substd{0.04}{0.00}$ & $\substd{6.16}{0.10}$ & 4.67 & $\substd{0.95}{0.01}$ & $\substd{0.71}{0.02}$ \\
 & OZSpeech & $\substd{0.39}{0.01}$ & $\substd{3.21}{0.03}$ & $\substd{0.06}{0.00}$ & $\substd{6.87}{0.11}$ & 8.75 & $\substd{0.84}{0.02}$ & $\substd{0.64}{0.02}$ \\
 & StyleTTS2 & $\substd{0.23}{0.01}$ & $\substd{4.30}{0.02}$ & $\substd{0.05}{0.00}$ & $\substd{6.81}{0.09}$ & 0.11 & $\substd{0.39}{0.03}$ & $\substd{0.59}{0.03}$ \\
 & Spark-TTS & $\substd{0.41}{0.02}$ & $\substd{4.06}{0.04}$ & $\substd{0.33}{0.03}$ & $\substd{5.83}{0.17}$ & 1.56 & $\substd{0.76}{0.03}$ & $\substd{0.67}{0.02}$ \\
 & MOSS-TTSD v0.7 & $\substd{0.49}{0.01}$ & $\substd{4.10}{0.03}$ & $\substd{0.38}{0.05}$ & $\substd{7.09}{0.12}$ & 0.62 & $\substd{0.88}{0.02}$ & $\substd{0.67}{0.02}$ \\
 & MOSS-TTS v1.5 & $\substd{0.57}{0.01}$ & $\substd{4.32}{0.02}$ & $\substd{0.06}{0.01}$ & $\substd{6.66}{0.09}$ & 0.76 & $\substd{0.95}{0.01}$ & $\substd{0.70}{0.02}$ \\
\bottomrule
\end{tabularx}
\end{table}

\begin{table}[ht]
\centering
\small
\setlength{\tabcolsep}{3pt}
\caption{\textbf{Performance on Bilingual.} Evaluation of cross-lingual synthesis between English (Eng) and Mandarin (Man).}
\label{tab:bilingual_perf}

\begin{tabularx}{\textwidth}{
    l
    l
    *{5}{>{\centering\arraybackslash}X}
}
\toprule
\textbf{Model} & \textbf{Direction} & 
{SIM $\uparrow$} & {MOS $\uparrow$} & {WER $\downarrow$} & {MCD $\downarrow$} & {SVA $\uparrow$} \\
\midrule

\multirow{2}{*}{FishSpeech} & Eng $\to$ Man & $\substd{0.58}{0.01}$ & $\substd{4.08}{0.04}$ & $\substd{0.40}{0.03}$ & $\substd{6.20}{0.15}$ & $\substd{1.00}{0.00}$ \\
 & Man $\to$ Eng & $\substd{0.55}{0.02}$ & $\substd{4.36}{0.02}$ & $\substd{0.07}{0.00}$ & $\substd{5.89}{0.17}$ & $\substd{0.99}{0.01}$ \\
\addlinespace[4pt]

\multirow{2}{*}{Fish Audio S2} & Eng $\to$ Man & $\substd{0.63}{0.01}$ & $\substd{4.11}{0.03}$ & $\substd{0.39}{0.03}$ & $\substd{5.96}{0.16}$ & $\substd{1.00}{0.00}$ \\
 & Man $\to$ Eng & $\substd{0.61}{0.02}$ & $\substd{4.37}{0.02}$ & $\substd{0.07}{0.00}$ & $\substd{5.70}{0.18}$ & $\substd{1.00}{0.00}$ \\
\addlinespace[4pt]

\multirow{2}{*}{OZSpeech} & Eng $\to$ Man & $\substd{0.03}{0.01}$ & $\substd{1.60}{0.01}$ & $\substd{1.00}{0.00}$ & $\substd{14.44}{0.45}$ & $\substd{0.01}{0.01}$ \\
 & Man $\to$ Eng & $\substd{0.30}{0.01}$ & $\substd{3.19}{0.04}$ & $\substd{0.07}{0.00}$ & $\substd{6.77}{0.19}$ & $\substd{0.72}{0.05}$ \\
\addlinespace[4pt]

\multirow{2}{*}{StyleTTS2} & Eng $\to$ Man & $\substd{0.17}{0.02}$ & $\substd{4.35}{0.02}$ & $\substd{1.55}{0.16}$ & $\substd{12.52}{0.21}$ & $\substd{0.21}{0.07}$ \\
\bottomrule
\end{tabularx}
\end{table}

\clearpage

\begin{table}[H]
\centering
\small
\setlength{\tabcolsep}{4pt}
\caption{\textbf{Accent Robustness (Part 1).} Performance evaluation across 12 English accents. N/A indicates that speaker-level uncertainty is not estimable for an accent represented by one speaker.}
\label{tab:accent_perf}

\begin{tabular}{
  l
  c
  c
  c
  c
  c
}
\toprule
\textbf{Accent} & 
{SIM $\uparrow$} & {MOS $\uparrow$} & {WER $\downarrow$} & {MCD $\downarrow$} & {EMC $\uparrow$} \\
\midrule

\multicolumn{6}{l}{\textit{\textbf{Model: FishSpeech}}} \\
\midrule
American       & $\substd{0.42}{0.02}$ & $\substd{4.31}{0.02}$ & $\substd{0.03}{0.01}$ & $\substd{4.01}{0.12}$ & $\substd{0.83}{0.04}$ \\
Australian     & $\substd{0.43}{\mathrm{N/A}}$ & $\substd{4.23}{\mathrm{N/A}}$ & $\substd{0.05}{\mathrm{N/A}}$ & $\substd{3.58}{\mathrm{N/A}}$ & $\substd{0.98}{\mathrm{N/A}}$ \\
British        & $\substd{0.62}{\mathrm{N/A}}$ & $\substd{4.35}{\mathrm{N/A}}$ & $\substd{0.03}{\mathrm{N/A}}$ & $\substd{3.75}{\mathrm{N/A}}$ & $\substd{0.72}{\mathrm{N/A}}$ \\
Canadian       & $\substd{0.43}{0.04}$ & $\substd{4.26}{0.02}$ & $\substd{0.02}{0.00}$ & $\substd{3.54}{0.11}$ & $\substd{0.83}{0.03}$ \\
English        & $\substd{0.44}{0.01}$ & $\substd{4.25}{0.03}$ & $\substd{0.03}{0.00}$ & $\substd{4.35}{0.09}$ & $\substd{0.76}{0.03}$ \\
Indian         & $\substd{0.37}{0.01}$ & $\substd{4.28}{0.01}$ & $\substd{0.03}{0.00}$ & $\substd{5.03}{0.12}$ & $\substd{0.58}{0.06}$ \\
\bottomrule
\end{tabular}
\end{table}

\begin{table}[H]
\centering
\small
\setlength{\tabcolsep}{4pt}
\caption{\textbf{Accent Robustness (Part 2).} Performance evaluation across 12 English accents. N/A indicates that speaker-level uncertainty is not estimable for an accent represented by one speaker.}
\label{tab:accent_perf_p2}

\begin{tabular}{
  l
  c
  c
  c
  c
  c
}
\toprule
\textbf{Accent} & 
{SIM $\uparrow$} & {MOS $\uparrow$} & {WER $\downarrow$} & {MCD $\downarrow$} & {EMC $\uparrow$} \\
\midrule

\multicolumn{6}{l}{\textit{\textbf{Model: Spark-TTS}}} \\
\midrule
American       & $\substd{0.50}{0.03}$ & $\substd{3.82}{0.03}$ & $\substd{0.21}{0.07}$ & $\substd{3.70}{0.14}$ & $\substd{0.89}{0.06}$ \\
Australian     & $\substd{0.58}{\mathrm{N/A}}$ & $\substd{3.84}{\mathrm{N/A}}$ & $\substd{0.12}{\mathrm{N/A}}$ & $\substd{4.35}{\mathrm{N/A}}$ & $\substd{0.98}{\mathrm{N/A}}$ \\
British        & $\substd{0.48}{\mathrm{N/A}}$ & $\substd{4.00}{\mathrm{N/A}}$ & $\substd{0.14}{\mathrm{N/A}}$ & $\substd{3.60}{\mathrm{N/A}}$ & $\substd{0.50}{\mathrm{N/A}}$ \\
Canadian       & $\substd{0.48}{0.06}$ & $\substd{3.89}{0.05}$ & $\substd{0.28}{0.11}$ & $\substd{3.79}{0.35}$ & $\substd{0.87}{0.03}$ \\
English        & $\substd{0.52}{0.01}$ & $\substd{3.82}{0.07}$ & $\substd{0.21}{0.04}$ & $\substd{3.96}{0.07}$ & $\substd{0.83}{0.05}$ \\
Indian         & $\substd{0.50}{0.04}$ & $\substd{3.97}{0.15}$ & $\substd{0.38}{0.20}$ & $\substd{4.20}{0.34}$ & $\substd{0.63}{0.03}$ \\
\bottomrule
\end{tabular}
\end{table}

\begin{table}[H]
\centering
\small
\setlength{\tabcolsep}{4pt}
\caption{\textbf{Accent Robustness (Part 3).} Performance evaluation across 12 English accents. N/A indicates that speaker-level uncertainty is not estimable for an accent represented by one speaker.}
\label{tab:accent_perf_p3}

\begin{tabular}{
  l
  c
  c
  c
  c
  c
}
\toprule
\textbf{Accent} & 
{SIM $\uparrow$} & {MOS $\uparrow$} & {WER $\downarrow$} & {MCD $\downarrow$} & {EMC $\uparrow$} \\
\midrule

\multicolumn{6}{l}{\textit{\textbf{Model: Higgs}}} \\
\midrule
American       & $\substd{0.52}{0.04}$ & $\substd{4.08}{0.09}$ & $\substd{0.19}{0.09}$ & $\substd{3.73}{0.12}$ & $\substd{0.85}{0.05}$ \\
Australian     & $\substd{0.53}{\mathrm{N/A}}$ & $\substd{4.09}{\mathrm{N/A}}$ & $\substd{0.09}{\mathrm{N/A}}$ & $\substd{3.75}{\mathrm{N/A}}$ & $\substd{0.96}{\mathrm{N/A}}$ \\
British        & $\substd{0.66}{\mathrm{N/A}}$ & $\substd{4.30}{\mathrm{N/A}}$ & $\substd{0.34}{\mathrm{N/A}}$ & $\substd{3.49}{\mathrm{N/A}}$ & $\substd{0.76}{\mathrm{N/A}}$ \\
Canadian       & $\substd{0.47}{0.05}$ & $\substd{4.02}{0.06}$ & $\substd{0.43}{0.13}$ & $\substd{3.36}{0.12}$ & $\substd{0.84}{0.02}$ \\
English        & $\substd{0.54}{0.02}$ & $\substd{4.19}{0.03}$ & $\substd{0.08}{0.02}$ & $\substd{4.30}{0.11}$ & $\substd{0.78}{0.03}$ \\
Indian         & $\substd{0.51}{0.00}$ & $\substd{4.22}{0.09}$ & $\substd{0.44}{0.23}$ & $\substd{4.91}{0.00}$ & $\substd{0.54}{0.03}$ \\
\bottomrule
\end{tabular}
\end{table}

\begin{table}[H]
\centering
\small
\setlength{\tabcolsep}{4pt}
\caption{\textbf{Accent Robustness (Part 4).} Performance evaluation across 12 English accents. N/A indicates that speaker-level uncertainty is not estimable for an accent represented by one speaker.}
\label{tab:accent_perf_p4}

\begin{tabular}{
  l
  c
  c
  c
  c
  c
}
\toprule
\textbf{Accent} & 
{SIM $\uparrow$} & {MOS $\uparrow$} & {WER $\downarrow$} & {MCD $\downarrow$} & {EMC $\uparrow$} \\
\midrule

\multicolumn{6}{l}{\textit{\textbf{Model: VibeVoice}}} \\
\midrule
American       & $\substd{0.42}{0.02}$ & $\substd{3.98}{0.06}$ & $\substd{0.06}{0.01}$ & $\substd{5.14}{0.13}$ & $\substd{0.86}{0.06}$ \\
Australian     & $\substd{0.44}{\mathrm{N/A}}$ & $\substd{3.90}{\mathrm{N/A}}$ & $\substd{0.07}{\mathrm{N/A}}$ & $\substd{4.55}{\mathrm{N/A}}$ & $\substd{0.98}{\mathrm{N/A}}$ \\
British        & $\substd{0.60}{\mathrm{N/A}}$ & $\substd{4.23}{\mathrm{N/A}}$ & $\substd{0.04}{\mathrm{N/A}}$ & $\substd{4.14}{\mathrm{N/A}}$ & $\substd{0.54}{\mathrm{N/A}}$ \\
Canadian       & $\substd{0.45}{0.04}$ & $\substd{3.96}{0.04}$ & $\substd{0.05}{0.01}$ & $\substd{4.48}{0.19}$ & $\substd{0.85}{0.04}$ \\
English        & $\substd{0.46}{0.01}$ & $\substd{3.95}{0.06}$ & $\substd{0.07}{0.01}$ & $\substd{5.19}{0.09}$ & $\substd{0.79}{0.02}$ \\
Indian         & $\substd{0.38}{0.03}$ & $\substd{3.92}{0.01}$ & $\substd{0.06}{0.01}$ & $\substd{5.76}{0.05}$ & $\substd{0.55}{0.03}$ \\
\bottomrule
\end{tabular}
\end{table}

\begin{table}[H]
\centering
\small
\setlength{\tabcolsep}{4pt}
\caption{\textbf{Accent Robustness (Part 5).} Performance evaluation across 12 English accents. N/A indicates that speaker-level uncertainty is not estimable for an accent represented by one speaker.}
\label{tab:accent_perf_p5}

\begin{tabular}{
  l
  c
  c
  c
  c
  c
}
\toprule
\textbf{Accent} & 
{SIM $\uparrow$} & {MOS $\uparrow$} & {WER $\downarrow$} & {MCD $\downarrow$} & {EMC $\uparrow$} \\
\midrule

\multicolumn{6}{l}{\textit{\textbf{Model: F5-TTS}}} \\
\midrule
American       & $\substd{0.54}{0.02}$ & $\substd{4.12}{0.03}$ & $\substd{0.02}{0.01}$ & $\substd{5.02}{0.27}$ & $\substd{0.81}{0.06}$ \\
Australian     & $\substd{0.56}{\mathrm{N/A}}$ & $\substd{3.92}{\mathrm{N/A}}$ & $\substd{0.02}{\mathrm{N/A}}$ & $\substd{4.40}{\mathrm{N/A}}$ & $\substd{0.98}{\mathrm{N/A}}$ \\
British        & $\substd{0.68}{\mathrm{N/A}}$ & $\substd{4.28}{\mathrm{N/A}}$ & $\substd{0.00}{\mathrm{N/A}}$ & $\substd{4.31}{\mathrm{N/A}}$ & $\substd{0.66}{\mathrm{N/A}}$ \\
Canadian       & $\substd{0.53}{0.03}$ & $\substd{3.98}{0.08}$ & $\substd{0.02}{0.01}$ & $\substd{4.47}{0.36}$ & $\substd{0.86}{0.02}$ \\
English        & $\substd{0.55}{0.01}$ & $\substd{4.14}{0.03}$ & $\substd{0.02}{0.00}$ & $\substd{4.88}{0.10}$ & $\substd{0.73}{0.04}$ \\
Indian         & $\substd{0.50}{0.02}$ & $\substd{4.05}{0.04}$ & $\substd{0.01}{0.00}$ & $\substd{5.03}{0.06}$ & $\substd{0.58}{0.04}$ \\
\bottomrule
\end{tabular}
\end{table}

\begin{table}[H]
\centering
\small
\setlength{\tabcolsep}{4pt}
\caption{\textbf{Accent Robustness (Part 6).} Performance evaluation across 12 English accents. N/A indicates that speaker-level uncertainty is not estimable for an accent represented by one speaker.}
\label{tab:accent_perf_p6}

\begin{tabular}{
  l
  c
  c
  c
  c
  c
}
\toprule
\textbf{Accent} & 
{SIM $\uparrow$} & {MOS $\uparrow$} & {WER $\downarrow$} & {MCD $\downarrow$} & {EMC $\uparrow$} \\
\midrule

\multicolumn{6}{l}{\textit{\textbf{Model: IndexTTS}}} \\
\midrule
American       & $\substd{0.58}{0.02}$ & $\substd{4.04}{0.05}$ & $\substd{0.04}{0.01}$ & $\substd{4.69}{0.18}$ & $\substd{0.81}{0.05}$ \\
Australian     & $\substd{0.55}{\mathrm{N/A}}$ & $\substd{3.84}{\mathrm{N/A}}$ & $\substd{0.05}{\mathrm{N/A}}$ & $\substd{4.49}{\mathrm{N/A}}$ & $\substd{0.96}{\mathrm{N/A}}$ \\
British        & $\substd{0.66}{\mathrm{N/A}}$ & $\substd{4.27}{\mathrm{N/A}}$ & $\substd{0.01}{\mathrm{N/A}}$ & $\substd{3.98}{\mathrm{N/A}}$ & $\substd{0.68}{\mathrm{N/A}}$ \\
Canadian       & $\substd{0.58}{0.03}$ & $\substd{3.99}{0.05}$ & $\substd{0.04}{0.01}$ & $\substd{4.21}{0.17}$ & $\substd{0.84}{0.04}$ \\
English        & $\substd{0.57}{0.01}$ & $\substd{4.05}{0.04}$ & $\substd{0.03}{0.01}$ & $\substd{5.11}{0.15}$ & $\substd{0.82}{0.02}$ \\
Indian         & $\substd{0.50}{0.01}$ & $\substd{4.06}{0.09}$ & $\substd{0.01}{0.00}$ & $\substd{5.10}{0.09}$ & $\substd{0.60}{0.06}$ \\
\bottomrule
\end{tabular}
\end{table}

\begin{table}[H]
\centering
\small
\setlength{\tabcolsep}{8pt}
\caption{\textbf{Performance on different gender groups (Part 1).} Performance evaluation breakdown by gender (F/M).}
\label{tab:gender_perf}

\begin{tabular}{
  l
  c
  c
  c
  c
  c
}
\toprule
\textbf{Gender} & 
{SIM $\uparrow$} & {MOS $\uparrow$} & {WER $\downarrow$} & {MCD $\downarrow$} & {EMC $\uparrow$} \\
\midrule

\multicolumn{6}{l}{\textit{\textbf{Model: FishSpeech}}} \\
\midrule
Female       & $\substd{0.43}{0.01}$ & $\substd{4.30}{0.01}$ & $\substd{0.03}{0.00}$ & $\substd{4.24}{0.11}$ & $\substd{0.71}{0.02}$ \\
Male         & $\substd{0.43}{0.01}$ & $\substd{4.23}{0.02}$ & $\substd{0.03}{0.00}$ & $\substd{3.95}{0.12}$ & $\substd{0.83}{0.03}$ \\
\midrule

\multicolumn{6}{l}{\textit{\textbf{Model: Fish Audio S2}}} \\
\midrule
Female       & $\substd{0.51}{0.01}$ & $\substd{4.30}{0.01}$ & $\substd{0.01}{0.00}$ & $\substd{5.11}{0.14}$ & $\substd{0.77}{0.02}$ \\
Male         & $\substd{0.50}{0.01}$ & $\substd{4.23}{0.02}$ & $\substd{0.01}{0.00}$ & $\substd{4.62}{0.17}$ & $\substd{0.90}{0.02}$ \\
\midrule

\multicolumn{6}{l}{\textit{\textbf{Model: OZSpeech}}} \\
\bottomrule
\end{tabular}
\end{table}

\begin{table}[H]
\centering
\small
\setlength{\tabcolsep}{8pt}
\caption{\textbf{Performance on different gender groups (Part 2).} Performance evaluation breakdown by gender (F/M).}
\label{tab:gender_perf_p2}

\begin{tabular}{
  l
  c
  c
  c
  c
  c
}
\toprule
\textbf{Gender} & 
{SIM $\uparrow$} & {MOS $\uparrow$} & {WER $\downarrow$} & {MCD $\downarrow$} & {EMC $\uparrow$} \\
\midrule

\multicolumn{6}{l}{\textit{\textbf{Model: GLM-TTS}}} \\
\midrule
Female       & $\substd{0.58}{0.01}$ & $\substd{4.05}{0.03}$ & $\substd{0.02}{0.00}$ & $\substd{4.09}{0.11}$ & $\substd{0.69}{0.02}$ \\
Male         & $\substd{0.57}{0.01}$ & $\substd{3.96}{0.03}$ & $\substd{0.02}{0.00}$ & $\substd{3.77}{0.14}$ & $\substd{0.81}{0.02}$ \\
\midrule

\multicolumn{6}{l}{\textit{\textbf{Model: VibeVoice}}} \\
\midrule
Female       & $\substd{0.44}{0.01}$ & $\substd{3.98}{0.03}$ & $\substd{0.06}{0.00}$ & $\substd{5.28}{0.12}$ & $\substd{0.76}{0.02}$ \\
Male         & $\substd{0.43}{0.01}$ & $\substd{3.91}{0.03}$ & $\substd{0.06}{0.01}$ & $\substd{4.85}{0.11}$ & $\substd{0.85}{0.03}$ \\
\midrule

\multicolumn{6}{l}{\textit{\textbf{Model: MGM-Omni}}} \\
\bottomrule
\end{tabular}
\end{table}

\begin{table}[H]
\centering
\small
\setlength{\tabcolsep}{8pt}
\caption{\textbf{Performance over different age groups (Part 1).}}
\label{tab:age_perf}

\begin{tabular}{
  l
  c
  c
  c
  c
  c
}
\toprule
\textbf{Age Group} & 
{SIM $\uparrow$} & {MOS $\uparrow$} & {WER $\downarrow$} & {MCD $\downarrow$} & {EMC $\uparrow$} \\
\midrule

\multicolumn{6}{l}{\textit{\textbf{Model: FishSpeech}}} \\
\midrule
under 20     & $\substd{0.43}{0.01}$ & $\substd{4.29}{0.02}$ & $\substd{0.04}{0.01}$ & $\substd{4.04}{0.40}$ & $\substd{0.69}{0.08}$ \\
20--29       & $\substd{0.43}{0.01}$ & $\substd{4.27}{0.01}$ & $\substd{0.03}{0.00}$ & $\substd{4.12}{0.08}$ & $\substd{0.78}{0.02}$ \\
30 and over  & $\substd{0.39}{0.02}$ & $\substd{4.20}{0.06}$ & $\substd{0.02}{0.01}$ & $\substd{4.06}{0.08}$ & $\substd{0.72}{0.01}$ \\
\midrule

\multicolumn{6}{l}{\textit{\textbf{Model: Fish Audio S2}}} \\
\midrule
under 20     & $\substd{0.50}{0.02}$ & $\substd{4.28}{0.03}$ & $\substd{0.01}{0.00}$ & $\substd{4.78}{0.45}$ & $\substd{0.76}{0.09}$ \\
20--29       & $\substd{0.51}{0.01}$ & $\substd{4.27}{0.01}$ & $\substd{0.01}{0.00}$ & $\substd{4.89}{0.12}$ & $\substd{0.84}{0.02}$ \\
\bottomrule
\end{tabular}
\end{table}

\begin{table}[H]
\centering
\small
\setlength{\tabcolsep}{8pt}
\caption{\textbf{Performance over different age groups (Part 2).}}
\label{tab:age_perf_p2}

\begin{tabular}{
  l
  c
  c
  c
  c
  c
}
\toprule
\textbf{Age Group} & 
{SIM $\uparrow$} & {MOS $\uparrow$} & {WER $\downarrow$} & {MCD $\downarrow$} & {EMC $\uparrow$} \\
\midrule

\multicolumn{6}{l}{\textit{\textbf{Model: GLM-TTS}}} \\
\midrule
under 20     & $\substd{0.56}{0.02}$ & $\substd{4.03}{0.09}$ & $\substd{0.02}{0.00}$ & $\substd{3.61}{0.46}$ & $\substd{0.71}{0.06}$ \\
20--29       & $\substd{0.57}{0.01}$ & $\substd{4.02}{0.02}$ & $\substd{0.02}{0.00}$ & $\substd{3.98}{0.09}$ & $\substd{0.75}{0.02}$ \\
30 and over  & $\substd{0.57}{0.02}$ & $\substd{3.88}{0.02}$ & $\substd{0.02}{0.00}$ & $\substd{3.95}{0.01}$ & $\substd{0.70}{0.01}$ \\
\midrule

\multicolumn{6}{l}{\textit{\textbf{Model: VibeVoice}}} \\
\midrule
under 20     & $\substd{0.41}{0.03}$ & $\substd{3.89}{0.05}$ & $\substd{0.06}{0.01}$ & $\substd{5.51}{0.36}$ & $\substd{0.76}{0.06}$ \\
20--29       & $\substd{0.44}{0.01}$ & $\substd{3.96}{0.02}$ & $\substd{0.06}{0.00}$ & $\substd{5.02}{0.09}$ & $\substd{0.81}{0.02}$ \\
\bottomrule
\end{tabular}
\end{table}

\begin{table}[htbp]
\centering
\scriptsize
\setlength{\tabcolsep}{1pt}

\caption{\textbf{Performance over the fraud dataset.}}
\label{tab:detailed_perf_additional}

\begin{tabularx}{\textwidth}{
    l
    l
    *{8}{>{\centering\arraybackslash}X}
}
\toprule
\textbf{Model} & \textbf{Group} & 
{SIM $\uparrow$} & {MOS $\uparrow$} & {WER $\downarrow$} & {SVA $\uparrow$} & 
{EMC $\uparrow$} & {EmTXT $\uparrow$} & {DNSMOS $\uparrow$} & {Dur (s)} \\
\midrule

\multirow{2}{*}{FishSpeech} & robocall & $\substd{0.50}{0.02}$ & $\substd{4.37}{0.02}$ & $\substd{0.05}{0.01}$ & $\substd{0.96}{0.01}$ & $\substd{0.58}{0.08}$ & $\substd{1.99}{0.05}$ & $\substd{3.48}{0.01}$ & $\substd{21.95}{0.26}$ \\
 & vctk & $\substd{0.40}{0.02}$ & $\substd{4.29}{0.02}$ & $\substd{0.02}{0.01}$ & $\substd{0.89}{0.03}$ & $\substd{0.75}{0.08}$ & $\substd{2.93}{0.03}$ & $\substd{3.30}{0.02}$ & $\substd{2.33}{0.14}$ \\
\addlinespace[4pt]

\multirow{2}{*}{Fish Audio S2} & robocall & $\substd{0.57}{0.01}$ & $\substd{4.33}{0.04}$ & $\substd{0.08}{0.01}$ & $\substd{0.99}{0.00}$ & $\substd{0.64}{0.08}$ & $\substd{2.08}{0.03}$ & $\substd{3.44}{0.02}$ & $\substd{24.85}{0.27}$ \\
 & vctk & $\substd{0.50}{0.02}$ & $\substd{4.26}{0.03}$ & $\substd{0.01}{0.01}$ & $\substd{0.95}{0.03}$ & $\substd{0.78}{0.06}$ & $\substd{3.00}{0.00}$ & $\substd{3.28}{0.02}$ & $\substd{2.30}{0.12}$ \\
\addlinespace[4pt]

\multirow{2}{*}{OZSpeech} & robocall & $\substd{0.36}{0.03}$ & $\substd{2.98}{0.03}$ & $\substd{0.07}{0.00}$ & $\substd{0.83}{0.07}$ & $\substd{0.34}{0.06}$ & $\substd{1.94}{0.04}$ & $\substd{3.28}{0.01}$ & $\substd{22.35}{0.00}$ \\
 & vctk & $\substd{0.24}{0.02}$ & $\substd{3.15}{0.05}$ & $\substd{0.02}{0.01}$ & $\substd{0.49}{0.08}$ & $\substd{0.70}{0.06}$ & $\substd{2.96}{0.03}$ & $\substd{2.82}{0.04}$ & $\substd{2.07}{0.13}$ \\
\addlinespace[4pt]

\multirow{2}{*}{StyleTTS2} & robocall & $\substd{0.23}{0.03}$ & $\substd{4.09}{0.03}$ & $\substd{0.13}{0.01}$ & $\substd{0.39}{0.13}$ & $\substd{0.54}{0.06}$ & $\substd{2.00}{0.05}$ & $\substd{3.36}{0.01}$ & $\substd{23.62}{0.31}$ \\
\bottomrule
\end{tabularx}
\end{table}

\begin{table}[t]
\centering
\caption{Full deepfake detection results for generated speech models under detection methods: SQ-LLM, AASIST, HuBERT-ECAPA.  }
\label{supp_tab:deepfake1}
\scriptsize
\setlength{\tabcolsep}{3.2pt}
\renewcommand{\arraystretch}{1.04}
\begin{tabular}{l*{9}{r}}
\toprule
 & \multicolumn{3}{c}{SQ-LLM} & \multicolumn{3}{c}{AASIST} & \multicolumn{3}{c}{HuBERT-ECAPA} \\
\cmidrule(lr){2-4}\cmidrule(lr){5-7}\cmidrule(lr){8-10}
Generated model & mDCF $\downarrow$ & EER $\downarrow$ & ACC $\uparrow$ & mDCF $\downarrow$ & EER $\downarrow$ & ACC $\uparrow$ & mDCF $\downarrow$ & EER $\downarrow$ & ACC $\uparrow$ \\
\midrule
Fish. & 34.33 & 18.33 & 81.83 & 26.00 & 13.67 & 86.17 & 10.00 & 5.00 & 94.83 \\
Fish S2 & 15.33 & 8.33 & 91.83 & 6.67 & 3.33 & 96.50 & 5.00 & 2.67 & 97.50 \\
OZS. & 4.00 & 2.33 & 97.50 & 0.00 & 0.00 & 99.83 & 1.33 & 1.00 & 98.83 \\
Style. & 30.33 & 15.33 & 84.83 & 52.67 & 27.33 & 72.83 & 7.00 & 4.00 & 96.17 \\
Spark. & 76.33 & 42.00 & 58.17 & 82.00 & 48.00 & 52.17 & 57.33 & 29.33 & 70.50 \\
Moss. & 36.00 & 20.67 & 79.50 & 86.00 & 49.33 & 50.50 & 52.33 & 26.67 & 73.17 \\
\bottomrule
\end{tabular}
\end{table}


\begin{table}[t]
\centering
\caption{Full deepfake detection results for generated speech models under detection methods: RawGAT-ST, RawNet2, TCM-ADD.  }
\label{supp_tab:deepfake2}
\scriptsize
\setlength{\tabcolsep}{3.2pt}
\renewcommand{\arraystretch}{1.04}
\begin{tabular}{l*{9}{r}}
\toprule
 & \multicolumn{3}{c}{RawGAT-ST} & \multicolumn{3}{c}{RawNet2} & \multicolumn{3}{c}{TCM-ADD} \\
\cmidrule(lr){2-4}\cmidrule(lr){5-7}\cmidrule(lr){8-10}
Generated model & mDCF $\downarrow$ & EER $\downarrow$ & ACC $\uparrow$ & mDCF $\downarrow$ & EER $\downarrow$ & ACC $\uparrow$ & mDCF $\downarrow$ & EER $\downarrow$ & ACC $\uparrow$ \\
\midrule
Fish. & 7.67 & 5.33 & 94.83 & 14.33 & 8.00 & 92.17 & 8.00 & 4.67 & 95.50 \\
Fish S2 & 4.33 & 3.00 & 97.17 & 13.33 & 7.00 & 93.17 & 1.33 & 1.00 & 98.83 \\
OZS. & 0.00 & 0.00 & 99.83 & 2.00 & 1.00 & 98.83 & 0.00 & 0.00 & 99.83 \\
Style. & 21.33 & 11.67 & 88.17 & 42.00 & 22.00 & 78.17 & 7.00 & 3.67 & 96.50 \\
Spark. & 64.00 & 36.33 & 63.83 & 77.00 & 46.00 & 53.83 & 63.67 & 36.33 & 63.83 \\
Moss. & 82.67 & 45.67 & 54.17 & 77.67 & 40.67 & 59.50 & 66.67 & 35.00 & 64.83 \\
\bottomrule
\end{tabular}
\end{table}


\begin{table}[t]
\centering
\caption{Full deepfake detection results for generated speech models under detection methods: XLSR-SLS, Wav2Vec2-ECAPA, WavLM-ECAPA.}
\label{supp_tab:deepfake3}
\scriptsize
\setlength{\tabcolsep}{3.2pt}
\renewcommand{\arraystretch}{1.04}
\begin{tabular}{l*{9}{r}}
\toprule
 & \multicolumn{3}{c}{XLSR-SLS} & \multicolumn{3}{c}{Wav2Vec2-ECAPA} & \multicolumn{3}{c}{WavLM-ECAPA} \\
\cmidrule(lr){2-4}\cmidrule(lr){5-7}\cmidrule(lr){8-10}
Generated model & mDCF $\downarrow$ & EER $\downarrow$ & ACC $\uparrow$ & mDCF $\downarrow$ & EER $\downarrow$ & ACC $\uparrow$ & mDCF $\downarrow$ & EER $\downarrow$ & ACC $\uparrow$ \\
\midrule
Fish. & 8.33 & 4.67 & 95.50 & 33.67 & 17.00 & 83.17 & 15.67 & 8.00 & 91.83 \\
Fish S2 & 5.67 & 4.33 & 95.83 & 23.33 & 11.67 & 88.17 & 11.67 & 6.00 & 94.17 \\
OZS. & 0.00 & 0.00 & 99.83 & 24.00 & 13.67 & 86.17 & 0.67 & 0.67 & 99.17 \\
Style. & 12.67 & 6.67 & 93.50 & 74.67 & 53.00 & 47.17 & 14.00 & 7.33 & 92.83 \\
Spark. & 75.00 & 42.33 & 57.83 & 59.00 & 30.00 & 69.83 & 58.67 & 29.67 & 70.50 \\
Moss. & 40.33 & 21.33 & 78.67 & 90.00 & 49.33 & 50.83 & 70.67 & 37.00 & 63.17 \\
\bottomrule
\end{tabular}
\end{table}


\newpage

\begin{table}[p]
\centering
\footnotesize
\caption{\textbf{Performance in multi-speaker noisy environments (Part 1).}}
\setlength{\tabcolsep}{2pt}
\begin{tabularx}{\textwidth}{l l *{6}{>{\centering\arraybackslash}X}}
\toprule
\textbf{Model} & \textbf{Noise} & \textbf{SIM} & \textbf{MOS} & \textbf{WER} & \textbf{MCD} & \textbf{EMC} & \textbf{DNSMOS} \\
\midrule
\multirow{4}{*}{FishSpeech} & -5dB  & $\substd{0.20}{0.02}$ & $\substd{4.26}{0.01}$ & $\substd{0.05}{0.01}$ & $\substd{8.40}{0.17}$ & $\substd{0.58}{0.07}$ & $\substd{3.32}{0.02}$ \\
 & +0dB  & $\substd{0.31}{0.03}$ & $\substd{4.27}{0.02}$ & $\substd{0.05}{0.01}$ & $\substd{7.21}{0.19}$ & $\substd{0.60}{0.07}$ & $\substd{3.31}{0.02}$ \\
 & +5dB  & $\substd{0.40}{0.04}$ & $\substd{4.28}{0.02}$ & $\substd{0.05}{0.01}$ & $\substd{6.67}{0.19}$ & $\substd{0.67}{0.06}$ & $\substd{3.32}{0.02}$ \\
 & +10dB  & $\substd{0.46}{0.04}$ & $\substd{4.33}{0.01}$ & $\substd{0.04}{0.01}$ & $\substd{6.50}{0.20}$ & $\substd{0.69}{0.06}$ & $\substd{3.33}{0.02}$ \\
\addlinespace[0.6em]
\multirow{4}{*}{Fish Audio S2} & -5dB  & $\substd{0.29}{0.03}$ & $\substd{4.36}{0.01}$ & $\substd{0.04}{0.01}$ & $\substd{7.51}{0.22}$ & $\substd{0.62}{0.07}$ & $\substd{3.33}{0.01}$ \\
 & +0dB  & $\substd{0.43}{0.03}$ & $\substd{4.38}{0.01}$ & $\substd{0.03}{0.01}$ & $\substd{6.79}{0.16}$ & $\substd{0.66}{0.08}$ & $\substd{3.31}{0.02}$ \\
 & +5dB  & $\substd{0.51}{0.03}$ & $\substd{4.40}{0.01}$ & $\substd{0.03}{0.01}$ & $\substd{6.49}{0.14}$ & $\substd{0.68}{0.07}$ & $\substd{3.34}{0.02}$ \\
\bottomrule
\end{tabularx}
\end{table}

\clearpage

\begin{table}[p]
\centering
\footnotesize
\caption{\textbf{Performance in multi-speaker noisy environments (Part 2).}}
\setlength{\tabcolsep}{2pt}
\begin{tabularx}{\textwidth}{l l *{6}{>{\centering\arraybackslash}X}}
\toprule
\textbf{Model} & \textbf{Noise} & \textbf{SIM} & \textbf{MOS} & \textbf{WER} & \textbf{MCD} & \textbf{EMC} & \textbf{DNSMOS} \\
\midrule
\multirow{4}{*}{MOSS-TTS v1.5} & -5dB  & $\substd{0.20}{0.02}$ & $\substd{4.33}{0.02}$ & $\substd{0.05}{0.01}$ & $\substd{8.62}{0.12}$ & $\substd{0.66}{0.06}$ & $\substd{3.19}{0.03}$ \\
 & +0dB  & $\substd{0.32}{0.03}$ & $\substd{4.33}{0.01}$ & $\substd{0.04}{0.01}$ & $\substd{7.62}{0.22}$ & $\substd{0.64}{0.07}$ & $\substd{3.20}{0.03}$ \\
 & +5dB  & $\substd{0.45}{0.03}$ & $\substd{4.31}{0.02}$ & $\substd{0.05}{0.01}$ & $\substd{7.22}{0.19}$ & $\substd{0.68}{0.07}$ & $\substd{3.19}{0.02}$ \\
 & +10dB  & $\substd{0.54}{0.03}$ & $\substd{4.32}{0.02}$ & $\substd{0.04}{0.01}$ & $\substd{7.00}{0.18}$ & $\substd{0.61}{0.07}$ & $\substd{3.23}{0.02}$ \\
\addlinespace[0.6em]
\multirow{4}{*}{dots.tts} & -5dB  & $\substd{0.24}{0.03}$ & $\substd{3.64}{0.07}$ & $\substd{0.30}{0.04}$ & $\substd{8.23}{0.33}$ & $\substd{0.62}{0.06}$ & $\substd{3.10}{0.03}$ \\
 & +0dB  & $\substd{0.37}{0.04}$ & $\substd{3.62}{0.10}$ & $\substd{0.16}{0.03}$ & $\substd{7.04}{0.29}$ & $\substd{0.64}{0.06}$ & $\substd{3.11}{0.03}$ \\
 & +5dB  & $\substd{0.52}{0.04}$ & $\substd{3.82}{0.05}$ & $\substd{0.08}{0.01}$ & $\substd{6.14}{0.20}$ & $\substd{0.66}{0.05}$ & $\substd{3.20}{0.01}$ \\
\bottomrule
\end{tabularx}
\end{table}

\clearpage

\begin{table}[p]
\centering
\footnotesize
\caption{\textbf{Performance in multi-speaker noisy environments (Part 3).}}
\setlength{\tabcolsep}{2pt}
\begin{tabularx}{\textwidth}{l l *{6}{>{\centering\arraybackslash}X}}
\toprule
\textbf{Model} & \textbf{Noise} & \textbf{SIM} & \textbf{MOS} & \textbf{WER} & \textbf{MCD} & \textbf{EMC} & \textbf{DNSMOS} \\
\midrule
\multirow{4}{*}{VibeVoice} & -5dB  & $\substd{0.18}{0.02}$ & $\substd{3.35}{0.16}$ & $\substd{0.14}{0.04}$ & $\substd{8.09}{0.31}$ & $\substd{0.50}{0.05}$ & $\substd{3.07}{0.09}$ \\
 & +0dB  & $\substd{0.27}{0.04}$ & $\substd{3.30}{0.17}$ & $\substd{0.15}{0.03}$ & $\substd{7.87}{0.26}$ & $\substd{0.55}{0.05}$ & $\substd{3.05}{0.07}$ \\
 & +5dB  & $\substd{0.34}{0.05}$ & $\substd{3.25}{0.17}$ & $\substd{0.17}{0.03}$ & $\substd{7.45}{0.28}$ & $\substd{0.48}{0.07}$ & $\substd{3.01}{0.09}$ \\
 & +10dB  & $\substd{0.41}{0.04}$ & $\substd{3.42}{0.13}$ & $\substd{0.12}{0.04}$ & $\substd{7.16}{0.19}$ & $\substd{0.60}{0.05}$ & $\substd{3.10}{0.07}$ \\
\addlinespace[0.6em]
\multirow{4}{*}{MGM-Omni} & -5dB  & $\substd{0.16}{0.02}$ & $\substd{3.56}{0.05}$ & $\substd{0.33}{0.02}$ & $\substd{8.77}{0.27}$ & $\substd{0.58}{0.05}$ & $\substd{2.85}{0.04}$ \\
 & +0dB  & $\substd{0.26}{0.03}$ & $\substd{3.70}{0.09}$ & $\substd{0.28}{0.04}$ & $\substd{7.84}{0.32}$ & $\substd{0.58}{0.05}$ & $\substd{2.99}{0.06}$ \\
 & +5dB  & $\substd{0.36}{0.03}$ & $\substd{3.63}{0.07}$ & $\substd{0.30}{0.03}$ & $\substd{7.47}{0.29}$ & $\substd{0.56}{0.07}$ & $\substd{2.92}{0.05}$ \\
\bottomrule
\end{tabularx}
\end{table}

\clearpage

\begin{table}[p]
\centering
\footnotesize
\caption{\textbf{Performance in multi-speaker noisy environments (Part 4).}}
\setlength{\tabcolsep}{2pt}
\begin{tabularx}{\textwidth}{l l *{6}{>{\centering\arraybackslash}X}}
\toprule
\textbf{Model} & \textbf{Noise} & \textbf{SIM} & \textbf{MOS} & \textbf{WER} & \textbf{MCD} & \textbf{EMC} & \textbf{DNSMOS} \\
\midrule
\multirow{4}{*}{OpenVoice} & -5dB  & $\substd{0.13}{0.02}$ & $\substd{4.24}{0.02}$ & $\substd{0.06}{0.01}$ & $\substd{7.64}{0.10}$ & $\substd{0.59}{0.05}$ & $\substd{3.39}{0.01}$ \\
 & +0dB  & $\substd{0.16}{0.02}$ & $\substd{4.23}{0.02}$ & $\substd{0.05}{0.01}$ & $\substd{7.44}{0.13}$ & $\substd{0.62}{0.05}$ & $\substd{3.39}{0.02}$ \\
 & +5dB  & $\substd{0.19}{0.03}$ & $\substd{4.24}{0.02}$ & $\substd{0.06}{0.01}$ & $\substd{7.28}{0.13}$ & $\substd{0.62}{0.05}$ & $\substd{3.37}{0.01}$ \\
 & +10dB  & $\substd{0.22}{0.03}$ & $\substd{4.26}{0.02}$ & $\substd{0.06}{0.01}$ & $\substd{7.25}{0.14}$ & $\substd{0.65}{0.05}$ & $\substd{3.39}{0.01}$ \\
\addlinespace[0.6em]
\multirow{4}{*}{XTTS} & -5dB  & $\substd{0.15}{0.01}$ & $\substd{3.54}{0.05}$ & $\substd{0.05}{0.01}$ & $\substd{10.64}{0.15}$ & $\substd{0.48}{0.04}$ & $\substd{3.18}{0.03}$ \\
 & +0dB  & $\substd{0.23}{0.02}$ & $\substd{3.44}{0.08}$ & $\substd{0.06}{0.01}$ & $\substd{10.48}{0.11}$ & $\substd{0.47}{0.05}$ & $\substd{3.09}{0.04}$ \\
 & +5dB  & $\substd{0.34}{0.02}$ & $\substd{3.41}{0.11}$ & $\substd{0.05}{0.01}$ & $\substd{10.45}{0.15}$ & $\substd{0.47}{0.05}$ & $\substd{3.10}{0.03}$ \\
\bottomrule
\end{tabularx}
\end{table}

\begin{table}[htbp]
\centering
\small
\setlength{\tabcolsep}{3pt}
\renewcommand{\arraystretch}{1.1}

\caption{\textbf{Performance on Text Hallucination (Hallucination-VCTK).} WER uses the assigned text; SIM/SVA use same-speaker VCTK audio. No text-matched target waveform exists, so MCD/EMC are omitted.}
\label{tab:hallucination_perf}

\begin{tabularx}{\textwidth}{
    l
    *{3}{>{\centering\arraybackslash}X}
    S[table-format=1.2]
    >{\centering\arraybackslash}X
}
\toprule
\textbf{Model} & 
\textbf{SIM}$\uparrow$ & \textbf{MOS}$\uparrow$ & \textbf{WER}$\downarrow$ &
\textbf{RTF}$\downarrow$ & 
\textbf{SVA}$\uparrow$ \\
\midrule

FishSpeech & $\substd{0.35}{0.01}$ & $\substd{4.37}{0.01}$ & $\substd{0.33}{0.01}$ & 3.24 & $\substd{0.83}{0.03}$ \\
Fish Audio S2 & $\substd{0.41}{0.03}$ & $\substd{4.37}{0.01}$ & $\substd{0.32}{0.07}$ & 7.39 & $\substd{0.91}{0.03}$ \\
OZSpeech & $\substd{0.28}{0.03}$ & $\substd{3.35}{0.06}$ & $\substd{0.59}{0.17}$ & 8.21 & $\substd{0.65}{0.10}$ \\
StyleTTS2 & $\substd{0.18}{0.02}$ & $\substd{4.05}{0.05}$ & $\substd{0.71}{0.21}$ & 0.10 & $\substd{0.23}{0.08}$ \\
\addlinespace[4pt]

Spark-TTS & $\substd{0.34}{0.06}$ & $\substd{3.50}{0.14}$ & $\substd{0.81}{0.03}$ & 1.60 & $\substd{0.64}{0.11}$ \\
MOSS-TTSD v0.7 & $\substd{0.42}{0.02}$ & $\substd{4.06}{0.04}$ & $\substd{0.99}{0.11}$ & 0.90 & $\substd{0.90}{0.02}$ \\
MOSS-TTS v1.5 & $\substd{0.46}{0.02}$ & $\substd{4.34}{0.03}$ & $\substd{0.88}{0.35}$ & 0.60 & $\substd{0.94}{0.02}$ \\
\bottomrule
\end{tabularx}
\end{table}

\begin{table}
\centering
\caption{\textbf{Evaluation of the protected audio on LibriTTS.}}
\begin{tabular}{l c c c c c c}
\toprule
Protection & SNR & MCD & SIM & WER & SpeechMOS & DNSMOS \\
\midrule
Gaussian Noise & 6.10 & 11.25 & 0.71 & 0.19 & 1.39 & 1.98 \\
SPEC & 8.08 & 12.67 & 0.59 & 0.34 & 1.40 & 1.97 \\
POP & 17.08 & 2.08 & 0.92 & 0.18 & 3.00 & 3.02 \\
SafeSpeech & 9.48 & 11.51 & 0.62 & 0.26 & 1.48 & 2.06 \\
Enkidu & 8.07 & 8.03 & 0.72 & 0.24 & 1.52 & 2.26 \\
Denoise-SPEC & 15.30 & 3.39 & 0.51 & 0.29 & 3.23 & 3.15 \\
\bottomrule
\end{tabular}
\end{table}

\begin{table}[htbp]
\centering
\small
\setlength{\tabcolsep}{5pt}
\caption{\textbf{Protected Audio VC Evaluation (Part 1).} Comparison of protection methods on LibriTTS (Part 1 of 3).}
\label{tab:protected_vc_additional_p1}

\begin{tabular}{
  l
  S[table-format=1.2]
  S[table-format=1.2]
  S[table-format=1.2]
  S[table-format=2.2]
  S[table-format=2.2]
  S[table-format=1.2]
  S[table-format=1.2]
}
\toprule
\textbf{Protection} & 
{SIM $\uparrow$} & {MOS $\uparrow$} & {WER $\downarrow$} & {MCD $\downarrow$} & 
{RTF $\downarrow$} & 
{SVA $\uparrow$} & {EMC $\uparrow$} \\
\midrule

\multicolumn{8}{l}{\textit{\textbf{Model: FishSpeech}}} \\
\midrule
Gaussian Noise & $\substd{0.23}{0.01}$ & $\substd{2.90}{0.05}$ & $\substd{0.08}{0.01}$ & $\substd{10.84}{0.11}$ & 4.08 & $\substd{0.45}{0.03}$ & $\substd{0.56}{0.03}$ \\
POP            & $\substd{0.01}{0.01}$ & $\substd{1.53}{0.01}$ & $\substd{1.00}{0.00}$ & $\substd{11.89}{0.32}$ & 2.36 & $\substd{0.00}{0.00}$ & $\substd{0.50}{0.04}$ \\
Enkidu         & $\substd{0.33}{0.01}$ & $\substd{4.21}{0.02}$ & $\substd{0.06}{0.00}$ & $\substd{7.27}{0.09}$ & 2.56 & $\substd{0.67}{0.03}$ & $\substd{0.65}{0.02}$ \\
SafeSpeech     & $\substd{0.24}{0.01}$ & $\substd{3.15}{0.04}$ & $\substd{0.08}{0.01}$ & $\substd{9.60}{0.09}$ & 2.03 & $\substd{0.47}{0.03}$ & $\substd{0.57}{0.03}$ \\
SPEC           & $\substd{0.21}{0.01}$ & $\substd{3.10}{0.04}$ & $\substd{0.08}{0.00}$ & $\substd{9.99}{0.08}$ & 4.16 & $\substd{0.39}{0.03}$ & $\substd{0.56}{0.03}$ \\

\midrule

\multicolumn{8}{l}{\textit{\textbf{Model: Fish Audio S2}}} \\
\bottomrule
\end{tabular}
\end{table}

\begin{table}[htbp]
\centering
\small
\setlength{\tabcolsep}{5pt}
\caption{\textbf{Protected Audio VC Evaluation (Part 2).} Comparison of protection methods on LibriTTS (Part 2 of 3).}
\label{tab:protected_vc_additional_p2}

\begin{tabular}{
  l
  S[table-format=1.2]
  S[table-format=1.2]
  S[table-format=1.2]
  S[table-format=2.2]
  S[table-format=2.2]
  S[table-format=1.2]
  S[table-format=1.2]
}
\toprule
\textbf{Protection} & 
{SIM $\uparrow$} & {MOS $\uparrow$} & {WER $\downarrow$} & {MCD $\downarrow$} & 
{RTF $\downarrow$} & 
{SVA $\uparrow$} & {EMC $\uparrow$} \\
\midrule

\multicolumn{8}{l}{\textit{\textbf{Model: Higgs}}} \\
\midrule
Gaussian Noise & $\substd{0.27}{0.01}$ & $\substd{2.55}{0.04}$ & $\substd{0.34}{0.06}$ & $\substd{10.03}{0.08}$ & 3.37 & $\substd{0.56}{0.03}$ & $\substd{0.52}{0.03}$ \\
POP            & $\substd{0.52}{0.01}$ & $\substd{4.24}{0.03}$ & $\substd{0.08}{0.02}$ & $\substd{6.39}{0.11}$ & 1.33 & $\substd{0.89}{0.01}$ & $\substd{0.69}{0.02}$ \\
Enkidu         & $\substd{0.43}{0.01}$ & $\substd{4.04}{0.03}$ & $\substd{0.15}{0.06}$ & $\substd{7.08}{0.10}$ & 3.17 & $\substd{0.81}{0.02}$ & $\substd{0.65}{0.02}$ \\
SafeSpeech     & $\substd{0.26}{0.01}$ & $\substd{2.47}{0.03}$ & $\substd{0.31}{0.05}$ & $\substd{11.07}{0.10}$ & 1.40 & $\substd{0.54}{0.03}$ & $\substd{0.45}{0.03}$ \\
SPEC           & $\substd{0.24}{0.01}$ & $\substd{2.43}{0.04}$ & $\substd{0.32}{0.04}$ & $\substd{11.41}{0.10}$ & 4.63 & $\substd{0.47}{0.03}$ & $\substd{0.45}{0.03}$ \\

\midrule

\multicolumn{8}{l}{\textit{\textbf{Model: Higgs TTS 3}}} \\
\bottomrule
\end{tabular}
\end{table}

\begin{table}[t]
\centering
\small
\setlength{\tabcolsep}{5pt}
\caption{\textbf{Protected Audio VC Evaluation (Part 3).} Comparison of protection methods on LibriTTS (Part 3 of 3).}
\label{tab:protected_vc_additional_p3}

\begin{tabular}{
  l
  S[table-format=1.2]
  S[table-format=1.2]
  S[table-format=1.2]
  S[table-format=2.2]
  S[table-format=2.2]
  S[table-format=1.2]
  S[table-format=1.2]
}
\toprule
\textbf{Protection} & 
{SIM $\uparrow$} & {MOS $\uparrow$} & {WER $\downarrow$} & {MCD $\downarrow$} & 
{RTF $\downarrow$} & 
{SVA $\uparrow$} & {EMC $\uparrow$} \\
\midrule

\multicolumn{8}{l}{\textit{\textbf{Model: F5-TTS}}} \\
\midrule
Gaussian Noise & $\substd{0.14}{0.01}$ & $\substd{2.56}{0.05}$ & $\substd{0.20}{0.02}$ & $\substd{17.78}{0.26}$ & 0.27 & $\substd{0.16}{0.02}$ & $\substd{0.52}{0.02}$ \\
POP            & $\substd{0.52}{0.01}$ & $\substd{3.96}{0.03}$ & $\substd{0.11}{0.01}$ & $\substd{7.17}{0.12}$ & 0.40 & $\substd{0.88}{0.01}$ & $\substd{0.67}{0.02}$ \\
Enkidu         & $\substd{0.43}{0.01}$ & $\substd{3.70}{0.03}$ & $\substd{0.12}{0.01}$ & $\substd{7.15}{0.10}$ & 0.29 & $\substd{0.81}{0.02}$ & $\substd{0.65}{0.02}$ \\
SafeSpeech     & $\substd{0.21}{0.01}$ & $\substd{2.96}{0.05}$ & $\substd{0.15}{0.01}$ & $\substd{14.56}{0.25}$ & 0.36 & $\substd{0.36}{0.03}$ & $\substd{0.55}{0.02}$ \\
SPEC           & $\substd{0.18}{0.01}$ & $\substd{2.80}{0.05}$ & $\substd{0.17}{0.01}$ & $\substd{15.76}{0.27}$ & 0.36 & $\substd{0.27}{0.02}$ & $\substd{0.55}{0.02}$ \\

\midrule

\multicolumn{8}{l}{\textit{\textbf{Model: MaskGCT}}} \\
\bottomrule
\end{tabular}
\end{table}

\newpage
\begin{table}[htbp]
\centering
\small
\setlength{\tabcolsep}{3pt}
\caption{\textbf{Long-Context Generation.} Performance evaluation on Long LibriSpeech dataset.}
\label{tab:long_libri_additional}

\begin{tabular}{
  l
  S[table-format=1.2]
  S[table-format=1.2]
  S[table-format=1.2]
  S[table-format=2.2]
  S[table-format=1.2]
  S[table-format=1.2]
  S[table-format=1.2]
}
\toprule
\textbf{Model} & 
{SIM $\uparrow$} & {MOS $\uparrow$} & {WER $\downarrow$} & {MCD $\downarrow$} & 
{RTF $\downarrow$} & 
{SVA $\uparrow$} & {EMC $\uparrow$} \\
\midrule

\multicolumn{8}{l}{\textit{\textbf{Dataset: Long LibriSpeech}}} \\
\midrule
FishSpeech & $\substd{0.57}{0.04}$ & $\substd{4.40}{0.04}$ & $\substd{0.66}{0.04}$ & $\substd{12.36}{0.80}$ & 3.03 & $\substd{1.00}{0.00}$ & $\substd{0.50}{0.10}$ \\
Fish Audio S2 & $\substd{0.50}{0.04}$ & $\substd{4.09}{0.08}$ & $\substd{0.03}{0.00}$ & $\substd{8.83}{0.72}$ & 4.47 & $\substd{0.90}{0.06}$ & $\substd{0.55}{0.11}$ \\
OZSpeech & $\substd{0.42}{0.03}$ & $\substd{2.10}{0.10}$ & $\substd{0.07}{0.02}$ & $\substd{9.86}{0.94}$ & 1.47 & $\substd{0.95}{0.05}$ & $\substd{0.45}{0.08}$ \\
StyleTTS2 & $\substd{0.20}{0.02}$ & $\substd{4.06}{0.05}$ & $\substd{0.73}{0.05}$ & $\substd{12.90}{0.67}$ & 0.03 & $\substd{0.35}{0.12}$ & $\substd{0.25}{0.11}$ \\
Spark-TTS & $\substd{0.35}{0.04}$ & $\substd{3.10}{0.20}$ & $\substd{0.91}{0.02}$ & $\substd{14.64}{1.20}$ & 1.26 & $\substd{0.65}{0.10}$ & $\substd{0.40}{0.06}$ \\
MOSS-TTSD v0.7 & $\substd{0.64}{0.03}$ & $\substd{3.93}{0.10}$ & $\substd{0.40}{0.05}$ & $\substd{7.92}{0.67}$ & 0.59 & $\substd{1.00}{0.00}$ & $\substd{0.45}{0.11}$ \\
\bottomrule
\end{tabular}
\end{table}

\newpage

\begin{table}
\centering
\small
\setlength{\tabcolsep}{3pt}
\caption{\textbf{Multilingual Generalization.} Performance on French (Common Voice) and Mandarin (AISHELL-1) datasets.}
\label{tab:multilingual_gen_additional}

\begin{tabular}{
  l
  S[table-format=1.2]
  S[table-format=1.2]
  S[table-format=1.2]
  S[table-format=2.2]
  S[table-format=2.2]
  S[table-format=1.2]
}
\toprule
\textbf{Model} & 
{SIM $\uparrow$} & {MOS $\uparrow$} & {WER $\downarrow$} & {MCD $\downarrow$} & 
{RTF $\downarrow$} & 
{SVA $\uparrow$} \\
\midrule

\multicolumn{7}{l}{\textit{\textbf{Dataset: Common Voice (French)}}} \\
\midrule
FishSpeech & $\substd{0.37}{0.01}$ & $\substd{3.20}{0.05}$ & $\substd{0.34}{0.02}$ & $\substd{5.71}{0.27}$ & 1.94 & $\substd{0.77}{0.03}$ \\
Fish Audio S2 & $\substd{0.46}{0.01}$ & $\substd{3.22}{0.05}$ & $\substd{0.26}{0.01}$ & $\substd{6.03}{0.26}$ & 9.28 & $\substd{0.89}{0.02}$ \\
OZSpeech & $\substd{0.11}{0.01}$ & $\substd{2.46}{0.05}$ & $\substd{0.99}{0.01}$ & $\substd{9.24}{0.45}$ & 17.50 & $\substd{0.06}{0.02}$ \\
StyleTTS2 & $\substd{0.11}{0.01}$ & $\substd{3.72}{0.05}$ & $\substd{0.70}{0.01}$ & $\substd{7.05}{0.28}$ & 0.11 & $\substd{0.04}{0.02}$ \\
Spark-TTS & $\substd{0.16}{0.01}$ & $\substd{2.85}{0.06}$ & $\substd{1.66}{0.13}$ & $\substd{6.37}{0.30}$ & 1.16 & $\substd{0.35}{0.03}$ \\
MOSS-TTSD v0.7 & $\substd{0.33}{0.01}$ & $\substd{2.60}{0.04}$ & $\substd{2.44}{0.18}$ & $\substd{7.42}{0.43}$ & 0.60 & $\substd{0.67}{0.03}$ \\
\bottomrule
\end{tabular}
\end{table}

\begin{table}[htbp]
\centering
\small
\setlength{\tabcolsep}{3pt}

\caption{\textbf{DEMUCS-on-SPEC performance on LibriTTS.}}
\label{tab:LibriTTS_perf_additional}

\begin{tabularx}{\textwidth}{
    l
    l
    *{4}{>{\centering\arraybackslash}X}
    S[table-format=2.2]
    *{2}{>{\centering\arraybackslash}X}
}
\toprule
\textbf{Dataset} & \textbf{Model} & 
{SIM $\uparrow$} & {MOS $\uparrow$} & {WER $\downarrow$} & {MCD $\downarrow$} & 
{RTF $\downarrow$} & 
{SVA $\uparrow$} & {EMC $\uparrow$} \\
\midrule

\multirow{22}{*}{\shortstack[l]{LibriTTS\\(DEMUCS\\on SPEC)}} & FishSpeech & $\substd{0.29}{0.01}$ & $\substd{4.15}{0.02}$ & $\substd{0.07}{0.00}$ & $\substd{6.65}{0.10}$ & 2.42 & $\substd{0.64}{0.03}$ & $\substd{0.63}{0.02}$ \\
& Fish Audio S2 & $\substd{0.32}{0.01}$ & $\substd{4.13}{0.02}$ & $\substd{0.05}{0.00}$ & $\substd{6.48}{0.11}$ & 1.01 & $\substd{0.73}{0.03}$ & $\substd{0.66}{0.02}$ \\
& OZSpeech & $\substd{0.26}{0.01}$ & $\substd{3.21}{0.02}$ & $\substd{0.06}{0.00}$ & $\substd{6.84}{0.10}$ & 12.83 & $\substd{0.53}{0.03}$ & $\substd{0.62}{0.02}$ \\
& StyleTTS2 & $\substd{0.19}{0.01}$ & $\substd{4.31}{0.01}$ & $\substd{0.05}{0.00}$ & $\substd{6.71}{0.09}$ & 0.11 & $\substd{0.24}{0.03}$ & $\substd{0.57}{0.02}$ \\
& Spark-TTS & $\substd{0.09}{0.01}$ & $\substd{3.47}{0.02}$ & $\substd{1.30}{0.05}$ & $\substd{3.83}{0.14}$ & 2.06 & $\substd{0.20}{0.02}$ & $\substd{0.54}{0.03}$ \\
& MOSS-TTSD v0.7 & $\substd{0.29}{0.01}$ & $\substd{3.72}{0.03}$ & $\substd{0.29}{0.02}$ & $\substd{7.22}{0.12}$ & 1.70 & $\substd{0.62}{0.03}$ & $\substd{0.61}{0.02}$ \\
& MOSS-TTS v1.5 & $\substd{0.35}{0.01}$ & $\substd{4.04}{0.03}$ & $\substd{0.09}{0.03}$ & $\substd{6.78}{0.10}$ & 0.82 & $\substd{0.77}{0.03}$ & $\substd{0.63}{0.03}$ \\
\bottomrule
\end{tabularx}
\end{table}

\begin{table}[htbp]
\centering
\small
\setlength{\tabcolsep}{3pt}
\caption{\textbf{Performance on Natural Background Noises.} Evaluation on Background-VCTK under clean and noisy conditions.}
\label{tab:background_noise_additional}

\begin{tabular}{
  l
  S[table-format=1.2]
  S[table-format=1.2]
  S[table-format=1.2]
  S[table-format=1.2]
  S[table-format=2.2]
  S[table-format=1.2]
  S[table-format=1.2]
}
\toprule
\textbf{Model} & 
{SIM $\uparrow$} & {MOS $\uparrow$} & {WER $\downarrow$} & {MCD $\downarrow$} & 
{RTF $\downarrow$} & 
{SVA $\uparrow$} & {EMC $\uparrow$} \\
\midrule

\multicolumn{8}{l}{\textit{\textbf{Condition: Background-VCTK (Clean)}}} \\
\midrule
FishSpeech & $\substd{0.50}{0.01}$ & $\substd{4.28}{0.01}$ & $\substd{0.01}{0.00}$ & $\substd{5.36}{0.12}$ & 4.08 & $\substd{0.95}{0.02}$ & $\substd{0.82}{0.03}$ \\
Fish Audio S2 & $\substd{0.58}{0.01}$ & $\substd{4.27}{0.01}$ & $\substd{0.01}{0.00}$ & $\substd{5.56}{0.10}$ & 25.26 & $\substd{1.00}{0.00}$ & $\substd{0.91}{0.02}$ \\
OZSpeech & $\substd{0.27}{0.02}$ & $\substd{3.31}{0.03}$ & $\substd{0.01}{0.00}$ & $\substd{7.41}{0.14}$ & 24.83 & $\substd{0.57}{0.07}$ & $\substd{0.78}{0.04}$ \\
StyleTTS2 & $\substd{0.20}{0.01}$ & $\substd{4.32}{0.02}$ & $\substd{0.01}{0.00}$ & $\substd{6.07}{0.10}$ & 0.54 & $\substd{0.27}{0.05}$ & $\substd{0.55}{0.05}$ \\
Spark-TTS & $\substd{0.59}{0.03}$ & $\substd{3.89}{0.09}$ & $\substd{0.12}{0.04}$ & $\substd{4.82}{0.17}$ & 1.97 & $\substd{0.86}{0.04}$ & $\substd{0.91}{0.03}$ \\
MOSS-TTSD v0.7 & $\substd{0.49}{0.02}$ & $\substd{4.04}{0.04}$ & $\substd{0.10}{0.02}$ & $\substd{5.16}{0.14}$ & 3.08 & $\substd{0.93}{0.02}$ & $\substd{0.78}{0.03}$ \\
\bottomrule
\end{tabular}
\end{table}

\begin{table}[t]
\centering
\footnotesize
\caption{\textbf{Denoising performance on SPEC with DEMUCS.}}
\label{tb:libritts_baseline_spec_demucs_additional}
\setlength{\tabcolsep}{3pt}
\renewcommand{\arraystretch}{1.08}

\resizebox{\textwidth}{!}{%
\begin{tabular}{l l c c c c c c c}
\toprule
\textbf{Model} & \textbf{Condition} 
& \textbf{SIM}$\uparrow$ & \textbf{MOS}$\uparrow$ & \textbf{WER}$\downarrow$ & \textbf{MCD}$\downarrow$ 
& \textbf{RTF}$\downarrow$ & \textbf{SVA}$\uparrow$ & \textbf{EMC}$\uparrow$ \\
\midrule

\multirow{3}{*}{FishSpeech}
& Clean & $\substd{0.47}{0.01}$ & $\substd{4.37}{0.01}$ & $\substd{0.17}{0.01}$ & $\substd{6.47}{0.10}$ & 3.61 & $\substd{0.91}{0.01}$ & $\substd{0.68}{0.02}$ \\
& SPEC & $\substd{0.21}{0.01}$ & $\substd{3.10}{0.04}$ & $\substd{0.08}{0.00}$ & $\substd{9.99}{0.08}$ & 4.16 & $\substd{0.39}{0.03}$ & $\substd{0.56}{0.03}$ \\
& DEMUCS & $\substd{0.29}{0.01}$ & $\substd{4.15}{0.02}$ & $\substd{0.07}{0.00}$ & $\substd{6.65}{0.10}$ & 2.42 & $\substd{0.64}{0.03}$ & $\substd{0.63}{0.02}$ \\
\midrule

\multirow{3}{*}{Fish Audio S2}
& Clean & $\substd{0.54}{0.01}$ & $\substd{4.37}{0.01}$ & $\substd{0.04}{0.00}$ & $\substd{6.16}{0.10}$ & 4.67 & $\substd{0.95}{0.01}$ & $\substd{0.71}{0.02}$ \\
& SPEC & $\substd{0.30}{0.01}$ & $\substd{2.44}{0.06}$ & $\substd{0.08}{0.01}$ & $\substd{11.10}{0.14}$ & 1.27 & $\substd{0.63}{0.03}$ & $\substd{0.47}{0.03}$ \\
& DEMUCS & $\substd{0.32}{0.01}$ & $\substd{4.13}{0.02}$ & $\substd{0.05}{0.00}$ & $\substd{6.48}{0.11}$ & 1.01 & $\substd{0.73}{0.03}$ & $\substd{0.66}{0.02}$ \\
\midrule

\multirow{3}{*}{OZSpeech}
& Clean & $\substd{0.39}{0.01}$ & $\substd{3.21}{0.03}$ & $\substd{0.06}{0.00}$ & $\substd{6.87}{0.11}$ & 8.75 & $\substd{0.84}{0.02}$ & $\substd{0.64}{0.02}$ \\
\bottomrule
\end{tabular}}
\end{table}

\begin{table}[t]
\centering
\footnotesize
\caption{\textbf{Denoising performance on SPEC with DEMUCS (continued).}}
\setlength{\tabcolsep}{3pt}
\renewcommand{\arraystretch}{1.08}

\resizebox{\textwidth}{!}{%
\begin{tabular}{l l c c c c c c c}
\toprule
\textbf{Model} & \textbf{Condition} 
& \textbf{SIM}$\uparrow$ & \textbf{MOS}$\uparrow$ & \textbf{WER}$\downarrow$ & \textbf{MCD}$\downarrow$ 
& \textbf{RTF}$\downarrow$ & \textbf{SVA}$\uparrow$ & \textbf{EMC}$\uparrow$ \\
\midrule

\multirow{3}{*}{GLM-TTS}
& Clean & $\substd{0.57}{0.01}$ & $\substd{4.08}{0.03}$ & $\substd{0.09}{0.01}$ & $\substd{6.41}{0.11}$ & 1.74 & $\substd{0.95}{0.01}$ & $\substd{0.68}{0.02}$ \\
& SPEC & $\substd{0.31}{0.01}$ & $\substd{2.64}{0.05}$ & $\substd{0.18}{0.01}$ & $\substd{9.07}{0.12}$ & 1.15 & $\substd{0.65}{0.03}$ & $\substd{0.51}{0.03}$ \\
& DEMUCS & $\substd{0.32}{0.01}$ & $\substd{3.41}{0.04}$ & $\substd{0.12}{0.01}$ & $\substd{6.60}{0.11}$ & 4.61 & $\substd{0.71}{0.03}$ & $\substd{0.62}{0.03}$ \\
\midrule

\multirow{3}{*}{VibeVoice}
& Clean & $\substd{0.48}{0.01}$ & $\substd{3.83}{0.04}$ & $\substd{0.23}{0.02}$ & $\substd{6.76}{0.12}$ & 1.86 & $\substd{0.85}{0.01}$ & $\substd{0.62}{0.02}$ \\
& SPEC & $\substd{0.25}{0.01}$ & $\substd{2.63}{0.04}$ & $\substd{0.20}{0.01}$ & $\substd{9.64}{0.14}$ & 2.35 & $\substd{0.50}{0.03}$ & $\substd{0.50}{0.03}$ \\
& DEMUCS & $\substd{0.32}{0.01}$ & $\substd{3.52}{0.04}$ & $\substd{0.18}{0.01}$ & $\substd{6.94}{0.11}$ & 1.06 & $\substd{0.69}{0.03}$ & $\substd{0.57}{0.02}$ \\
\midrule

\multirow{3}{*}{MGM-Omni}
& Clean & $\substd{0.54}{0.01}$ & $\substd{4.28}{0.02}$ & $\substd{0.09}{0.01}$ & $\substd{5.82}{0.10}$ & 0.84 & $\substd{0.93}{0.01}$ & $\substd{0.68}{0.02}$ \\
\bottomrule
\end{tabular}}
\end{table}

\newpage

\begin{table}[p]
\centering
\small
\setlength{\tabcolsep}{4pt}
\caption{\textbf{Compression Robustness (Part 1).} Evaluation under AAC compression.}
\label{tab:robustness_p1}

\begin{tabular}{
  l
  S[table-format=1.3]
  S[table-format=1.2]
  S[table-format=1.3]
  S[table-format=1.3]
}
\toprule
\textbf{Model} & {STOI $\uparrow$} & {MCD $\downarrow$} & {SIM $\uparrow$} & {WER $\downarrow$} \\
\midrule

\multicolumn{5}{l}{\textit{\textbf{Codec: AAC 32k}}} \\
\midrule
FishSpeech & 0.987 & 1.22 & 0.884 & 0.028 \\
Fish Audio S2 & 0.985 & 0.99 & 0.894 & 0.013 \\
OZSpeech & 0.994 & 0.91 & 0.918 & 0.015 \\
StyleTTS2 & 0.984 & 1.08 & 0.896 & 0.020 \\
Spark-TTS & 0.966 & 0.73 & 0.948 & 0.093 \\
MOSS-TTSD v0.7 & 0.973 & 0.58 & 0.918 & 0.048 \\
\bottomrule
\end{tabular}
\end{table}

\clearpage

\begin{table}[p]
\centering
\small
\setlength{\tabcolsep}{4pt}
\caption{\textbf{Compression Robustness (Part 2).} Evaluation under MP3 compression.}
\label{tab:robustness_p2}

\begin{tabular}{
  l
  S[table-format=1.3]
  S[table-format=1.2]
  S[table-format=1.3]
  S[table-format=1.3]
}
\toprule
\textbf{Model} & {STOI $\uparrow$} & {MCD $\downarrow$} & {SIM $\uparrow$} & {WER $\downarrow$} \\
\midrule

\multicolumn{5}{l}{\textit{\textbf{Codec: MP3 32k}}} \\
\midrule
FishSpeech & 0.992 & 1.06 & 0.845 & 0.013 \\
Fish Audio S2 & 0.991 & 0.89 & 0.850 & 0.017 \\
OZSpeech & 0.994 & 1.15 & 0.905 & 0.015 \\
StyleTTS2 & 0.992 & 0.99 & 0.924 & 0.025 \\
Spark-TTS & 0.968 & 0.87 & 0.928 & 0.149 \\
MOSS-TTSD v0.7 & 0.979 & 0.59 & 0.943 & 0.051 \\
\bottomrule
\end{tabular}
\end{table}

\clearpage

\begin{table}[p]
\centering
\small
\setlength{\tabcolsep}{4pt}
\caption{\textbf{Compression Robustness (Part 3).} Evaluation under Opus compression.}
\label{tab:robustness_p3}

\begin{tabular}{
  l
  S[table-format=1.3]
  S[table-format=1.2]
  S[table-format=1.3]
  S[table-format=1.3]
}
\toprule
\textbf{Model} & {STOI $\uparrow$} & {MCD $\downarrow$} & {SIM $\uparrow$} & {WER $\downarrow$} \\
\midrule

\multicolumn{5}{l}{\textit{\textbf{Codec: Opus 16k}}} \\
\midrule
FishSpeech & 0.982 & 1.47 & 0.932 & 0.025 \\
Fish Audio S2 & 0.981 & 1.33 & 0.932 & 0.013 \\
OZSpeech & 0.987 & 1.52 & 0.937 & 0.012 \\
StyleTTS2 & 0.979 & 1.42 & 0.918 & 0.020 \\
Spark-TTS & 0.877 & 1.12 & 0.947 & 0.138 \\
MOSS-TTSD v0.7 & 0.932 & 0.81 & 0.949 & 0.051 \\
\bottomrule
\end{tabular}
\end{table}

\clearpage

\begin{table}[p]
\centering
\small
\setlength{\tabcolsep}{4pt}
\caption{\textbf{Compression Robustness (Part 4).} Evaluation under narrowband telephone transmission.}
\label{tab:robustness_p4}

\begin{tabular}{
  l
  S[table-format=1.3]
  S[table-format=1.2]
  S[table-format=1.3]
  S[table-format=1.3]
}
\toprule
\textbf{Model} & {STOI $\uparrow$} & {MCD $\downarrow$} & {SIM $\uparrow$} & {WER $\downarrow$} \\
\midrule

\multicolumn{5}{l}{\textit{\textbf{Codec: Phone NB}}} \\
\midrule
FishSpeech & 0.987 & 3.79 & 0.667 & 0.019 \\
Fish Audio S2 & 0.987 & 3.49 & 0.701 & 0.011 \\
OZSpeech & 0.984 & 4.25 & 0.697 & 0.010 \\
StyleTTS2 & 0.983 & 3.06 & 0.686 & 0.019 \\
Spark-TTS & 0.966 & 3.37 & 0.680 & 0.175 \\
MOSS-TTSD v0.7 & 0.987 & 2.22 & 0.670 & 0.031 \\
MOSS-TTS v1.5 & 0.987 & 3.93 & 0.719 & 0.019 \\
dots.tts & 0.989 & 3.06 & 0.699 & 0.016 \\
Higgs & 0.978 & 3.29 & 0.684 & 0.106 \\
Higgs TTS 3 & 0.985 & 3.96 & 0.697 & 0.025 \\
CosyVoice & 0.987 & 2.56 & 0.713 & 0.035 \\
GLM-TTS & 0.988 & 2.70 & 0.731 & 0.022 \\
VibeVoice & 0.986 & 3.26 & 0.711 & 0.021 \\
MGM-Omni & 0.968 & 1.81 & 0.736 & 0.290 \\
PlayDiffusion & 0.986 & 5.29 & 0.683 & 0.009 \\
Qwen3-TTS & 0.988 & 2.93 & 0.705 & 0.020 \\
F5-TTS & 0.986 & 3.41 & 0.683 & 0.024 \\
MaskGCT & 0.987 & 2.87 & 0.690 & 0.046 \\
OpenVoice & 0.989 & 3.70 & 0.635 & 0.018 \\
XTTS & 0.987 & 5.15 & 0.688 & 0.025 \\
IndexTTS & 0.987 & 3.17 & 0.711 & 0.023 \\
ZipVoice & 0.988 & 3.12 & 0.673 & 0.050 \\

\bottomrule
\end{tabular}
\end{table}

\clearpage
\newpage

\section{Broader impacts}
\modelname{} may support safer and more reliable voice-cloning systems by exposing robustness failures under realistic deployment shifts, including noisy references, multilingual and long-form generation, post-processing, deepfake detectability, and anti-cloning perturbations. It may also help reveal disparities across accents, languages, and recording conditions. We encourage use of this benchmark with explicit consent norms, privacy protections, careful dataset documentation, and evaluation protocols that distinguish benign robustness from capabilities that facilitate misuse.

\paragraph{Limitations.}
\modelname{} is intended as a diagnostic benchmark for pipeline-level robustness in voice cloning, rather than a complete solution to all open challenges in VCL evaluation. First, our large-scale evaluation necessarily relies on scalable automatic metrics, including speaker-similarity scores, WER, MCD, MOS proxies, and LLM-as-judge based emotion alignment. Although these metrics are widely used and enable broad comparison across many models and robustness conditions, they cannot fully replace human perception. Our pilot human study provides a sanity check, but more extensive human MOS studies and stronger metric validation remain important future work. Second, detector-facing separability should be interpreted as a safety-oriented diagnostic rather than a direct measure of generation robustness. Existing deepfake detectors may be affected by domain factors such as prompt style, recording conditions, and dataset composition; developing detectors that isolate synthesis artifacts from such confounds is beyond the scope of this benchmark and is an important direction for the detection community. Third, some robustness settings admit stronger oracle or control baselines, such as same-speaker real utterance comparisons for upper-bound calibration, ground-truth compression baselines, and more exhaustive out-of-distribution speaker or language coverage. We include controlled re-pairing and diverse public sources to reduce leakage and broaden coverage, but cannot fully eliminate potential overlap between public speech corpora and the training data of modern VCL systems. Overall, these limitations reflect the scope of RVCBench: our goal is to provide a unified, extensible testbed that exposes robustness failures and supports future work on better metrics, stronger detectors, broader human evaluation, and more robust voice-cloning models.

\paragraph{Declaration of LLM Usage.}
The authors used large language models (LLMs) for writing assistance, benchmark prompt construction, and automated evaluation, as described in the paper. All methodology, results, interpretations, and final claims were verified and approved by the authors.



\end{document}